\documentclass[5p]{elsarticle}

\usepackage{lineno,hyperref}
\modulolinenumbers[5]

\journal{Journal of Atmospheric and Solar-Terrestrial Physics}

\biboptions{authoryear}

\begin{document}

\begin{frontmatter}

\title{15\,years of VLT/UVES OH intensities and temperatures in comparison with
TIMED/SABER data}

\author[innsbruck]{Stefan Noll\corref{snoll}}
\cortext[snoll]{Corresponding author}
\ead{stefan.noll@uibk.ac.at}
\author[antofagasta,innsbruck]{Stefan Kimeswenger}
\author[innsbruck,goettingen]{Bastian Proxauf}
\author[innsbruck]{Stefanie Unterguggenberger}
\author[vienna,innsbruck]{Wolfgang Kausch}
\author[garching]{Amy M. Jones}

\address[innsbruck]{Institut f\"ur Astro- und Teilchenphysik, Universit\"at
  Innsbruck, Technikerstr.~25/8, 6020 Innsbruck, Austria}
\address[antofagasta]{Instituto de Astronom\'ia, Universidad Cat\'olica del
  Norte, Avenida Angamos 0610, Antofagasta, Chile}
\address[goettingen]{Max Planck Institute for Solar System Research,
  Justus-von-Liebig-Weg 3, 37077 G\"ottingen, Germany}
\address[vienna]{Department of Astrophysics, University of Vienna,
  T\"urkenschanzstrasse 17, 1180 Vienna, Austria}
\address[garching]{Max Planck Institute for Astrophysics,
  Karl-Schwarzschild-Str.~1, 85748 Garching, Germany}

\begin{abstract}
The high-resolution echelle spectrograph UVES of the Very Large Telescope at
Cerro Paranal in Chile has been regularly operated since April 2000. Thus,
UVES archival data originally taken for astronomical projects but also
including sky emission can be used to study airglow variations on a time scale
longer than a solar cycle. Focusing on OH emission and observations until March
2015, we considered about 3,000 high-quality spectra from two instrumental
set-ups centred on 760 and 860\,nm, which cover about 380\,nm each. These data
allowed us to measure line intensities for several OH bands in order to derive
band intensities and rotational temperatures for different upper vibrational
levels as a function of solar activity and observing date. The results were
compared with those derived from emission and temperature profile data of the
radiometer SABER on the TIMED satellite taken in the Cerro Paranal area between
2002 and 2015. In agreement with the SABER data, the long-term variations in OH
intensity and temperature derived from the UVES data are dominated by the solar
cycle, whereas secular trends appear to be negligible. Combining the UVES and
SABER results, the solar cycle effects for the OH intensity and temperature are
about 12 to 17\% and 4 to 5\,K per 100\,sfu and do not significantly depend on
the selected OH band. The data also reveal that variations of the effective OH
emission layer height and air density can cause significant changes in the OH
rotational temperatures due to a varying ratio of OH thermalising collisions
by air molecules and OH radiation, deactivation, and destruction processes
which impede the rotational relaxation. However, this effect appears to be of
minor importance for the explanation of the rotational temperature variations
related to the solar activity cycle, which causes only small changes in the OH
emission profile.
\end{abstract}


\end{frontmatter}


\section{Introduction}\label{sec:intro}

One of the most prominent nighttime airglow emissions of the Earth's mesopause
region is caused by excited hydroxyl (OH) \citep{meinel50}. It originates at a
typical altitude of 87\,km with a full width at half maximum (FWHM) of about 8
to 9\,km \citep{baker88}. The emission is related to a reaction of atomic
hydrogen and ozone \citep{bates50}, which mainly populates the vibrational
levels $v = 8$ and 9 of the OH electronic ground state \citep[e.g.][]{adler97}.
The airglow radiation originates from these and lower levels, which causes
dozens of molecular emission bands in the optical and near-IR regime
\citep[e.g.][]{rousselot00,cosby07,noll15}. Apart from cascades of radiative
transitions, the redistribution of the vibrational and rotational level
populations is also caused by collisions
\citep{dodd94,adler97,savigny12,xu12,kalogerakis16}. In terms of the
vibrational relaxation, molecular oxygen (O$_2$) is most important. For the
rotational level populations, O$_2$ and the predominating atmospheric
constituent molecular nitrogen (N$_2$) are essential. Collisions involving
atomic oxygen mostly lead to a multi-quantum deactivation of the vibrational
excitation or even destruction of OH. They are crucial for the observed spread
of the emission layer depending on the vibrational level $v'$ of the upper
state of the transition. Higher $v'$ bands emit at higher altitudes. For
$\Delta v' = 1$, the differences amount to several hundred metres
\citep{savigny12,noll16}. 

Apart from the $v'$-dependent OH intensity, which is an important tracer of
atmospheric chemistry and dynamics, OH emission lines are also very valuable
for estimates of the kinetic temperature in the mesopause region. The
widely-used approach \citep[e.g.][]{beig03,schmidt13,reisin14} assumes that the
population distribution over the lowest rotational levels $N$ for a fixed $v$
equals a Boltzmann distribution that corresponds to the ambient temperature
\citep[e.g.][]{beig03,cosby07,khomich08}. This requires a sufficient number of
thermalising collisions in comparison to the $v$-dependent radiative lifetimes
between 5 ($v = 9$) and 58\,ms ($v = 1$) \citep{xu12}. However, a local
thermodynamic equlibrium (LTE) is no longer guaranteed in the highest parts of
the emission layer, where the N$_2$ and O$_2$ densities are low and the
counteracting oxygen atoms are numerous. This can lead to non-LTE contributions
of several Kelvins to the measured OH rotational temperatures $T_\mathrm{rot}$
\citep{cosby07,noll15,noll16}. The exact amount depends on $v'$ and the line
set used. It also varies with time due to changes in the effective height of
the emission layer and the corresponding chemical composition. \citet{noll16}
found that mean nocturnal $T_\mathrm{rot}$ variations for OH bands with high
$v'$ could even be dominated by non-LTE effects.

Investigations of long-term variations of OH intensities
\citep{clemesha05,reid14,gao16} and $T_\mathrm{rot}$
\citep{beig03,beig08,beig11a,beig11b,huang16,kalicinsky16} are particularly
interesting since they provide important information on the influence of the
solar activity and anthropogenic greenhouse gas emissions on the mesopause
region. Such studies require time series which should ideally cover more than
a solar activity cycle of 11\,years. While remote sensing of OH emission is
the approach that is most frequently used to investigate the mesopause region,
suitable data sets are still rare. Therefore, any additional time series of
sufficient length is very valuable, especially in regions of the world with a
lack of data. Moreover, most time series only include a single OH band
\citep[e.g.][]{beig03,schmidt13} and well-calibrated OH intensities are not
always available. Finally, it is questionable how the solar cycle effects and
long-term trends for OH $T_\mathrm{rot}$ are affected by changes in the OH
emission layer structure. This is critical since studies of the long-term
temperature variations at OH layer heights rely on the suitability of OH
$T_\mathrm{rot}$ as a tracer of the kinetic temperature. As discussed before,
this might not be fulfilled.

The Very Large Telescope (VLT) of the European Southern Observatory (ESO) at
Cerro Paranal in Chile (24$^{\circ}$38$^{\prime}$\,S, 70$^{\circ}$24$^{\prime}$\,W)
has regularly been operating the Ultraviolet and Visual Echelle Spectrograph
(UVES) \citep{dekker00} since April 2000. The archival data of this
astronomical instrument allow us to investigate the long-term variations of
several OH bands at a high spectral resolution in parallel. Therefore, we have
selected a UVES sample of 3,113 high-quality spectra comprising the first
15\,years of operation. This data set is the first one covering more than a
solar cycle originating from a site located west of the Andes. In South
America, longer OH time series were only published for El Leoncito
(32$^{\circ}$\,S, 69$^{\circ}$\,W) in Argentina \citep{reisin02,scheer13} and
Cachoeira Paulista (23$^{\circ}$\,S, 45$^{\circ}$\,W) in Brazil
\citep{clemesha05}. Apart from a general characterisation of the solar cycle
effects and long-term trends in the OH intensity and $T_\mathrm{rot}$, our study
focuses on the contribution of varying non-LTE effects to the $T_\mathrm{rot}$
long-term variations. For this purpose and a comparison of complementary
independent data sets, we also investigated OH limb-sounding data
and CO$_2$-based kinetic temperature profiles from the multi-channel radiometer
SABER on the TIMED satellite \citep{russell99}. With the start of the archive
in January 2002, this data set only has a slightly shorter time coverage than
our sample of UVES spectra.

In the following, we will describe the UVES (Section~\ref{sec:uves}) and SABER
data sets (Section~\ref{sec:saber}) for this study in detail. Then, we will
address the measurements of OH line intensities and temperatures
(Section~\ref{sec:measure}), the analysis of the long-term variations for both
data sets (Section~\ref{sec:analvar}), and the estimation of non-LTE effects
related to the measured UVES OH $T_\mathrm{rot}$ (Section~\ref{sec:estimnlte}).
The results for the long-term variations in intensity and temperature,
their relation to other observations, and the impact of non-LTE effects on the
$T_\mathrm{rot}$ trends will be discussed in Sections~\ref{sec:resvar} to
\ref{sec:resnlte}. Finally, we will draw our conclusions
(Section~\ref{sec:conclusions}).

\section{Data sets}\label{sec:data}

\subsection{UVES}\label{sec:uves}

The high-resolution echelle spectrograph UVES \citep{dekker00} mounted at the
VLT covers a maximum wavelength range from 300 to 1100\,nm, where up
to 500\,nm can be observed in parallel by a blue and a red arm, which are fed
by means of a dichroic beam splitter. The exact wavelength coverage depends on
the selected beam splitter and the order-separating cross disperser. For the
investigation of OH emission, we focus on the red arm and set-ups with central
wavelengths of 760 and 860\,nm covering about 380\,nm each. Since the spectra
of the red arm are registered by a mosaic of two charge-coupled devices (CCDs),
there is a gap of the spectral coverage around the central wavelength, which
amounts to about 14\,nm. The combined range of both set-ups between about 570
and 1040\,nm comprises 15 OH bands with $v'$ between 3 and 9
\citep[see][]{noll15}. The overlapping range of the 760 and 860\,nm set-ups
covers six OH bands with $v'$ between 5 and 9. Our study of long-term
variations therefore focuses on the \mbox{OH($v'$-$v''$)} bands \mbox{(5-1)},
\mbox{(6-2)}, \mbox{(7-3)}, \mbox{(8-3)}, and \mbox{(9-4)}, which neglects the
weaker $v' = 7$ band \mbox{OH(7-2)}. \mbox{OH(4-0)} cannot be included since it
is located in the central gap of the 760\,nm set-up. 

\begin{figure}[t]
\begin{center}
\includegraphics[width=88mm]{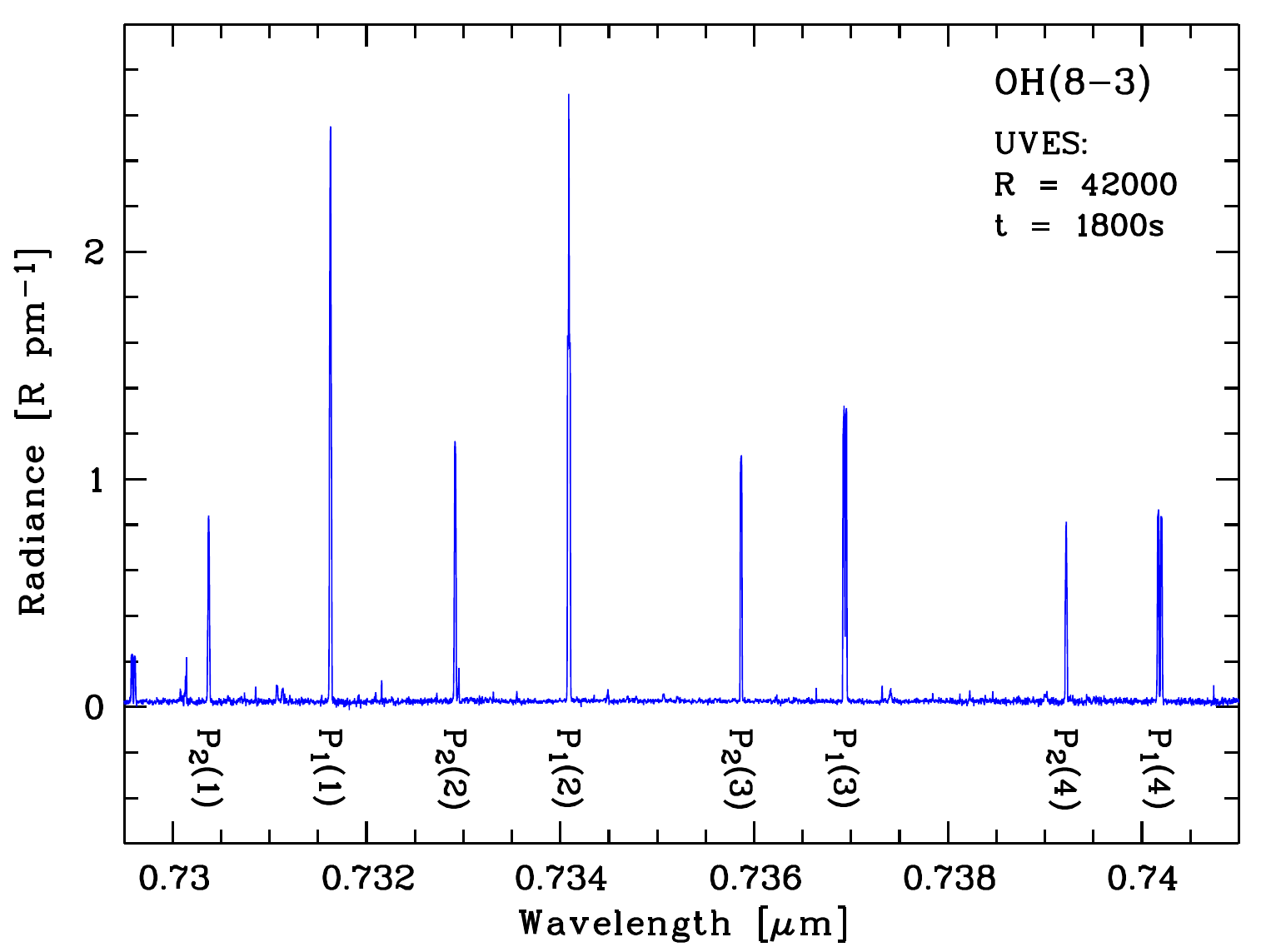}
\end{center}
\caption{P-branch lines of the \mbox{OH(8-3)} band for a typical UVES sample
  spectrum with a standard resolving power of 42,000 and nearly average
  exposure time of 1,800\,s. Radiance and wavelength are given in Rayleighs per
  picometre and micrometres, respectively. Note that the $\Lambda$ doublets
  of the higher P$_1$-branch lines are partially resolved, which makes these
  lines broader and flatter.
\label{fig:spec}}
\end{figure}

The spatial coverage of the sky along the slit axis is $8''$ for the 760\,nm
and $12''$ for the 860\,nm set-up. The slit width can be set freely. For a
standard width of $1''$, the resolving power of the spectrograph is about
42,000. The maximum is 110,000. Therefore, the OH emission lines are usually
well separated, as Fig.~\ref{fig:spec} illustrates for the lower P-branch of
the weakest selected OH band \mbox{(8-3)} in a typical UVES spectrum. Even
the $\Lambda$ doublet structure of the OH lines \citep[e.g.][]{rousselot00} is
partially resolved. Line blends are rare. For a comprehensive overview of UVES
airglow spectra, see \citet{hanuschik03} and \citet{cosby06}.

\subsubsection{UVES data reduction}\label{sec:uvesreduct}

For this study, we used the UVES-related so-called Phase\,3 data products from
the ESO Science Archive Facility. These files were produced by means of the
ESO data reduction pipeline for UVES and contain wavelength-calibrated
one-dimensional spectra with merged echelle orders derived from the initial
two-dimensional raw frames. We only considered those data products which
include a sky-subtracted spectrum of the astronomical target with and without
flux calibration and a spectrum of the sky background derived by the sky
subtraction procedure of the pipeline. The latter exhibits the airglow emission
during the exposure of the spectrum and is therefore the basis for our
analysis. 

The sky data from the UVES Phase\,3 products are not flux calibrated. First, we
corrected these spectra for the variable slit area on the sky in order to be
representative of a unit solid angle. Then, we used the object spectra with and
without flux calibration to obtain instrument response curves. For this, we
also had to consider that the object spectra were corrected for atmospheric
extinction using the extinction curve of \citet{tueg77}. We did not apply this
correction to the sky spectra since losses by scattering and absorption differ
from those related to point-like continuum sources and are also variable with
time. For the investigated wavelength regime, the extinction by atmospheric
scattering can be neglected for the extended airglow emission at Cerro Paranal
\citep{noll12}. The correction of the absorption is described in
Section~\ref{sec:measure}. Depending on instrumental set-up and observing
period, different response curves were applied to the UVES Phase\,3 products.
We identified two response curves for the 760\,nm set-up spectra and six for
those with a central wavelength of 860\,nm. We did not find any significant
long-term trend in the curve shape pointing to instrument and telescope
efficiency changes. The structure of all curves is very similar independent of
the set-up. An exception are wavelength ranges which are affected by strong
molecular absorption by O$_2$ and water vapour \citep[e.g.][]{noll12}. This
suggests that the response curves are not reliable in these ranges due to an
inadequate handling of atmospheric molecular absorption in the
spectrophotometric standard star spectra used for flux calibration. Therefore,
we scaled the individual response curves to a reference one in a safe
wavelength range and then built a master response curve by
reliability-dependent weighted averaging and interpolation in the most affected
regions. The average shape correction for all covered wavelengths and all
spectra is 1.2\%. The differences in the flux calibration for lines of the same
OH band should be significantly smaller.

For flux-calibrated astronomical spectra, the absolute flux uncertainties are
typically of the order of 5 to 10\%. This expectation is in agreement with the
results of a comparison of the OH band intensities of the UVES data (see
Section~\ref{sec:measure}) and those of a sample of spectra taken with the VLT
echelle spectrograph \mbox{X-shooter} \citep{vernet11} discussed in
\citet{noll15,noll16} for the overlap period from October 2009 to March 2013.
For the investigation of long-term variations in intensity, the relative flux
uncertainties have to be distinctly lower. We therefore checked the
flux-calibrated spectra from standard star observations which were Phase\,3
science targets and compared them with each other and to the corresponding
reference fluxes from literature \citep{hamuy94,moehler14}. In general, there
seems to be a good agreement. However, the response curve correction for the
Phase\,3 products does not consider flux differences related to the use of
observing modes with single and double pixel binning. Taking CCD\,1 (with
wavelengths shorter than the central gap) of the more popular mode without
pixel merging as reference, we found a correction factor of $0.944$ for CCD\,2
and $1.038$ for both chips operated with two-pixel binning. Moreover, 760\,nm
spectra taken until 2009 show a conspicuous difference in the flux levels
related to the two CCDs. Since the Phase\,3 products do not contain standard
star observations for this set-up, we compared non-standard star spectra that
were calibrated with both response curves. As a consequence, we multiplied the
fluxes for CCD\,2 by a factor, which was $0.875$ for two-pixel binning. These
corrections reduced the systematic errors significantly. As described in
Section~\ref{sec:regress}, we used the OH line measurements to check the
quality of the calibration of the data related to the different observing modes
and response curves. The resulting relative flux calibration accuracy is about
2\%, which is sufficiently good for studies of long-term variations. Note that
this accuracy estimate is mainly based on averages considering several hundred
spectra. For individual observations or rarely applied combinations of
observing mode and response curve, the uncertainty is probably higher.

\subsubsection{UVES sample selection}\label{sec:uvessample}

This study focuses on the first 15\,years of UVES data, i.e.~spectra taken
until March 2015 were considered. For this period, the number of 760 and
860\,nm spectra exceeds 10,000. Since the analysis of long-term variations for
different OH bands requires a high-quality sample, we reduced the data set by
applying several selection criteria. First, we only considered spectra with a
minimum exposure time of 5\,min to avoid noisy OH lines. This criterion also
rejects the spectra with the brightest astronomical targets. The contamination
of the sky spectra by emission of astronomical objects is critical since the
sky integration procedure of the UVES pipeline only excludes the slit positions
which are closer than 25\% of the slit length to the target centre. For the 760
and 860\,nm set-ups, this limit corresponds to 2 and $3''$, respectively.
Hence, bright targets, extended objects, or a strong blurring of point-like
stellar sources by high atmospheric turbulence can cause a significant
contamination of the sky spectrum, which increases the statistical and
systematic uncertainties related to the OH line measurements depending on the
shape of the object spectrum. For this reason, we applied an additional
selection limit based on the average continuum flux at the positions of the
measured OH lines. Only spectra with less than 2,000
photons\,s$^{-1}$m$^{-2}\mu$m$^{-1}$arcsec$^{-2}$, i.e.~about 11\,R\,nm$^{-1}$,
were considered. This limit still allows data taken in moonlit nights to be
selected. Furthermore, we rejected spectra observed with very wide slits of
more than $2''$. This limit corresponds to a minimum resolving power of about
20,000 and avoids possible issues related to the measurement of broad
box-shaped lines.

The selection criteria listed so far result in a sample of 3,475 spectra. This
data set was then checked for outliers in the derived $T_\mathrm{rot}$ and their
uncertainties (Section~\ref{sec:measure}). For this, we performed an iterative
$\sigma$-clipping procedure, where we applied $5\,\sigma$ limits for the five
OH bands \mbox{(5-1)}, \mbox{(6-2)}, \mbox{(7-3)}, \mbox{(8-3)}, and
\mbox{(9-4)} relevant for the trend analysis in terms of $T_\mathrm{rot}$, the
$T_\mathrm{rot}$ difference compared to \mbox{OH(6-2)}, and the $T_\mathrm{rot}$
uncertainty. The resulting sample optimised for the five bands comprises 1,463
spectra centred on 760\,nm and 1,650 spectra centred on 860\,nm, i.e.~3,113
spectra in total. Note that the identification of outliers is not a simple task
for the investigated sample since the statistical and systematic errors of the
individual $T_\mathrm{rot}$ measurements vary due to differences in the exposure
time, object contamination, spectral resolution, and line intensity. Therefore,
we also performed an alternative selection with the same criteria except for a
$4\,\sigma$ limit for the $T_\mathrm{rot}$ uncertainty. This slight change
reduced the sample to 2,885 spectra, which illustrates that the data
distribution cannot be described by a single Gaussian. In order to investigate
the effect of the sample selection on the analysis of long-term variations, we
performed the entire analysis for both samples. The differences were distinctly
smaller than the estimated errors, which indicates that our results do not
critically depend on the selection criteria. In the following, we will
therefore focus on our main sample of 3,113 spectra.

\begin{figure}[t]
\begin{center}
\includegraphics[width=88mm]{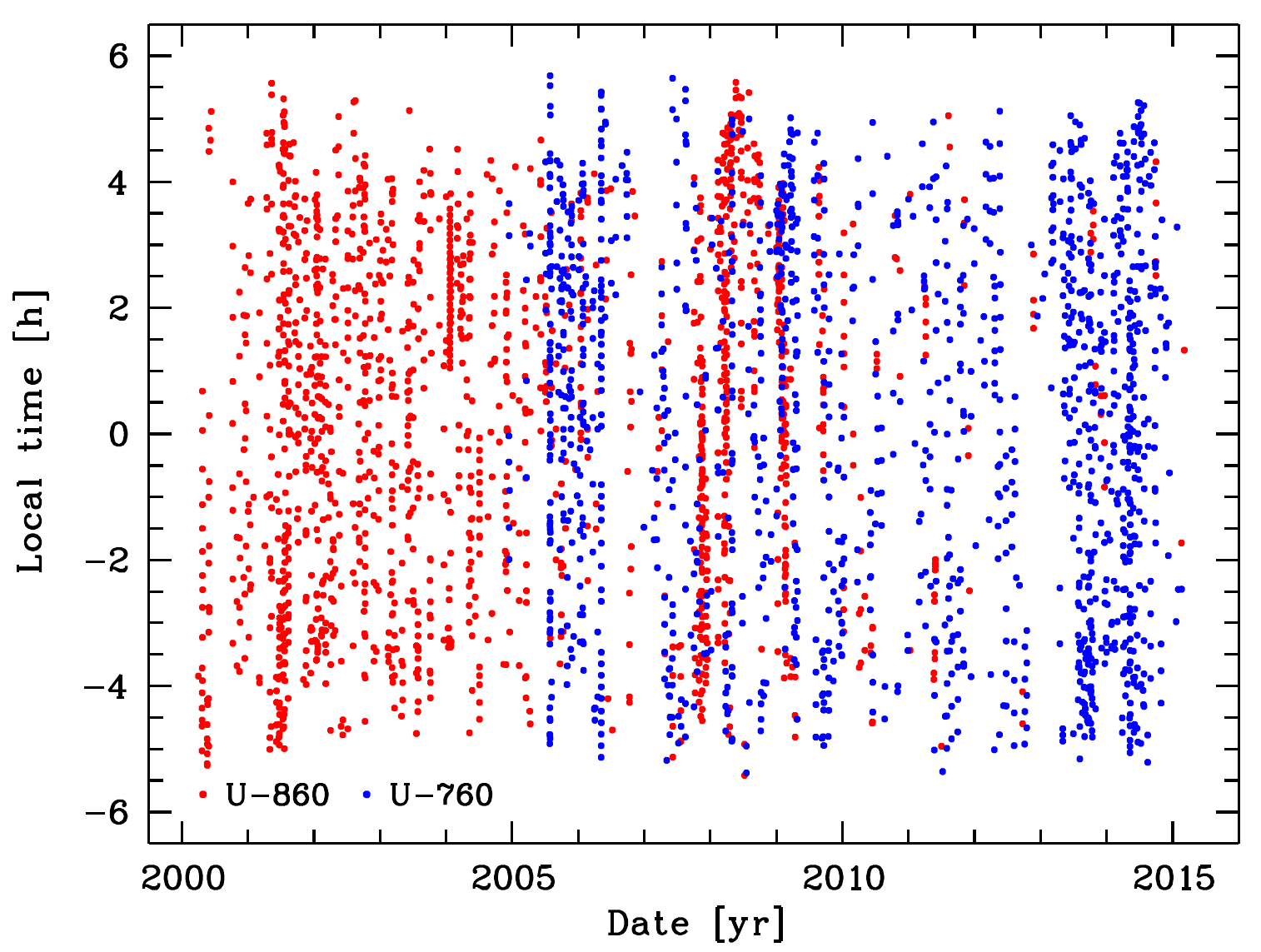}
\end{center}
\caption{Sample of UVES data used for this study. The local time 
  (i.e.~Universal Time corrected for the longitude of Cerro Paranal) of the
  mid-exposure in hours and the date in years are shown for 1,463 spectra
  centred on 760\,nm (blue) and 1,650 spectra centred on 860\,nm (red). 
\label{fig:sample_uves}}
\end{figure}

Fig.~\ref{fig:sample_uves} shows the local solar mean time in hours and the
observing date in years for the mid-exposures of our UVES sample spectra. For
short time scales, the density of data points is highly variable due to the
various possible UVES observing modes, where only a minor fraction fulfils our
selection criteria. Moreover, ESO usually operates two to three competitive
instruments at the same telescope. For longer periods, the variations are
smaller. On an annual basis calculated from April to March, the number of
spectra varies between 83 and 385. The standard deviation is about 50\% of the
mean value. A comparison of the two halves of the covered period shows a
difference of less than 5\% of the sample size. While the full sample indicates
a good coverage of the 15\,years, this is not the case for the subsamples of
the 760 and 860\,nm set-ups. Before 2005, very few 760\,nm spectra were taken.
Then, ESO started to officially offer this set-up, which became eventually more
popular than the 860\,nm mode. Since the calibrations of the spectra of both
set-ups appear to agree very well (see also Section~\ref{sec:regress}), this
small sample overlap does not seem to be critical for the investigation of the
long-term variations. The distribution of local times reflect the fact that
astronomical observations are usually carried out at nighttime. The minimum
solar zenith angle of our sample is 102$^{\circ}$. The few twilight spectra in
the full data set were rejected by our continuum flux limit.

Note that we selected additional set-up-specific subsamples, which were used in
the course of the investigation of non-LTE effects in
Sections~\ref{sec:estimnlte} and \ref{sec:resnlte}. For these subsamples, the
$\sigma$-clipping was performed for a higher number of OH bands than we
considered for our main sample. We will describe these data sets when they are
used.

\subsection{SABER}\label{sec:saber}

\begin{figure}[t]
\begin{center}
\includegraphics[width=88mm]{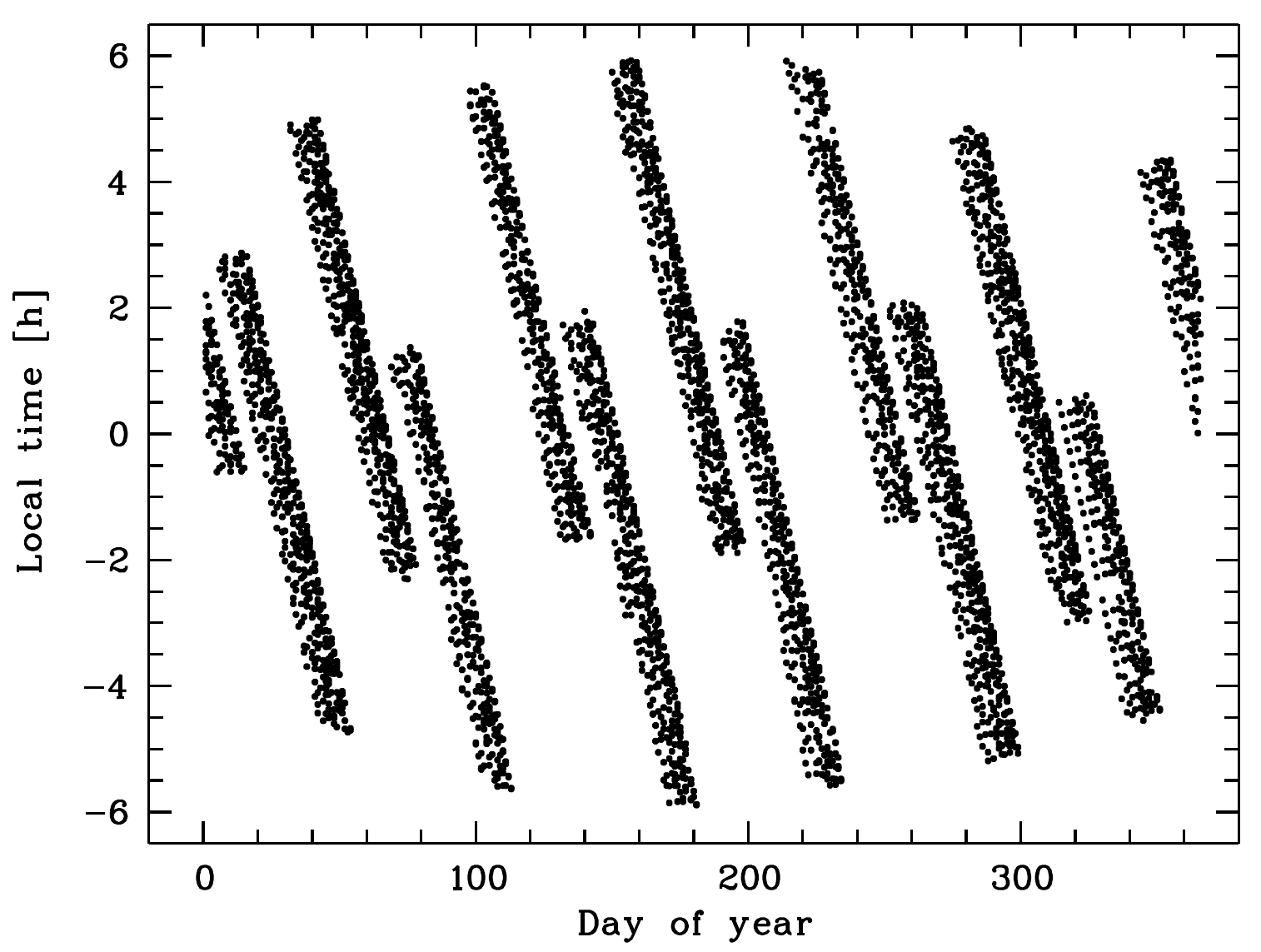}
\end{center}
\caption{Sample of SABER data used for this study. The local time in hours for
  the profile data point at 87\,km and the day of year are shown for
  4,496 profiles. The jumps in local time around midnight for the covered
  stripes are related to the regular yaw manoeuvres of TIMED \citep{russell99}.
\label{fig:sample_saber}}
\end{figure}

The TIMED satellite launched in December 2001 has a slowly-precessing low Earth
orbit with an inclination of 74$^{\circ}$ and a period of 1.6\,h and achieves
global coverage (except for high latitudes) within 60\,days \citep{russell99}.
SABER performs limb-sounding in ten channels between 1.27 and 15\,$\mu$m with a
vertical resolution of 2\,km. For this study, we focus on the two OH-related
channels centred on 1.64 and 2.06\,$\mu$m. Since each channel covers lines of
three OH bands, the effective $v'$ correspond to $4.6$ and $8.3$
\citep{noll16}. Moreover, we consider the CO$_2$-related kinetic temperature
$T_\mathrm{kin}$ products. Their elaborate retrieval involves non-LTE radiative
transfer calculations \citep{rezac15}. For the OH volume emission rate (VER)
and $T_\mathrm{kin}$ profiles, we use the version 2.0 products from the SABER
data archive for the period from January 2002 (start of the archive) to
December 2015. Hence, the SABER time coverage is slighter shorter and shifted
to later dates compared to our UVES sample (Section~\ref{sec:uves}). Moreover,
only nocturnal profiles corresponding to a minimum solar zenith angle of
100$^{\circ}$ were considered. In order to focus on the VLT region, the profile
selection was restricted to the latitude range from 23 to 26$^{\circ}$\,S and
plus/minus 10$^{\circ}$ around the longitude of Cerro Paranal of
70.4$^{\circ}$\,W. This selection results in a total number of 4,496 profiles.
More details on the selection criteria and the SABER profile characteristics
are given by \citet{noll16}, who performed the same analysis for a SABER data
set comprising the years from 2009 to 2013.

Fig.~\ref{fig:sample_saber} shows the local times and days of year for our
SABER sample. The diagram area is only sparsely filled due to the orbital
parameters of TIMED. For a specific day of year, one or two narrow time windows
are available. The width of the ranges depends on the number of included years.
Despite the large gaps, the SABER data are promising for the analysis of
long-term variations in comparison to our UVES data set
(Fig.~\ref{fig:sample_uves}) since the pattern of coverage is very regular with
the mean local time exactly at midnight and a mean day of year of 180. The
shift of the annual pattern, which results in $-12$ to $-11$\,days for the
covered period of 14\,years, should not be critical.

\section{Analysis}\label{sec:analysis}

\subsection{Measurements of intensities and temperatures}
\label{sec:measure}

The derivation of UVES OH intensities and $T_\mathrm{rot}$ was carried out
in a similar way as described in \citet{noll15}, where 343 VLT \mbox{X-shooter}
spectra with a lower resolving power (3,000 to 18,000) but a wider wavelength
coverage ($0.3$ to $2.5$\,$\mu$m) were investigated in terms of average
nocturnal and seasonal variations.

The line measurements focus on the first four lines ($N^{\prime} \le 4$) of the
OH P$_1$ and P$_2$ branches (see Fig.~\ref{fig:spec}). The transitions of both
branches cause an increase of the total angular momentum and involve the OH
$X^2{\Pi}_{3/2}$ and $X^2{\Pi}_{1/2}$ substates, respectively. The considered
lines are relatively strong, well separated, and usually used for
$T_\mathrm{rot}$ determinations \citep[e.g.][]{beig03,cosby07}. It is not
possible to measure reliable intensities of all eight lines for each OH band
due to contamination by lines of other bands or severe absorption of the
emission at lower altitudes of the atmosphere. In principle, the former issue
is less critical for UVES than for X-shooter. However, we tried to use the same
line selection as described in \citet{noll15} for a better comparison and an
easier analysis (see also Section~\ref{sec:corrtemp}). For this reason, we did
not consider bands where the UVES spectral coverage was not sufficient to
measure all required lines. This criterion resulted in nine OH bands for each
of the two set-ups and six bands present in the spectra of both set-ups (see
Section~\ref{sec:uves}). The OH bands \mbox{(6-1)}, \mbox{(8-2)}, \mbox{(9-3)}
could only be measured in the 760\,nm spectra, whereas \mbox{(3-0)},
\mbox{(4-0)}, and \mbox{(8-4)} were only accessible in the spectra taken with
the 860\,nm set-up. \mbox{OH(9-5)} was rejected due to the location of the
P$_1$($N^{\prime} = 1$) line in a gap between two echelle orders, which can
happen in the reddest part of the UVES wavelength range. Using the line sets of
\citet{noll15}, the number of considered lines is between 4 and 8. On average,
$6.4$ P-branch lines were measured.

While strongly absorbed emission lines were excluded from the analysis, lines
with a moderate absorption up to several per cent were corrected. As discussed
in \citet{noll15}, the derivation of the effective line-specific absorption
requires the calculation of OH line profiles and high-resolution atmospheric
transmission spectra. For the former, Doppler-broadened Gaussian profiles for
a typical temperature of 190\,K \citep[c.f.][]{noll16} were assumed. The OH
line positions originate from the HITRAN2012 database \citep{rothman13}. HITRAN
is also used as the input for the transmission calculations via the radiative
transfer code LBLRTM \citep{clough05}, where Cerro Paranal atmospheric mean
conditions were assumed \citep{noll12}. The resulting line transmissions were
corrected for the strongly varying water vapour \citep[see][]{noll15}. The
column density of this trace gas was derived for each spectrum individually.
The continua of the astronomical objects observed in parallel with the airglow
emission were sufficiently bright to apply the fitting code {\tt molecfit}
\citep{smette15,kausch15}. The fit for the water vapour absorption was
performed in the wavelength range from 814.0 to 818.3\,nm, which is covered by
both UVES set-ups. The absorption correction also considers the dependence of
the optical depth of the absorber on the zenith angle of the telescope, which
varied between 1 and 69$^{\circ}$ with a mean value of 36$^{\circ}$. The formula
of \citet{rozenberg66} for air was applied for this purpose. The corresponding
correction for the OH emission is based on the formula by \citet{vanrhijn21}
for thin layers and a reference altitude of 87\,km. For the latter, the mean
and maximum horizontal distances of the observed emission from Cerro Paranal
are about 70 and 220\,km, respectively. Hence, our UVES-based results are
related to a geographical area which includes oceanic and mountainous desert
regions.

The integration of the line intensities was performed as follows. After the
subtraction of the continuum by means of a wide median filter, the line
profiles were integrated within spectral intervals with a width of about
3\,FWHM and centred on the HITRAN wavelengths. The separation of the $\Lambda$
doublet was considered for the derivation of the integration limits. In the
case of a clear separation of the lines, both components were integrated
separately. The final line intensities were then derived by applying the
corrections described above to the individual $\Lambda$ doublet components.
For unresolved doublets, the measured intensity was halved for this procedure.
For the analysis of the intensity long-term variations, the intensities of all
measured lines of an OH band were summed up.    

The intensities $I$ of all measured $\Lambda$ doublets of an OH band were used
to calculate band-specific $T_\mathrm{rot}$. Assuming a Boltzmann distribution
of the population distribution, this can be done by performing a regression
analysis for the expression $\ln(I/(Ag'))$, including the Einstein~$A$
coefficient and the statistical weight of the upper state $g'$, as a function
of the energy of the upper state of the transition $E'$ \citep{mies74,noll15}.
The reciprocal of the resulting slope corresponds to $T_\mathrm{rot}$ times a
constant. For $A$, $g'$, and $E'$, we used the OH data in the HITRAN2012
database \citep{rothman13}, which are essentially based on the calculations of
\citet{goldman98}. Note that the molecular parameters for OH are still
relatively uncertain \citep{goldman98,pendleton02,khomich08,brooke16}.
Depending on the data source, the P-branch-related $T_\mathrm{rot}$ results can
differ by several Kelvins \citep{cosby07,noll15,liu15,parihar17}. Thus, the
derived non-LTE contributions to these $T_\mathrm{rot}$ and their change with
$v'$ (see Section~\ref{sec:resnlte}) are also affected. According to
\citet{noll15,noll16}, the HITRAN data provide promising results if only
P-branch lines of the same OH band are considered. For this reason and the
comparability with the X-shooter-based results, we focused on HITRAN-related
molecular data for this study. Moreover, the possible constant temperature
offsets are not an issue for the analysis of temporal $T_\mathrm{rot}$
variations. Therefore, the conclusions drawn from this study on long-term
variations do not depend on the set of molecular parameters used. 

SABER OH VER profile data are available for two channels with effective $v'$ of
$4.6$ and $8.3$ (Section~\ref{sec:saber}). For a comparison with the long-term
variations of the UVES-related OH intensities, we performed a vertical
integration of the VER profiles for altitude ranges limited by a maximum
deviation of 15\,km from the mean of both interpolated half-maximum VER
heights. For the OH-related temperatures, we used the CO$_2$-based kinetic
temperature $T_\mathrm{kin}$ profiles and weighted them with the corresponding
VERs of the two OH-related channels. We considered the same variable altitude
limits as for the VER integration for this calculation in order to avoid
systematic errors due to non-zero SABER OH VERs far away from the emission
layer \citep[see also][]{noll16}. The resulting effective temperatures
$T_\mathrm{eff}$ show systematic uncertainties of a few Kelvins \citep{rezac15}.
Similar to $T_\mathrm{eff}$, we calculated the effective emission height
$h_\mathrm{eff}$ and decadal logarithm of the number density of air molecules
$(\log n)_\mathrm{eff}$, which are used for the discussion of $T_\mathrm{rot}$
non-LTE effects in Section~\ref{sec:resnlte}.

\subsection{Analysis of long-term variations}\label{sec:analvar}

\begin{figure*}[t]
\begin{center}
\mbox{    
\includegraphics[width=90mm]{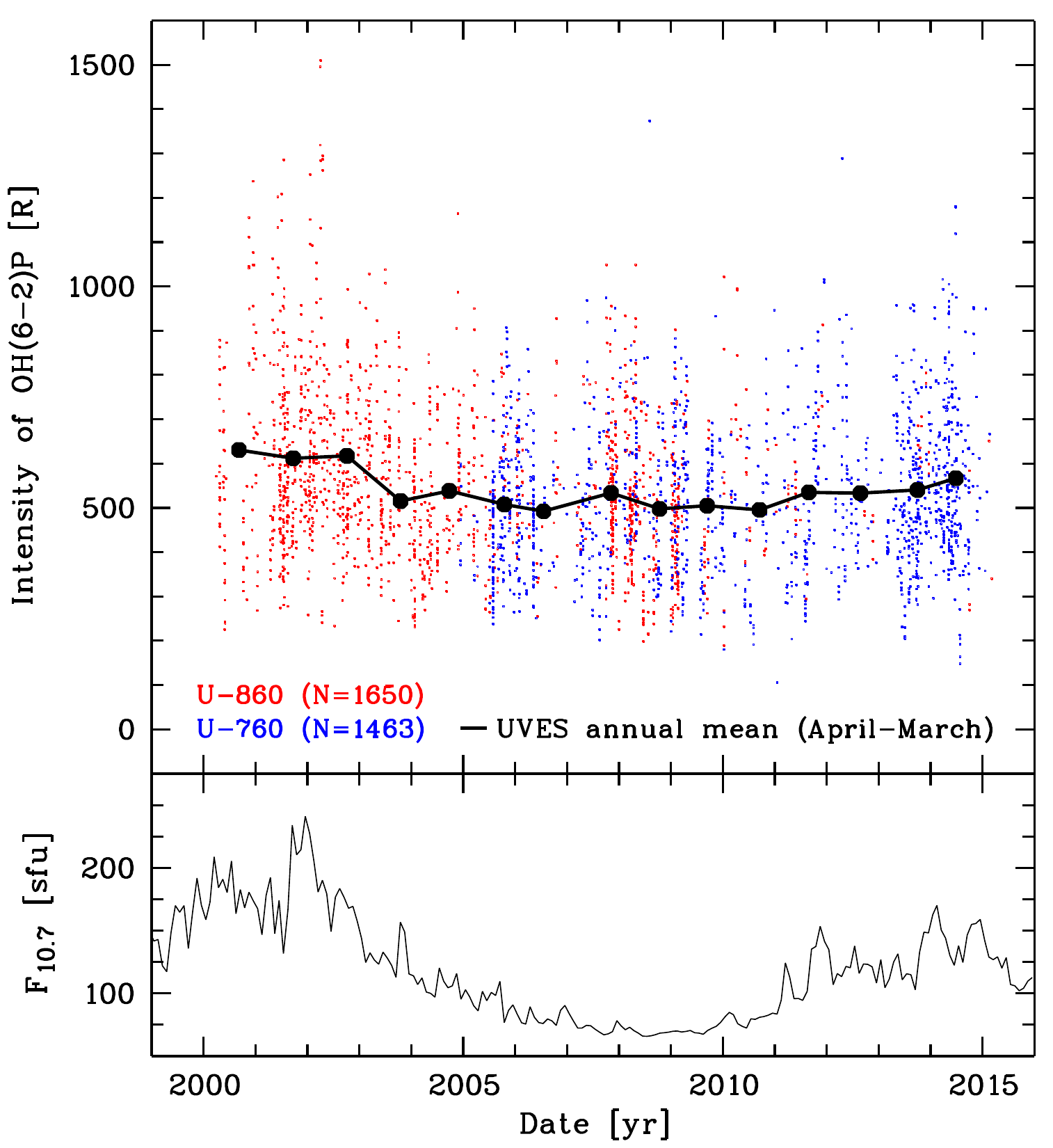}
\includegraphics[width=90mm]{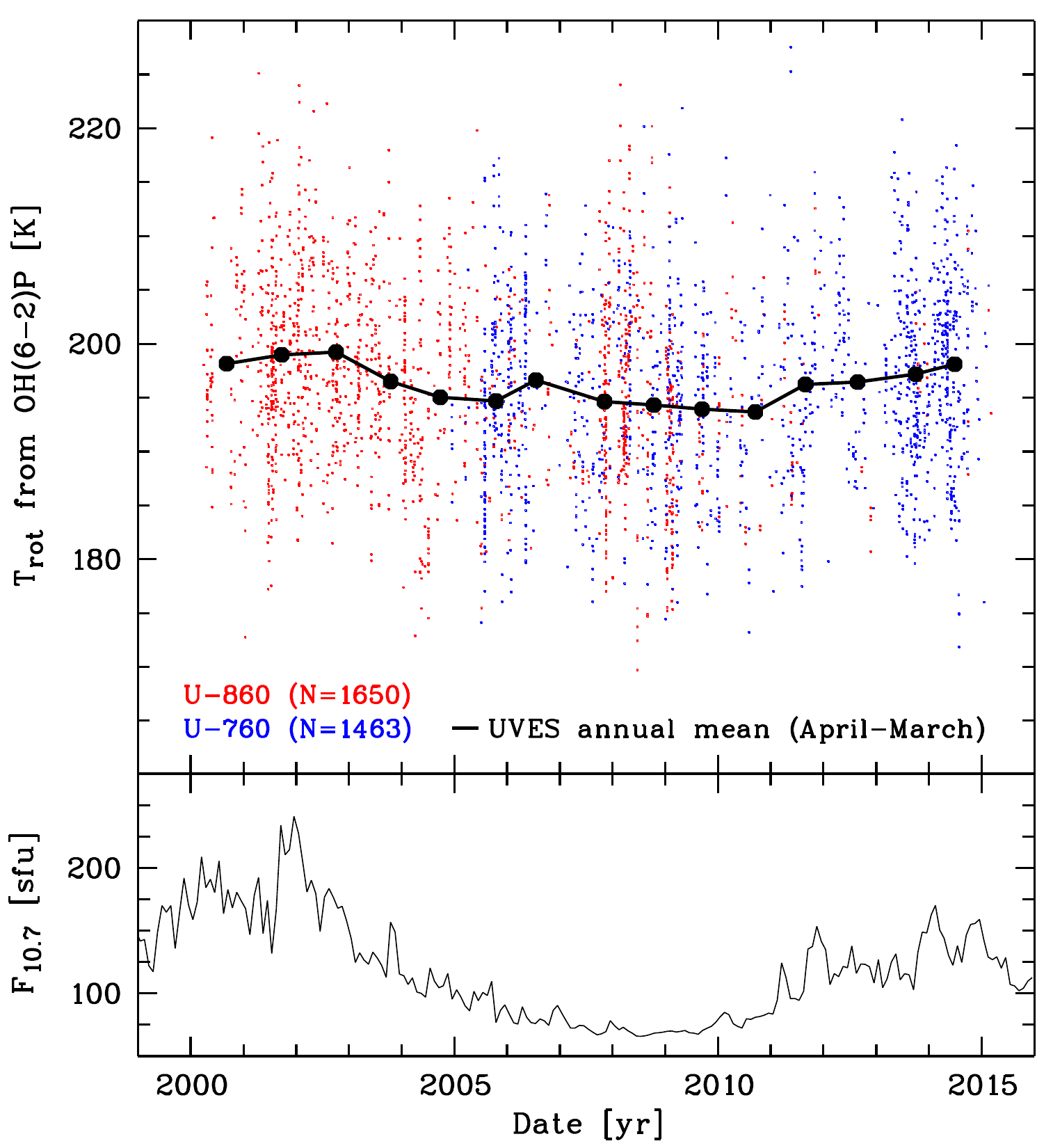}}
\end{center}
\caption{Integrated intensity in R (left) and $T_\mathrm{rot}$ in K (right)
  for the four lowest P$_1$ and P$_2$ branch lines of \mbox{OH(6-2)}. The UVES
  sample is the same as in Fig.~\ref{fig:sample_uves}. Both subfigures also
  show annual mean values averaged for the period from April to March.
  Inhomogeneities in the intra-annual data distribution (e.g.~in the year 2006)
  were not corrected. The lower panels display the monthly mean solar radio
  flux at 10.7\,cm in sfu for the period covered by the UVES spectra.
\label{fig:val_date}}
\end{figure*}

\begin{figure*}[t]
\begin{center}
\mbox{    
\includegraphics[width=90mm]{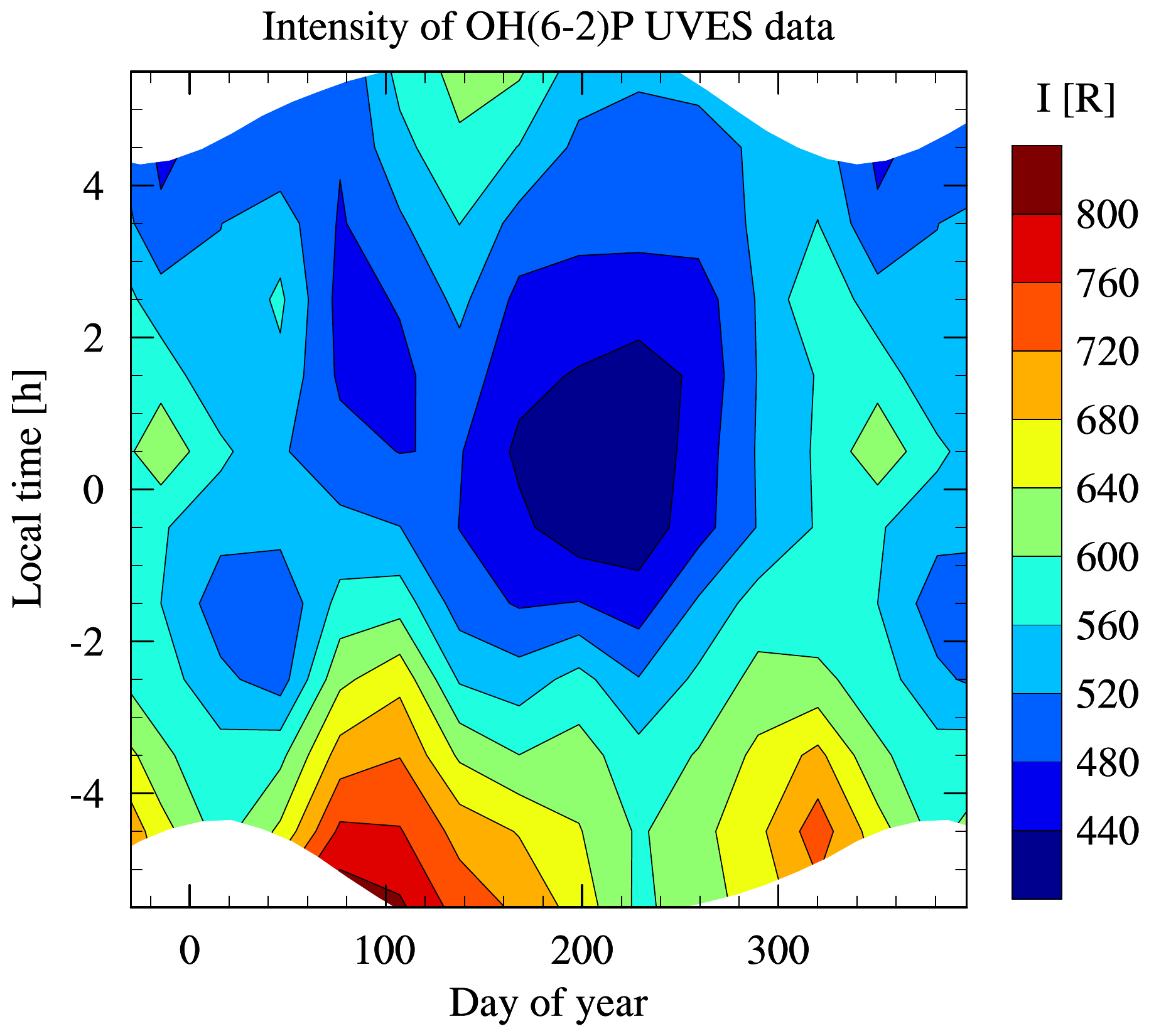}
\includegraphics[width=90mm]{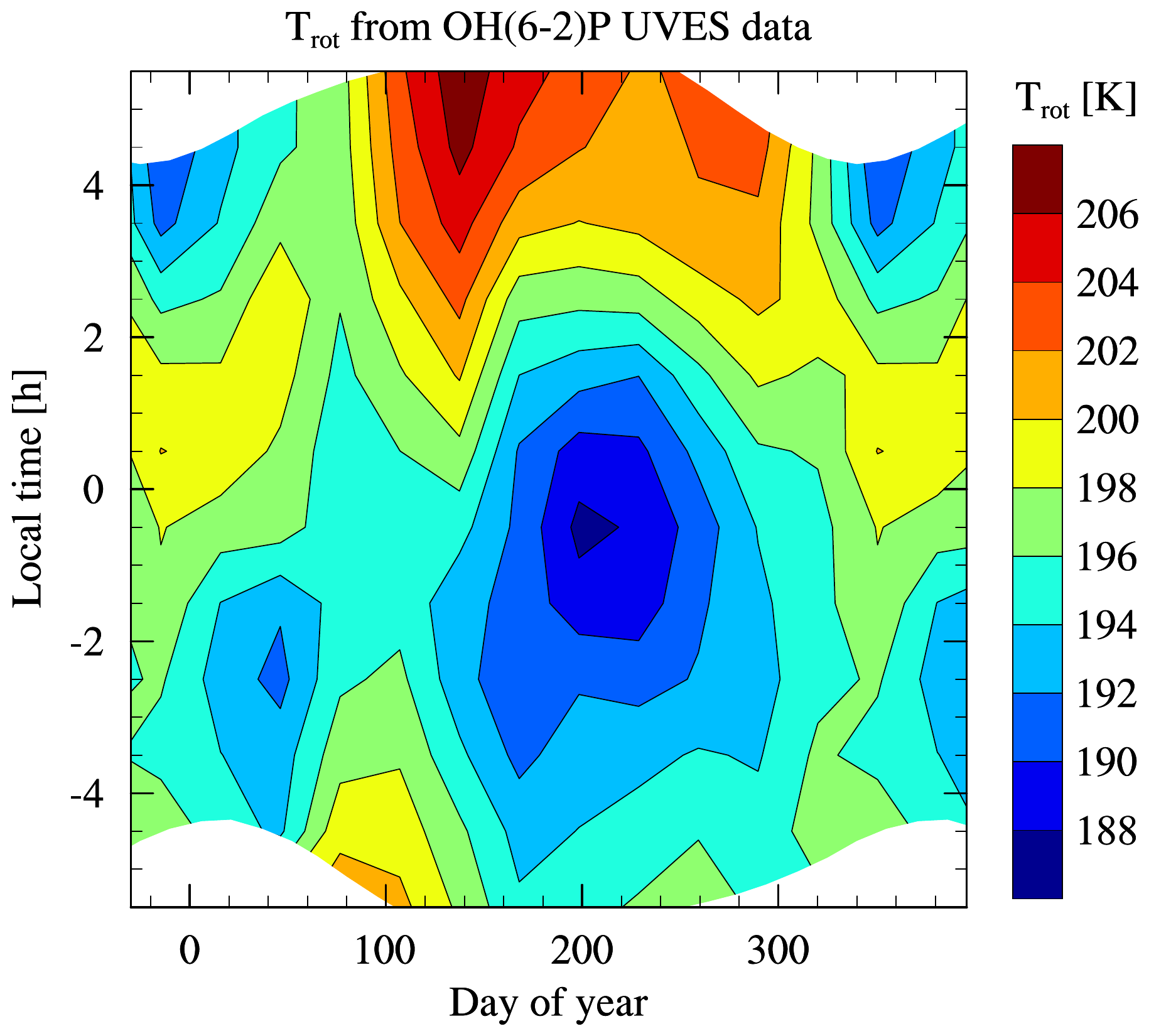}}
\end{center}
\caption{Contour plots of the integrated intensity in R (left) and
  $T_\mathrm{rot}$ in K (right) of the measured \mbox{OH(6-2)} P-branch lines
  (see Fig.~\ref{fig:val_date}) for a $12 \times 12$ grid of local times in
  hours and days of year. The grid points centred on the middle of month and
  hour were derived by means of smoothing of the 3,113 UVES data points
  (Fig.~\ref{fig:sample_uves}) via a two-dimensional Gaussian with $\sigma$ of
  half a month and half an hour. The contours are only shown for nighttime
  conditions with solar zenith angles larger than 100$^{\circ}$, which are
  present in the UVES data set.
\label{fig:doy_lt}}
\end{figure*}

\begin{figure*}[t]
\begin{center}
\mbox{    
\includegraphics[width=90mm]{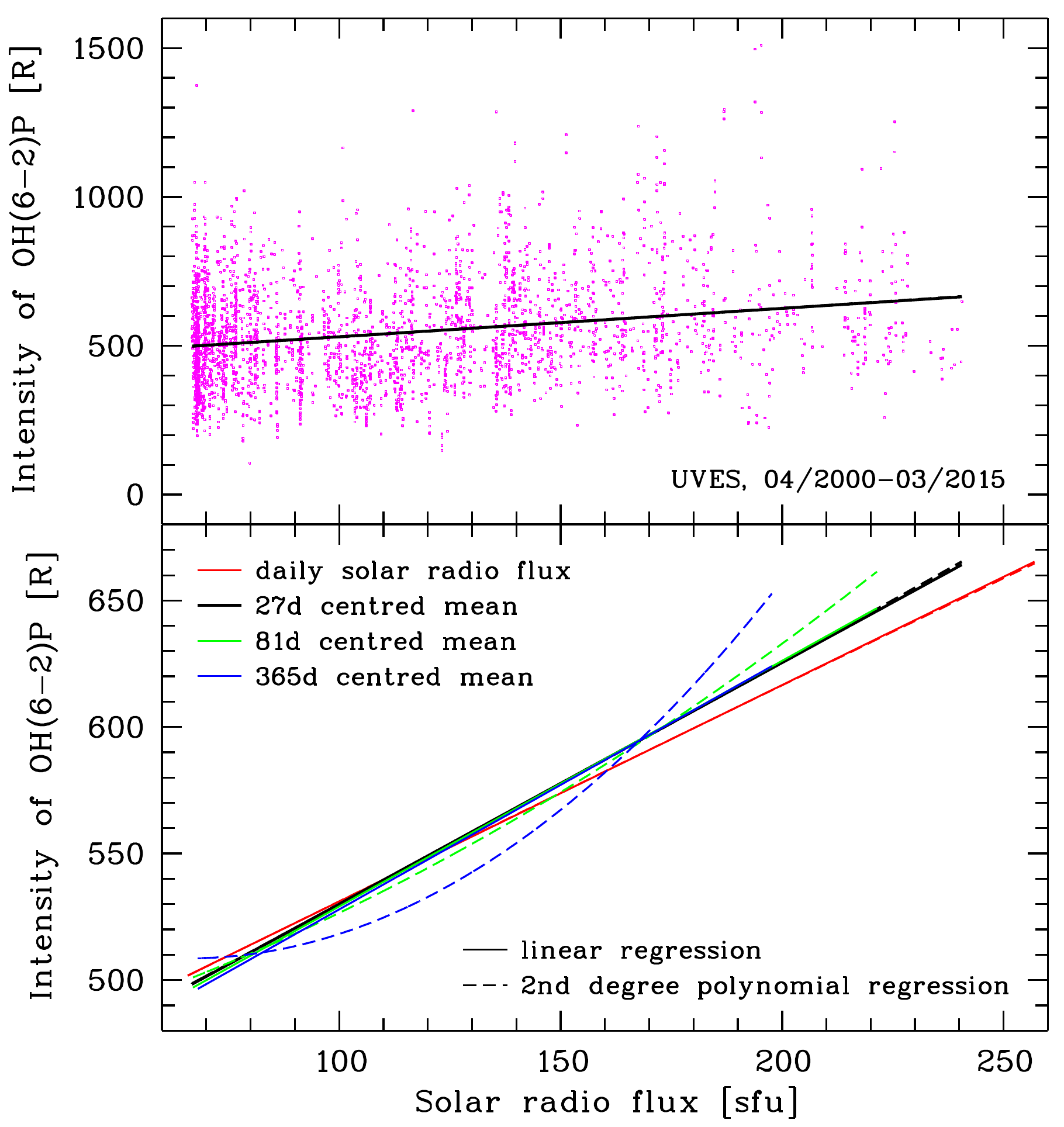}
\includegraphics[width=90mm]{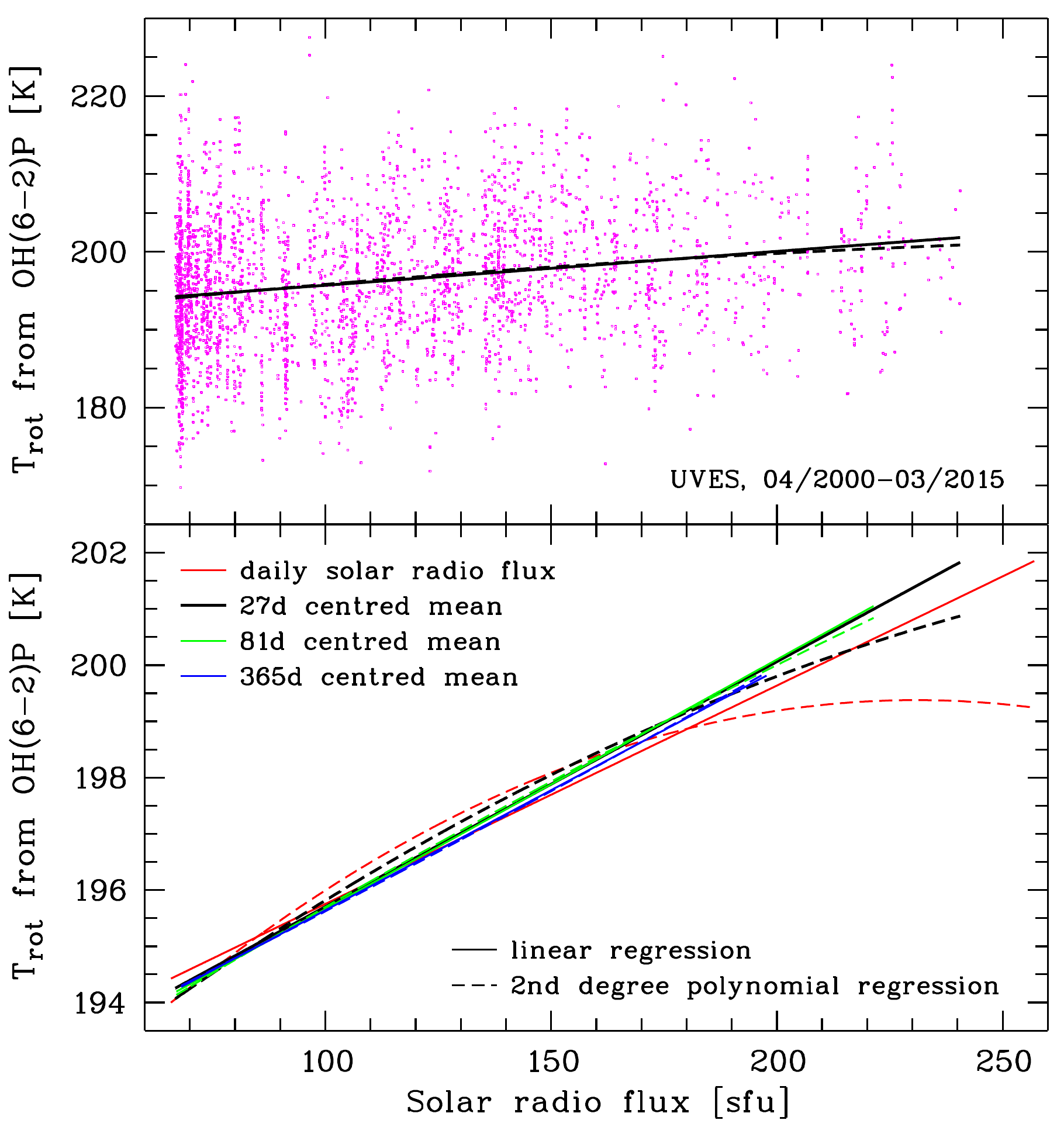}}
\end{center}
\caption{Regression analysis of the relation of the integrated intensity in R
  (left) and $T_\mathrm{rot}$ in K (right) of the measured \mbox{OH(6-2)}
  P-branch lines (see Fig.~\ref{fig:val_date}) with the solar radio flux in
  sfu for the full UVES sample (Fig.~\ref{fig:sample_uves}). The individual
  data points for a centred 27-day mean of the solar radio flux $F_{10.7}$
  ($S_\mathrm{27d}$) are plotted in the upper panels, which also show the
  corresponding regression line (solid) and best second degree polynomial
  (dashed). The lower panels compare the first and second degree regression
  results for centred mean $F_{10.7}$ for 1, 27, 81, and 365\,days.
\label{fig:val_srf}}
\end{figure*}

Example time series of the intensity and $T_\mathrm{rot}$ measurements
discussed in Section~\ref{sec:measure} are shown in Fig.~\ref{fig:val_date} for
\mbox{OH(6-2)} and the main sample of 3,113 spectra
(Section~\ref{sec:uvessample}). For \mbox{OH(6-2)}, the full set of eight
P-branch lines can be used. This results in a mean integrated intensity of
540\,R and a standard deviation of 170\,R. The corresponding values for
$T_\mathrm{rot}$ are $196.3$ and $8.6$\,K. The mean error related to the
regression-based $T_\mathrm{rot}$ derivation for \mbox{OH(6-2)} is $2.5$\,K,
which should be considered as an upper limit for statistical uncertainties
because of deviations of the rotational level populations from a Boltzmann
distribution due to non-LTE effects (see Section~\ref{sec:corrtemp}). In any
case, the time-dependent variations in Fig.~\ref{fig:val_date} are distinctly
stronger than the measurement uncertainties. Moreover, the scatter plots
indicate a trend with time for both quantities. The values tend to be higher at
the beginning and the end of the time series. This is convincingly confirmed by
annual mean values where we considered periods from April to March in order to
avoid incomplete years. For the intensity and $T_\mathrm{rot}$, the four years
with the highest mean values are 2000, 2001, 2002, and 2014. These are also
those years with the highest solar activity as measured by the solar radio flux
at 10.7\,cm \citep{tapping13}, which is a good tracer of the long-term
variations of the OH-relevant solar extreme UV (EUV) flux
\citep{dudok08,dudok09}. It is most frequently used for studies of solar cycle
effects in the mesopause region \citep{beig08,beig11b}. The monthly mean values
of this $F_{10.7}$ index are shown in the lower panels of
Fig.~\ref{fig:val_date}. They are given in solar flux units (sfu), which
correspond to 10\,kJy. Neglecting a possible long-term trend, a simple linear
regression analysis for the annual mean values of the \mbox{OH(6-2)} intensity
divided by the 15-year mean and the $F_{10.7}$ index results in $19 \pm 2$\,\%
per 100\,sfu. For $T_\mathrm{rot}$, a similar analysis provides $4.1 \pm 0.5$\,K
per 100\,sfu. Both results show a significant solar cycle effect for
\mbox{OH(6-2)}.

In the following subsections, we discuss a more elaborate approach to study the
solar cycle effect, the long-term trend, and measurement uncertainties in
parallel. Our procedure uses the individual observations instead of annual mean
values since these data contain more information on the variation with respect
to the solar activity. Moreover, the inhomogeneities in the data samples can
better be handled in this way.

\subsubsection{Correction of data gaps}\label{sec:detrend}

For the UVES sample, we performed a detrending of the data set in terms of
nocturnal and intra-annual variations as a first step in order to avoid
possible biases due to data gaps (see Fig.~\ref{fig:sample_uves}). For this
purpose, we created a $12 \times 12$ grid of hours and months. The intensities
and $T_\mathrm{rot}$ for the grid points in the bin centres were then calculated
by weighted averaging of the values for the individual measurements of the
whole sample, which neglects second order effects by year-to-year changes of
the nocturnal and seasonal variability pattern. The weights were derived by
means of a two-dimensional Gaussian with $\sigma$ of half an hour and half a
month. In this way, data points within a $1\,\sigma$ distance from the grid
point have similar weight, which is comparable with the result for simple
binning and averaging if the data density is high. For sparsely populated
cells, the approach makes sure that a sufficient number of values from
neighbouring cells contribute to avoid possible issues with low number
statistics. Scaling the weight of the centre of the Gaussian to 1, the average
weight sum for the relevant nighttime grid points (excluding twilight) was 39
with a standard deviation of 13 for the UVES main sample. 

As an example, Fig.~\ref{fig:doy_lt} shows the grid-based climatologies for
the \mbox{OH(6-2)} intensity and $T_\mathrm{rot}$, which exhibit a dynamical
range of the nocturnal variations of about 80\% of the mean and 20\,K,
respectively. These are effects which are distinctly larger than the amplitude
of the long-term variations. Therefore, differences in the distribution of the
data over the 15-year period for the individual grid points could be neglected
for the detrending. Note that the seasonal variability pattern significantly
depends on the local time for both quantities, which justifies our
two-dimensional detrending approach. While an annual oscillation in intensity
and $T_\mathrm{rot}$ with southern hemisphere winter minimum dominates around
midnight, a semi-annual oscillation (SAO) tends to be more relevant in the
evening and morning. The high impact of the SAO with maxima around the
equinoxes and the dependence of the amplitude and phase of the nocturnal
variability pattern on season are in good agreement with other OH intensity and
$T_\mathrm{rot}$ measurements at similar latitudes as Cerro Paranal
\citep{takahashi95,takahashi98,gelinas08}. 

The detrending of the UVES data was carried out based on the differences
between the closest climatology grid point and the annual nighttime mean. The
analysis of the long-term variations with and without detrending for the
different OH bands then showed the impact of this correction, which is about
$-0.5$\% and $+0.3$\,K per 100\,sfu for the solar cycle effect on average. For
the long-term trends, the mean shifts are $+0.8$\% and $-0.1$\,K per decade.
The change in the $T_\mathrm{rot}$ solar cycle effect is the most significant.
It is of the same order as the uncertainties of the regression analysis
(Sections~\ref{sec:regress} and \ref{sec:resvar}). The relatively high
influence of the removal of the nocturnal and seasonal variations in this case
is probably related to the importance of the relatively small fraction of the
observations at high $F_{10.7}$, which is more affected by inhomogeneities than
the full sample.

We did not perform a similar correction for the SABER sample since the data set
is more homogeneous despite the significant gaps (Fig.~\ref{fig:sample_saber}),
which suggests that the average nocturnal and seasonal variations have only a
small influence if data from different years are compared.

\subsubsection{Solar cycle proxy}\label{sec:proxy}

While the parameter for studying the long-term trend is simply the date of
observation, the situation is less clear for the solar cycle effect. As already
stated, we focus on the solar radio flux index $F_{10.7}$. The use of individual
observations instead of annual mean values as discussed before requires to
average this daily index in an optimal way. The daily measurements are less
suitable to trace the solar EUV flux between 10 and 121\,nm than averages over
a longer period. According to \citet{wintoft11}, centred averages covering at
least half a solar rotational period of 27\,days appear to be of similarly good
quality. In order to keep the averaged period short, we therefore selected the
centred 27-day mean of $F_{10.7}$ ($S_\mathrm{27d}$) as our proxy for the solar
cycle effect.

In Fig.~\ref{fig:val_srf}, we show the effect of this decision on the analysis
of the UVES \mbox{OH(6-2)} intensity and $T_\mathrm{rot}$ data as given
in Fig.~\ref{fig:val_date}. For a comparison, the linear regression lines for
averages of $F_{10.7}$ over 1, 27, 81, and 365\,days are plotted. As expected,
the lines for the daily $F_{10.7}$ exhibit the largest discrepancy. They are
also those with the highest rms (root mean square). The second highest rms are
found for the 365-day mean ($S_\mathrm{365d}$). However, the differences in the
slope compared to $S_\mathrm{27d}$ are only about $+0.3$\% and $-0.1$\,K per
100\,sfu. Very small discrepancies are also found for $S_\mathrm{81d}$. Hence,
the choice of $S_\mathrm{27d}$ is sufficient and not critical for the analysis
of the solar cycle effect. Fig.~\ref{fig:val_srf} also illustrates the
reliability of the assumption of a linear relation by showing the results for a
second order polynomial regression. In general, these curves are close to those
for the linear relations, which supports the linear regression approach.
Exceptions are present for $T_\mathrm{rot}$ versus $S_\mathrm{1d}$ and intensity
versus $S_\mathrm{365d}$. The latter might be related to the non-Gaussian
distribution of intensities due to a high intensity tail and the relatively
narrow range of $S_\mathrm{365d}$ values due to the smoothing of the high solar
activity peaks. In this respect, a linear regression is certainly more robust.

\subsubsection{Regression analysis}\label{sec:regress}

Our analysis of the OH intensity and temperature long-term variations in
our UVES and SABER samples is based on bilinear regressions involving
the parameters observing date and $S_\mathrm{27d}$. In the case of UVES, we
also considered the impact of flux calibration uncertainties on the results.

In order to handle systematic errors related to the UVES $T_\mathrm{rot}$
measurements for the five investigated OH bands, we assumed that there are
possible deviations in the response curves for spectra of the 760 and 860\,nm
set-ups (see Sect.~\ref{sec:uvessample}). If this assumption was correct, it
would be critical for the analysis, especially for the long-term trend, due to
the very different periods covered. The offset between the $T_\mathrm{rot}$ of
both set-ups was derived for the best overlapping years from 2005 to 2010, when
742 spectra centred on 760\,nm and 638 spectra centred on 860\,nm were taken.
After the correction of the derived $T_\mathrm{rot}$ offsets, the regression
analysis was performed for the full sample. The resulting solar cycle effects
and long-term trends were then used to detrend the 2005-to-2010 data for
another refined derivation of the set-up-related $T_\mathrm{rot}$ differences.
Then, the regression analysis was repeated again for the offset-corrected
$T_\mathrm{rot}$ of the full sample. The convergence of this iterative procedure
was quick due to the low solar activity variations in the comparison period.

The final offsets of the 760\,nm $T_\mathrm{rot}$ with respect to the 860\,nm
ones are between $-0.7$ and $+0.4$\,K with a sample-related uncertainty of
$0.5$\,K for all bands, i.e.~the differences are not significant. Their impact
on the results for the solar cycle effect is negligible. The long-term trends
for the five OH bands studied increase by about $0.2$\,K per decade on average.
The mean absolute difference of $0.3$\,K per decade is about half the
regression uncertainties. This is also the impact of the uncertainty of the
$T_\mathrm{rot}$ correction of $0.5$\,K. Consequently, the combination of 760
and 860\,nm data only causes a minor increase in the uncertainties related to
the long-term trends in $T_\mathrm{rot}$.

The handling of systematic flux calibration errors relevant for the OH
intensity analysis is more complex since eight instrumental response curves
(two of them for the 760\,nm data) were used for the calibration of the UVES
Phase\,3 products in different periods. Moreover, there are the UVES operation
modes with single and double pixel binning. Finally, the red-arm spectra are
registered by two independent CCDs (see Section~\ref{sec:uves}). This results
in 32 different flux calibration cases containing 7 to 585 spectra of the UVES
main sample. We can assume that the deviations are constant factors since
wavelength-dependent variations were already corrected (see
Section~\ref{sec:uvesreduct}). In order to estimate the flux calibration
accuracy and the corresponding uncertainties for the analysis of long-term
intensity variations, we performed the bilinear regression analysis for the
intensities of the five selected UVES OH bands. Then, we measured the detrended
mean intensities for the different combinations of response curve and binning.
We especially focused on the bright \mbox{OH(7-3)} band, which is always on
CCD\,2. For CCD\,1, we measured \mbox{OH(7-2)}, which originates from the same
upper vibrational level $v' = 7$. Therefore, it could be detrended with the
results for \mbox{OH(7-3)} and the intensity ratio of both bands should be
constant, which makes it relatively easy to detect flux calibration errors.

The resulting flux error weighted for the number of spectra in each subsample
is 2.0\% for CCD\,1 and 1.8\% for CCD\,2, which verifies that our flux
calibration corrections discussed in Section~\ref{sec:uvesreduct} were
sufficient. The deviations could even be lower since the statistical
uncertainties due to the sample intensity variance also contribute, especially
in the case of subsamples with a small number of observations. Moreover, the
assumption of a linear model for the solar cycle effect and the long-term trend
has some influence (see Section~\ref{sec:proxy}). Next, we used the differences
between the mean intensities for each subsample and an (uncritical) arbitrary
reference to study their impact on the trend analysis. For this purpose, we
subtracted them from and added them to the corresponding individual intensity
measurements for each considered band. The differences in the results of a
subsequent regression analysis for both modified data sets then indicated
the influence of the flux calibration uncertainties on the trend parameters.
In general, there is a non-neglibible contribution to the final uncertainties
in the solar cycle effect and long-term trend related to the intensities of
the five investigated OH bands (see Section~\ref{sec:resvar}). The amount of
these systematic effects ranges from 0.35 to 1.76 times the corresponding
errors of the regression analysis. The only ratio which is distinctly higher
than unity is related to the solar cycle effect for \mbox{OH(8-3)}. This could
be related to the fact that this band is the only one that is always on CCD\,1,
which suggests that the results for this chip and the 760\,nm set-up appear to
be particularly sensitive to flux calibration uncertainties.

\subsection{Estimation of non-LTE effects}\label{sec:estimnlte}

Apart from the characterisation of the long-term variations in UVES and SABER
OH data, this study also focuses on an analysis of the contribution of varying
non-LTE effects to the UVES OH $T_\mathrm{rot}$ solar cycle effects and
long-term trends.

\subsubsection{Correction of systematic temperature errors}\label{sec:corrtemp}

In order to quantify the $v'$-dependent non-LTE effects, it is required to
correct systematic errors in the measured $T_\mathrm{rot}$ of different OH
bands. There are instrument-, reduction-, and analysis-related uncertainties.
Moreover, the measurements were carried out with band-dependent line sets (see
Section~\ref{sec:measure}). In this case, the resulting $T_\mathrm{rot}$ differ
due to specific non-LTE contributions for each OH line. For the line sets used,
this effect can easily amount to several Kelvins \citep{noll15}. Based on
\mbox{X-shooter} data, \citet{noll15} could correct the different systematics
by studying 25 OH bands with various possible line sets simultaneously. The
UVES data allow us to investigate 12 bands, i.e.~nine bands in each set-up. A
$\sigma$-clipping approach as described in Section~\ref{sec:uvessample}
resulted in 1,333 and 1,471 spectra centred on 760 and 860\,nm with reliable
$T_\mathrm{rot}$ for all measured bands. Studying systematic $T_\mathrm{rot}$
errors only based on the 12 OH bands in the UVES data would probably lead to
less accurate results than the \mbox{X-shooter}-related analysis. Therefore, we
combined the UVES and \mbox{X-shooter} data to obtain improved reference
$T_\mathrm{rot}$ for each $v'$.

A joint analysis of both data sets requires the correction of $T_\mathrm{rot}$
differences by discrepancies in the time coverage. The 343 \mbox{X-shooter}
spectra were taken between October 2009 and March 2013, which is much shorter
than the 15\,years of UVES. Therefore, we took the results of our analysis of
long-term variations (Sections~\ref{sec:analvar} and \ref{sec:resvar}) to
correct the solar cycle effects and long-term trends in the \mbox{X-shooter}
and two UVES data sets. Since the $v'$ of the bands to be corrected range from
2 to 9 but the trend parameters only cover the range from 5 to 9, we used
$v'$-independent average coefficients. The reference $S_\mathrm{27d}$ index and
date for the results are 100\,sfu and New Year of 2011. The nocturnal and
seasonal variations can be corrected as described in
Section~\ref{sec:detrend}. Since the variability pattern significantly depends
on $v'$ \citep[see][]{noll15}, we prepared a sample of 1,526 spectra of the
860\,nm set-up optimised for the OH bands \mbox{(3-0)}, \mbox{(4-0)},
\mbox{(5-1)}, \mbox{(6-2)}, \mbox{(7-3)}, \mbox{(8-3)}, and \mbox{(9-4)}.
Assuming that the nocturnal and seasonal variations are similar for the
different samples, we directly corrected the OH bands with $v'$ between 3 and
9. For the \mbox{OH(2-0)} band, which is only present in the \mbox{X-shooter}
data, we applied the \mbox{OH(3-0)}-related climatology. For the correction of
the differences in the band-specific line sets, we used the $T_\mathrm{rot}$
shifts derived by \citet{noll15}, which are between $-10.4$ and $-3.4$\,K for
the OH bands covered by the UVES data. Moreover, we changed the reference line
set from the first three P$_1$- and P$_2$-branch lines to P$_1$($N' = 1$),
P$_1$(2), and P$_1$(3) only. According to \citet{noll16}, this is related to a
general $T_\mathrm{rot}$ increase of $3.0$\,K. Afterwards, we derived the
sample- and band-related mean $T_\mathrm{rot}$.

\begin{figure}[t]
\begin{center}
\includegraphics[width=88mm]{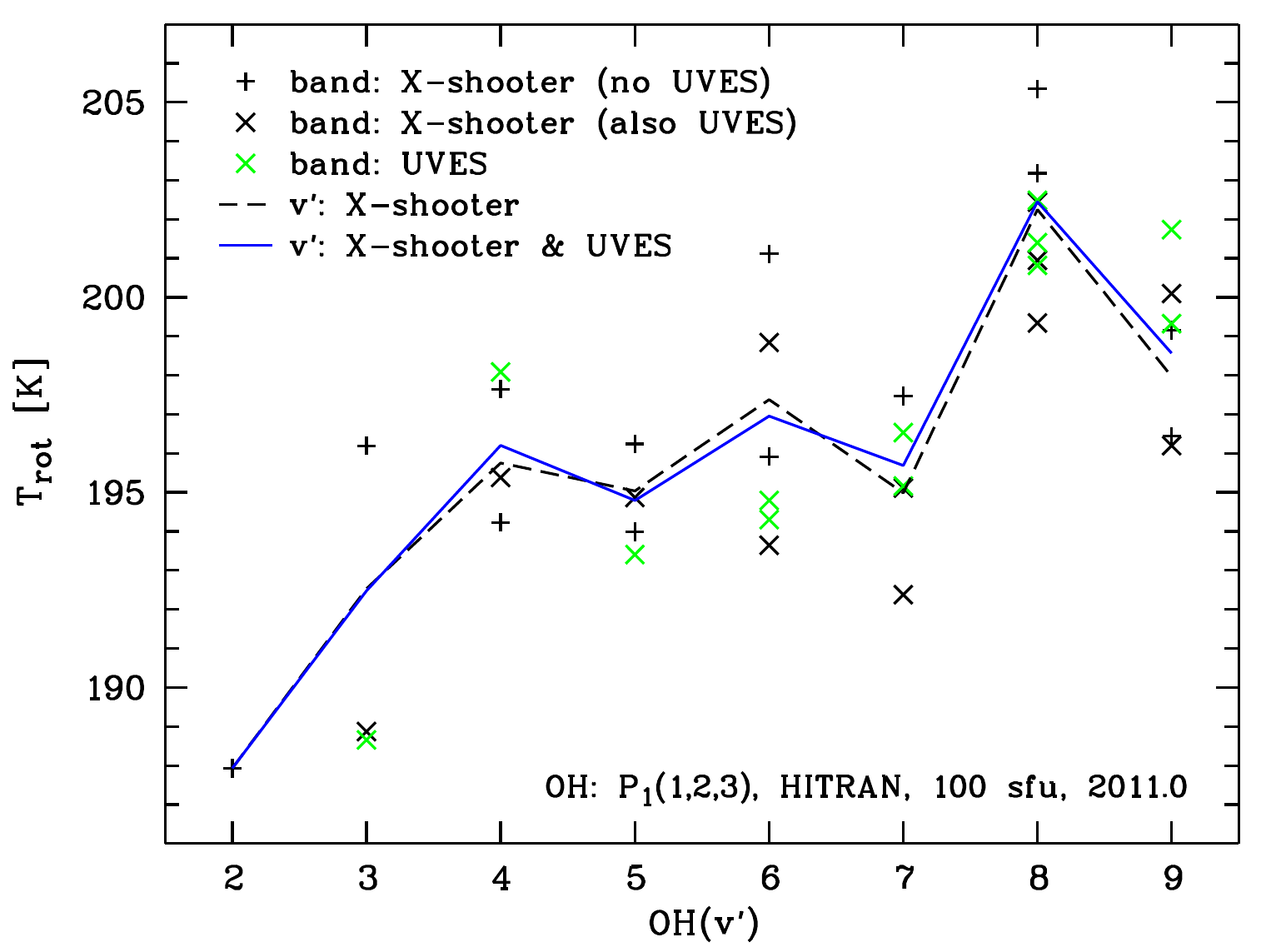}
\end{center}  
\caption{Mean $T_\mathrm{rot}$ in K for 25 OH bands based on the UVES data set
  of this study (green $\times$ symbols) and the \mbox{X-shooter} data set of
  \citet{noll15} (black $\times$ and + symbols, the latter indicating bands
  without UVES data). All mean values refer to the first three P$_1$-branch
  lines, HITRAN molecular parameters, a reference solar radio flux of 100\,sfu,
  and the start of year 2011. The plot also shows average $T_\mathrm{rot}$ for
  fixed $v'$. While the dashed curve connects mean values only considering
  \mbox{X-shooter} data, the solid curve is related to mixed UVES and
  \mbox{X-shooter} results, where first the band-specific and then the
  $v'$-related averages were calculated.  
\label{fig:tbandmean_vp}}
\end{figure}

Fig.~\ref{fig:tbandmean_vp} shows the results of these calculations.
$T_\mathrm{rot}$ for the 25 \mbox{X-shooter}- and 12 UVES-related OH bands are
indicated. For the six bands which are present in the spectra of both UVES
set-ups, we give the arithmetic mean of the two measurements. In the next step,
we averaged the band-specific $T_\mathrm{rot}$ for both instruments, which
indicate a mean absolute difference of 1.9\,K. Finally, we derived the mean
values of all bands with the same $v'$. The scatter of the band-specific
$T_\mathrm{rot}$ is only 2.2\,K. The resulting $T_\mathrm{rot}$($v'$) function
reveals a trend of increasing $T_\mathrm{rot}$ with increasing $v'$ and higher
temperatures for even $v'$ compared to the results for adjacent odd $v'$. A
clear maximum is found for $v' = 8$. This remarkable pattern was first
identified by \citet{cosby07} and further studied by \citet{noll15,noll16}
based on the discussed \mbox{X-shooter} data. It can only be explained by
considering non-LTE effects. Adding the UVES data to the \mbox{X-shooter} data
does not significantly change the structure of the function. The deviation is
less than 1\,K for each $v'$. We can use the final $T_\mathrm{rot}$($v'$)
function to correct the systematic uncertainties in the band-specific
$T_\mathrm{rot}$. This works quite well at least for $v' \ge 4$, where data of
several bands can be combined. The resulting absolute corrections for the 12
UVES OH bands were 1.7\,K on average.

\subsubsection{Linking UVES with SABER temperature data}\label{sec:linking}

\citet{noll16} showed that the non-LTE effects vary with time and the
variations are weaker for OH bands with lower $v'$. Therefore, the comparison
of $T_\mathrm{rot}$ for very different $v'$ can be used to study the variability
of the non-LTE contributions. For this purpose, it is necessary to eliminate
the differences due to the $v'$-dependent VER-related weighting of the
altitude-dependent kinetic temperatures $T_\mathrm{kin}$. This can be achieved
by using the SABER OH VER and $T_\mathrm{kin}$ profiles
(Section~\ref{sec:saber}). Following \citet{noll16}, we calculated
$v'$-dependent normalised OH VER profiles based on the two SABER channels with
effective $v'$ of $4.6$ and $8.3$ by assuming a positive linear relation
between emission height and $v'$ \citep[see][]{savigny12} and by averaging the
profile shapes of both OH VER profiles. For our SABER sample, the mean
difference in the VER-weighted effective emission height $h_\mathrm{eff}$ is
$0.39$\,km for $\Delta v' = 1$. The scatter amounts to $0.15$\,km. Next, we
derived VER-weighted effective temperatures $T_\mathrm{eff}$ for each $v'$ by
combining OH emission and $T_\mathrm{kin}$ profiles. The resulting mean
$T_\mathrm{eff}$ difference for $\Delta v' = 1$ based on the $T_\mathrm{eff}$ for
$v' = 2$ and 9 is only $0.01$\,K, i.e.~there is no significant temperature
gradient for the OH emission altitudes on average. Although the scatter of
$0.56$\,K is distinctly higher, this means that for long-term averages the
effect of the $T_\mathrm{kin}$ gradient on the OH $T_\mathrm{rot}$ measurements
seems to be relatively small. Consequently, the structures in
Fig.~\ref{fig:tbandmean_vp} should be dominated by non-LTE effects.

The difference of the $T_\mathrm{eff}$ for two given $v'$ can be used to remove
the effect of the ambient temperature profile on the difference of the
$T_\mathrm{rot}$ for OH bands with the same two $v'$. For this, the individual
UVES observations have to be combined with suitable SABER profiles. Since an
exact match is not possible, each UVES $T_\mathrm{rot}$ was linked to SABER
$T_\mathrm{eff}$ data at similar observing times and days of year. Following
\citet{noll16}, this was done as described for the climatology grid in
Section~\ref{sec:detrend} but with the times and dates of the individual UVES
observations as reference instead of a regular grid. Unlike \citet{noll16},
where only 3.5\,years of data were studied, we also considered the solar cycle
effect. For this, the results of a linear regression of the SABER
$T_\mathrm{eff}$ data and $S_\mathrm{27d}$ were used. The solar cycle correction
was calculated by deriving the $S_\mathrm{27d}$ difference between a UVES
observation and the corresponding SABER-weighted mean and applying the
regression slope. Apart from $T_\mathrm{eff}$, the SABER-based parameters
$h_\mathrm{eff}$ and $(\log n)_\mathrm{eff}$ (see Section~\ref{sec:measure}) were
processed in the same way to link UVES and SABER data. It is clear that
the agreement between the considered quantities of both data sets cannot be
particularly good if single UVES observations are studied. The weighted
averaging of the SABER data cancels out aperiodic variations. For the weight
sum defined in Section~\ref{sec:detrend}, we obtain a mean of 60 and a scatter
of 27. Therefore, the described procedure is only reliable for studies of large
samples. This criterion is fulfilled for the discussion in
Section~\ref{sec:resnlte}.

\section{Results and discussion}\label{sec:results}

\subsection{Long-term variations}\label{sec:resvar}

\begin{figure*}[t]
\begin{center}
\mbox{    
\includegraphics[width=90mm]{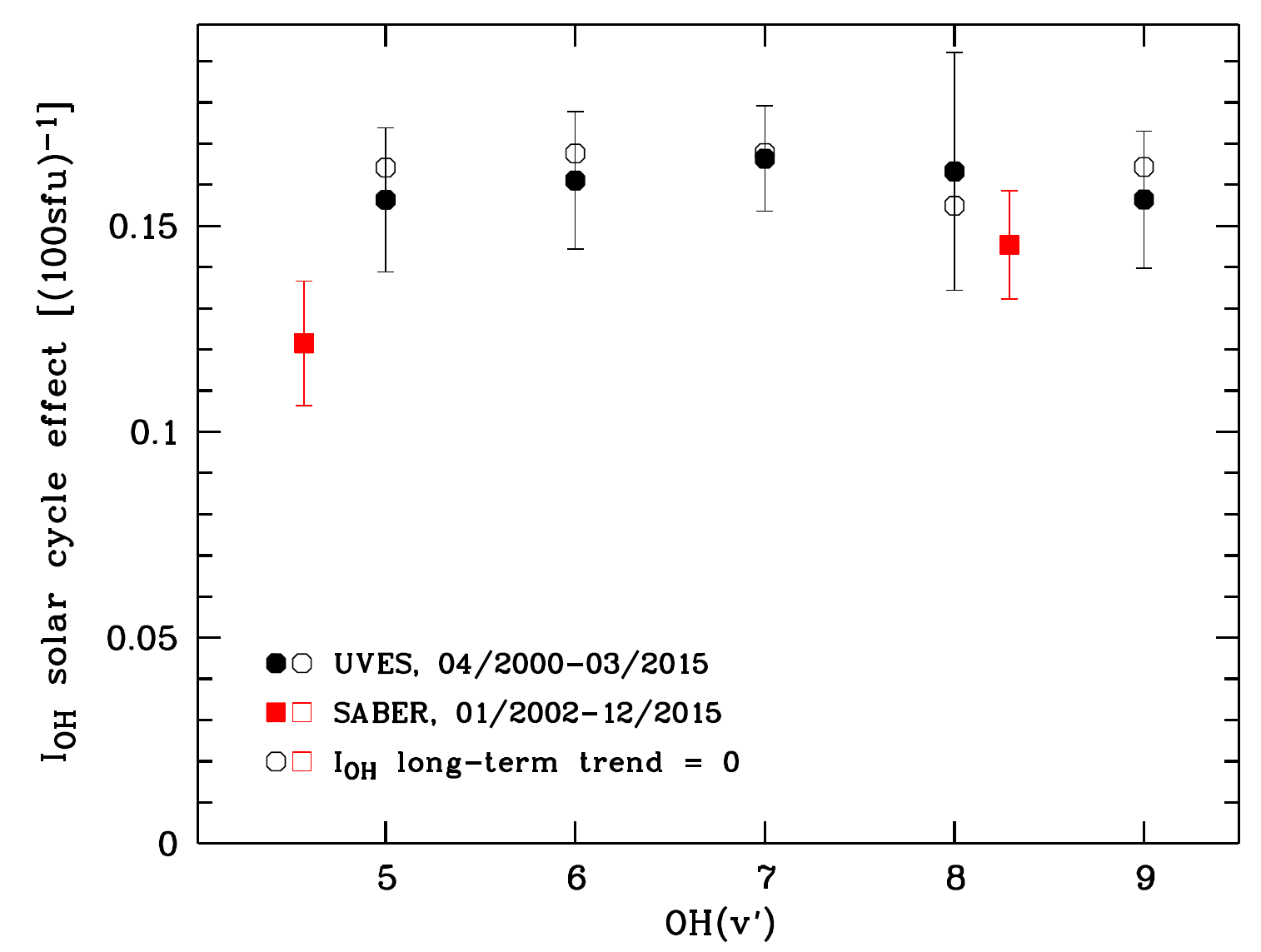}
\includegraphics[width=90mm]{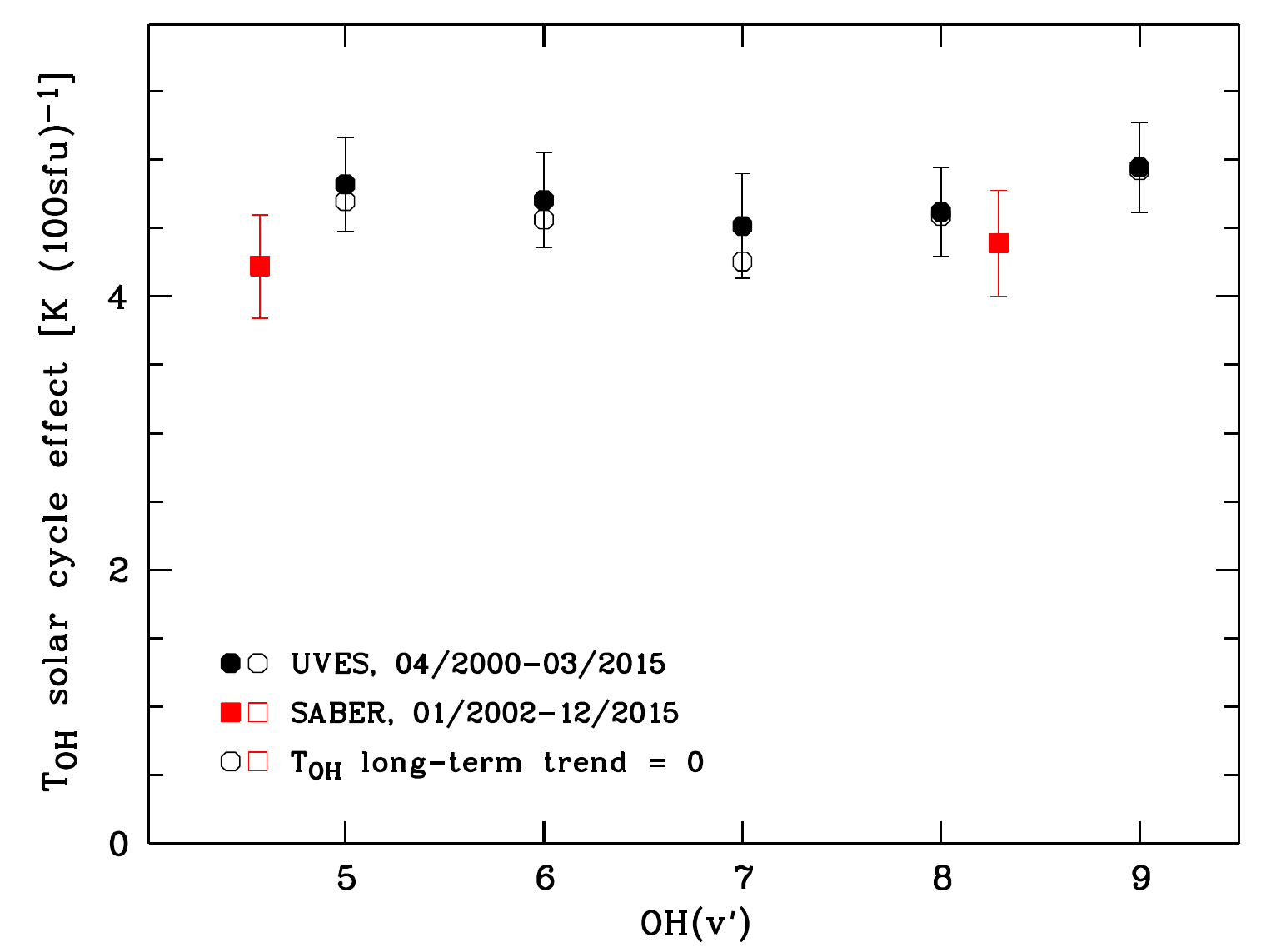}}
\mbox{
\includegraphics[width=90mm]{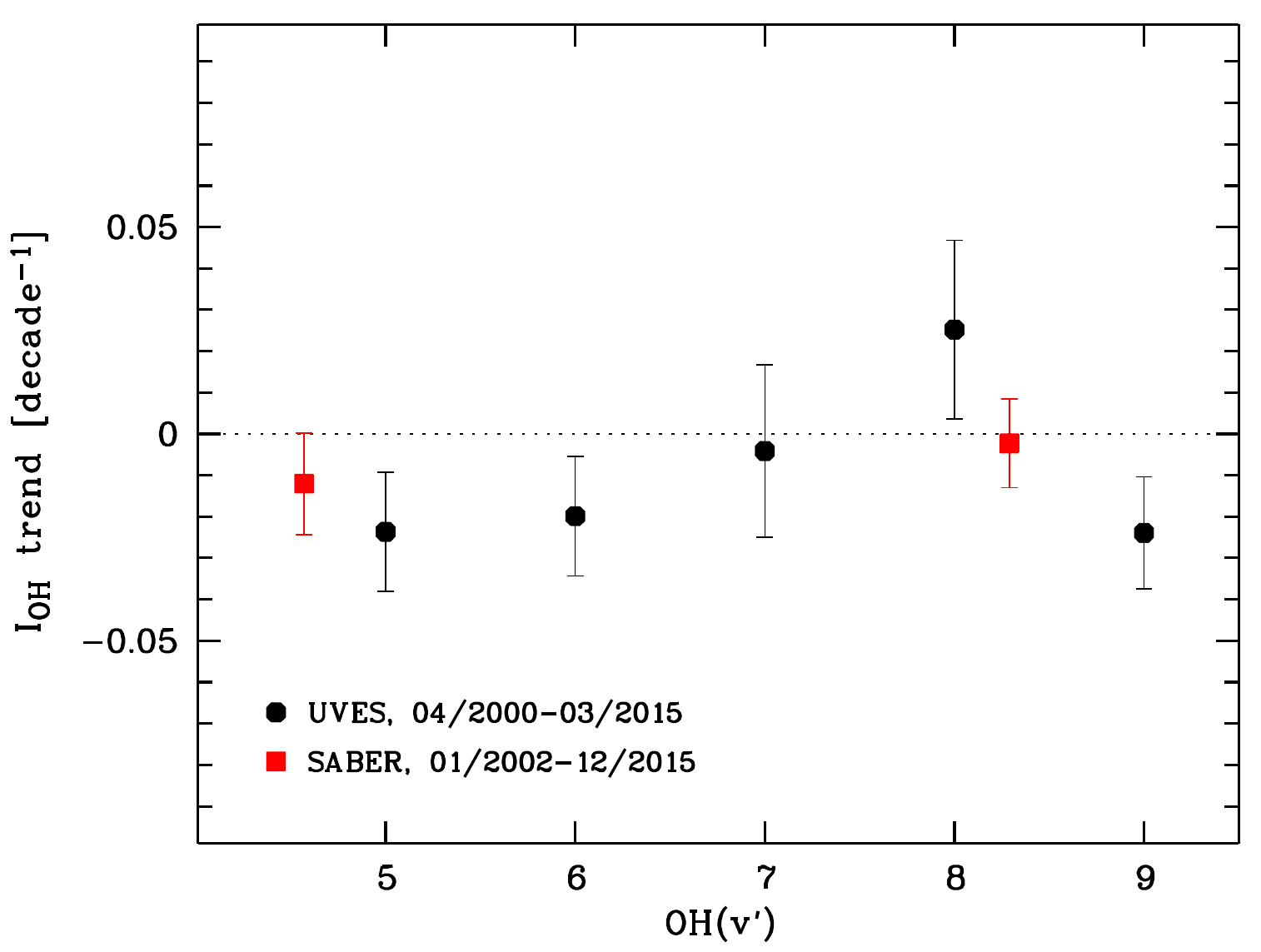}
\includegraphics[width=90mm]{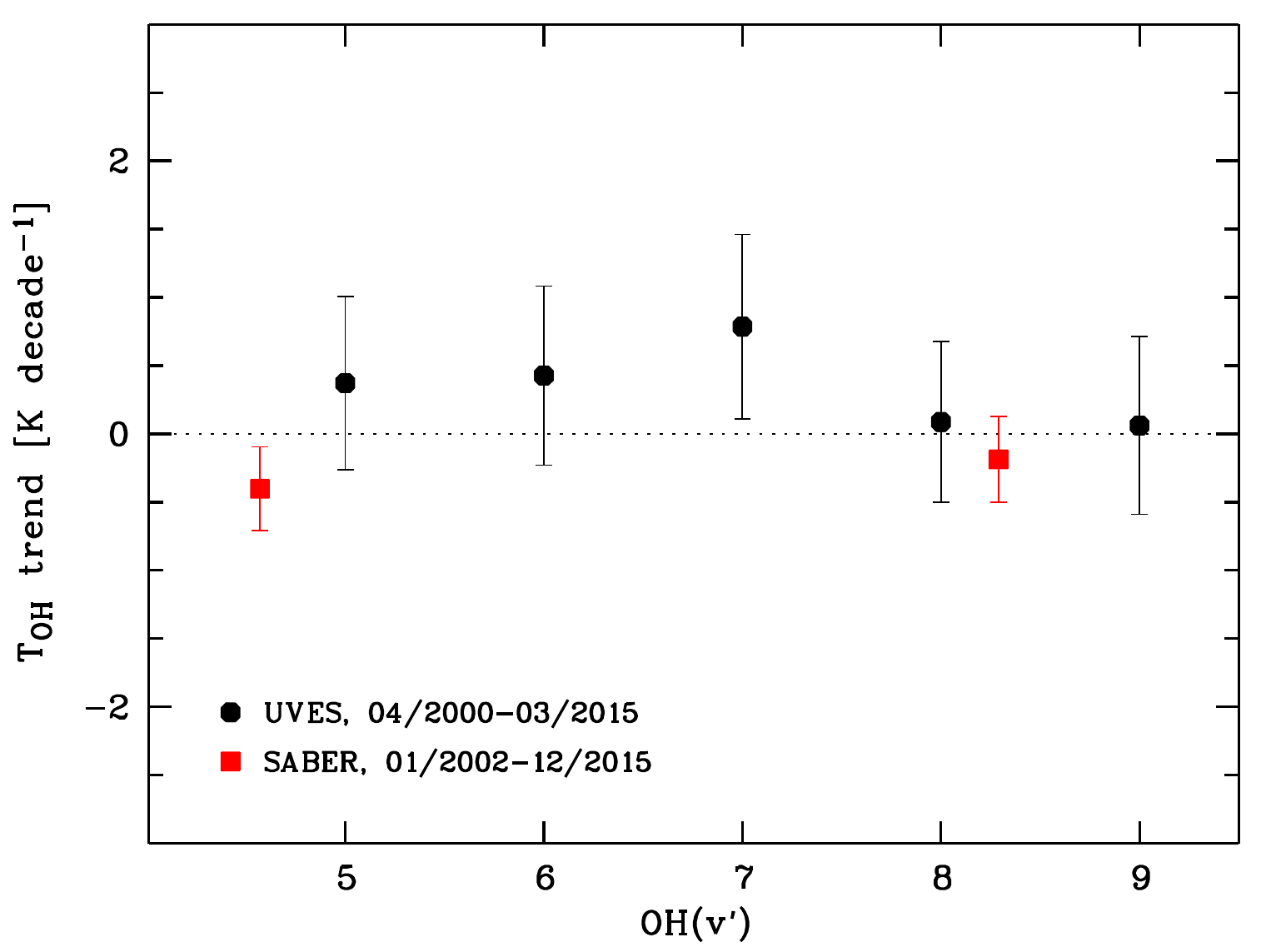}}
\end{center}
\caption{Solar cycle effects per 100\,sfu (top panels) and long-term trends per
  decade (bottom) for the relative OH intensity (left) and $T_\mathrm{rot}$ in K
  (right) for five OH bands measured in the UVES data (circles) and both SABER
  OH channels (squares). The filled symbols are related to a linear regression
  analysis involving the solar radio flux $S_\mathrm{27d}$ and the observing
  date. For the open symbols, possible long-term trends were neglected. The
  error bars (only shown for the filled symbols) consider regression and for
  the UVES data also flux calibration uncertainties. Note that averaging of
  results for different OH bands does not significantly decrease the errors.
  However, the uncertainties are lower if results of different OH bands are
  compared.
\label{fig:val_vp}}
\end{figure*}

We investigated the solar cycle effect and long-term trend in intensity and
$T_\mathrm{rot}$ for five OH bands with $v' = 5$ to 9 measured with UVES at
Cerro Paranal. The results for our sample of 3,113 spectra taken between April
2000 and March 2015 are based on a bilinear regression analysis, which was also
carried out for a sample of 4,496 SABER OH VER profiles representative of
$v' \approx 4.6$ and $8.3$. The selection includes nocturnal profiles taken
close to the Cerro Paranal region and within the period from January 2002 to
December 2015. The investigated SABER-related quantities were column-integrated
VER and $T_\mathrm{eff}$. The resulting slopes of the regression analysis for
both samples are shown in Fig.~\ref{fig:val_vp}.

The OH band intensities relative to the UVES sample mean indicate solar cycle
effects between $+15.6$ and $+16.6$\% per 100\,sfu for the individual bands.
The uncertainties of these percentages are about $1.6$\% per 100\,sfu, except
for \mbox{OH(8-3)}, where $2.9$\% per 100\,sfu is found. The latter value is
dominated by flux calibration errors, whereas the other uncertainties are
mainly related to the bilinear regression (see Section~\ref{sec:regress}).
Although the real uncertainties could even be higher due to possible additional
systematics, there appears to be a highly significant solar cycle effect. The
results for the five OH bands do not show any clear trend with $v'$. This
statement remains valid if it is taken into account that the uncertainties for
comparisons of intensities for different OH bands are smaller than those for
the total effect. Consequently, we can average the effects for the different
bands, which results in $+16.1 \pm 1.9$\,\% per 100\,sfu. The solar cycle
effects of $+12.1 \pm 1.5$ and $+14.5 \pm 1.3$\,\% per 100\,sfu for the two
SABER OH channels agree quite well with the UVES-related findings in terms of
the total amount of the effect and the absence of a significant $v'$-dependent
trend. The deviations from the higher UVES-related values are less than
$2\,\sigma$.

The UVES data do not show a significant long-term trend. The mean effect is
about $-1 \pm 2$\,\% per decade and the significance for the individual
measurements is below $2\,\sigma$. The long-term trend in the OH intensity is
the most uncertain parameter of our analysis. It is very sensitive to data
calibration issues, as the offset in the trend for \mbox{OH(8-3)} indicates.
Nevertheless, the SABER-related results fully agree. The trends for both OH
channels are consistent with zero, although the errors are distinctly smaller
than those for the UVES data. As illustrated by Fig.~\ref{fig:val_vp},
neglecting a long-term trend does not significantly change the results for
the solar cycle effect. The mean absolute difference for the UVES-related
trends would only be $0.6$\,\% per 100\,sfu. For SABER, no effect is visible.

The results for the temperature are very similar. There is no significant
long-term trend. The mean values for $T_\mathrm{rot}$ and $T_\mathrm{eff}$ are
$+0.3 \pm 0.7$ and $-0.3 \pm 0.3$\,K per decade. The solar cycle effect in
$T_\mathrm{rot}$ is the most robust trend parameter that we could determine for
the UVES data set. The effects for the individual OH bands vary between $+4.5$
and $+4.9$\,K per 100\,sfu with a mean value of $+4.7 \pm 0.4$\,K per 100\,sfu.
We do not find a dependence of the solar cycle effect on $v'$. The
SABER-related data sets for $v' \approx 4.6$ and $8.3$ show $+4.2 \pm 0.4$ and
$+4.4 \pm 0.4$\,K per 100\,sfu. These values are lower than the UVES-related
mean effect but with a significance of less than $1\,\sigma$.

In conclusion, 15\,years of UVES and 14\,years of SABER data show a significant
positive solar cycle effect but no discernible long-term trend in OH intensity
and temperature for the Cerro Paranal region. Based on the more precise SABER
results for the long-term trend, effects beyond the intervals from $-5$ to
$+3$\% and $-1.4$ to $+0.8$\,K per decade can be excluded with $3\,\sigma$
confidence. The derived numbers certainly depend on the covered period. An
interval of 14 to 15\,years is relatively short for studying long-term trends.
It is possible that the trends cannot fully be separated from solar cycle
effects. In our case, the very high solar activity at the beginning of the
time series, especially for UVES, is an issue since $F_{10.7}$ values typical
of the maximum of solar cycle 23 are not found in cycle 24 (see
Fig.~\ref{fig:val_date}). As illustrated by Fig.~\ref{fig:val_vp}, the solar
cycle effects of about 15\,\% and 4 to 5\,K per 100\,sfu are almost unaffected
by the uncertainty in the long-term trends. 

Our results for the solar influence can depend on the range of $F_{10.7}$ values
covered by the sample. In order to investigate this effect, we split the UVES
sample between March and April 2008 as well as the SABER sample between
December 2007 and January 2008 to separate the two solar cycles covered and to
keep periods of whole years. Then, we measured the solar cycle effects
(neglecting long-term trends) in intensity and temperature for the two
subsamples. Here, we focus on the results for the UVES \mbox{OH(6-2)} band and
the $v' \approx 4.6$ SABER channel since we do not see a significant $v'$
dependence. There are remarkable differences in the slopes. The solar cycle
temperature effects in the UVES and SABER data for cycle 24 are about
$101 \pm 27$ and $50 \pm 28$\,\% stronger than for cycle 23. The slope for the
UVES data taken in cycle 24 indicates a very large value of $+7.2 \pm 0.6$\,K
per 100\,sfu. In the case of the intensity, the cycle 24 data of both
instruments show effects which are about 20\% stronger, but these differences
are not significant. The temperature-related results could be explained by the
weak possible flattening of the $T_\mathrm{rot}$--$S_\mathrm{27d}$ relation for
high $S_\mathrm{27d}$ indicated in Fig.~\ref{fig:val_srf}. However, note that
the mean $S_\mathrm{27d}$ indices for both UVES subsamples are still similar
with 122 and 102\,sfu. As an additional check, we calculated the solar cycle
effects in cycle 23 for $S_\mathrm{27d} \le 166$\,sfu, which corresponds to the
highest value found in our cycle 24 samples. The percentages for the change of
the solar impact in cycle 24 relative to cycle 23 are then $28 \pm 20$ and
$56 \pm 40$\,\%. The effect for the UVES sample decreased a lot but the
SABER-related difference is even slightly higher. Therefore, non-linearities in
the $T_\mathrm{rot}$--$S_\mathrm{27d}$ relation could be an explanation. It is
also possible that the solar cycles 23 and 24 could be different in terms of
the response of the mesopause temperature to solar forcing. In any case, these
results show that the derived solar cycle effect appears to depend on the
covered period.

\subsection{Comparison to literature}\label{sec:compliterat}

\subsubsection{Temperatures}\label{sec:comptrot}

The UVES-related mean solar cycle effect for $T_\mathrm{rot}$ from five OH bands
is $+4.7 \pm 0.4$\,K per 100\,sfu. This value can be compared with results
from independent measurements to better understand the impact of geographical
location, covered period, and measurement approach. For SABER archival data,
we have already done this. The mean for the two OH-related channels is
$+4.3 \pm 0.4$\,K per 100\,sfu, which is only slightly lower. This is in good
agreement despite possible remaining systematic effects due to differences in
the period (see Section~\ref{sec:resvar}), observational gaps, the relatively
large area where the SABER profiles originate from, and that $T_\mathrm{rot}$
are compared with non-LTE-corrected $T_\mathrm{kin}$ (for the latter see
Section~\ref{sec:resnlte}).

The impact of the geographical location on our results can be estimated by
comparing our SABER-related trends with those of other more global
investigations of SABER data. \citet{huang16} studied zonal average
$T_\mathrm{kin}$ profiles for the years from 2002 to 2014. For 25$^{\circ}$\,S,
i.e.~the latitude of Cerro Paranal, their altitude--latitude map for the solar
cycle effect shows values increasing from about $2.5$ to 4\,K per 100\,sfu for
OH-relevant heights from 80 to 100\,km. Hence, our SABER-related response
appears to be somewhat stronger. \citet{nath14} investigated the average effect
in the latitude range from 10 to 15$^{\circ}$\,N for the period from 2002 to
2012. Their regression analysis resulted in a slope of 4 to 5\,K per 100\,sfu
for OH emission altitudes, which is in better agreement with our values. The
slope could even be higher by a few tenths of a Kelvin for Cerro Paranal if the
latitude-dependent results of \citet{huang16} are used for an extrapolation of
the \citet{nath14} values. According to \citet{huang16}, the discrepancies
between both SABER-based publications can be explained by differences in the
regression model used. In contrast to \citet{huang16}, \citet{nath14} did not
include seasonal and nocturnal variations. Since this is also the case for our
analysis of SABER data, this could explain the better agreement.

There are further reasons for discrepancies between our and the published data.
We only considered nocturnal SABER measurements with solar zenith angles above
100$^{\circ}$. This limitation can cause systematic deviations if there is a
significant dependence of the $T_\mathrm{kin}$ long-term variations on local
time. We did not include quasi-biennial oscillations in our analysis. However,
these variations should be sufficiently independent of the long-term
variations. Finally, both publications cover a shorter period than our SABER
sample and do not consider longitudinal variations, which could be as high as
latitudinal variations. Therefore, it is difficult to explain the differences
in detail.

The SABER-related studies also contain information on the long-term trend.
\citet{nath14} included this effect in their regression model and found trends
between $-1$ and 0\,K per decade for the OH emission heights, with smaller
absolute values at higher altitudes. \citet{huang16} neglected a long-term
trend for their final analysis since they did not find a significant impact on
their results for the solar cycle effect, which is in good agreement with our
findings. 

For a direct evaluation of our UVES-based results, other ground-based OH
measurements are best suited. In terms of latitude, the \mbox{OH(6-2)}
observations at Cachoeira Paulista (23$^{\circ}$\,S) are the closest. For the
years from 1987 to 2000, \citet{clemesha05} found a $T_\mathrm{rot}$ 11-year
oscillation with an amplitude of $+6.0 \pm 1.3$\,K, which corresponds to about
9\,K per 100\,sfu \citep{beig08}. Their long-term trend is highly negative with
$-10.8 \pm 1.5$\,K per decade. However, the difference in $T_\mathrm{rot}$ for
years of maximum and minimum solar activity is only about 5\,K for their data
set, which well agrees with our observations (see Fig.~\ref{fig:val_date}).
Therefore, it appears that solar cycle effect and long-term trend influence
each other in the analysis of \citet{clemesha05}. Values consistent with our
results could have been possible as well. Nevertheless, an exact match is not
expected because of the very different period studied. A more negative
long-term trend for their study would even be consistent with the observations
of a weakening of the cooling trend over several decades \citep{beig11a}.

Another station with a long \mbox{OH(6-2)} $T_\mathrm{rot}$ time series in South
America is El Leoncito (32$^{\circ}$\,S). Considering only the years from 1998
to 2002, \citet{scheer05} found $+0.9 \pm 0.3$\,K per 100\,sfu for a fitted
trend of $-2.7$\,K per decade. Without a trend, the result was
$-0.1 \pm 0.3$\,K per 100\,sfu, i.e.~a solar cycle effect was not significant.
This divergent finding could be related to the different and also very short
period covered. As discussed in Section~\ref{sec:resvar}, this can be critical.
\citet{scheer13} also considered data taken between 2006 and 2011. Under the
assumption of no solar cycle effect, they derived a long-term trend of
$-2.1 \pm 0.6$\,K per decade. As in the case of \citet{clemesha05}, their data
can also be interpreted in a different way. Their annual mean $T_\mathrm{rot}$
for years of high and low solar activity differ by about 3\,K. This is smaller
than the 5\,K that we obtain for the same years in the UVES sample
(Fig.~\ref{fig:val_date}). If we assume that the long-term trend is close to
zero, as the UVES and SABER data indicate, this would suggest a solar cycle
effect of 2 to 3\,K per 100\,sfu at El Leoncito. The difference in the response
to solar forcing compared to Cerro Paranal might be related to the observing
site. The latitude difference is 7$^{\circ}$ and El Leoncito is located on the
other side of the Andes.

For further comparisons to OH-related data sets with a good time coverage, we
have to consider data from northern mid-latitudes. The corresponding results
have to be taken with care since the analysis of SABER data by \citet{huang16}
suggests that the solar cycle effect at their limiting latitude of
48$^{\circ}$\,N could be stronger by about 1\,K per 100\,sfu than at
25$^{\circ}$\,S. A very long continuous ground-based time series of
\mbox{OH(3-1)} data starting in 1988 is from Wuppertal (51$^{\circ}$\,N) in
Germany. \citet{kalicinsky16} derived $4.2 \pm 0.9$\,K per 100\,sfu and
$-0.9 \pm 0.6$\,K per decade for these data and a linear regression approach.
However, they noticed that a linear long-term trend does not fit the data well.
A better model appears to be an oscillation with a period of $25 \pm 3$\,years
and an amplitude of $2.0 \pm 0.4$\,K. It results in a response to solar
activity of $4.1 \pm 0.8$\,K per 100\,sfu, which is in good agreement with our
findings. Their solar cycle effect appears to be relatively stable over the
three activity maxima covered, which does not support the possible difference
for cycle 23 and 24 found in our data (see Section~\ref{sec:resvar}). If the
oscillation model was realistic, it would cause a negative long-term trend for
the period covered by our data. There could also be site-related effects, which
could explain why we do not see such a trend. In general, the published studies
related to different measurement approaches and mid-latitude observing sites
reveal long-term trends which are slightly negative or consistent with zero
\citep{beig11a}. The corresponding solar cycle effects are of the order of
several Kelvins \citep{beig11b}. As another example, \citet{pertsev08} derived
an effect of about $4.5$\,K per 100\,sfu based on \mbox{OH(6-2)} measurements
at Zvenigorod (56$^{\circ}$\,N) near Moscow in Russia for the years 2000 to 2006.

All these results suggest that the spatial and temporal differences in the
solar cycle effects and long-term trends for temperatures at OH emission
heights for low to mid-latitudes are only moderate and that our Cerro Paranal
results are in good agreement with them.

\subsubsection{OH intensities}\label{sec:compint}

Our UVES data for five OH bands show a mean relative solar cycle effect of
$+16.1 \pm 1.9$\,\% per 100\,sfu. This is only slightly higher than our result
for the two SABER OH channels, which is $+13.3 \pm 1.9$\,\% per 100\,sfu on
average. \citet{gao16} studied the response of solar radiation to OH
intensities also based on SABER data. For the period from 2002 to 2014, local
times within a radius of 3\,h around midnight, and latitudes up to 50$^{\circ}$
in both hemispheres, they found $11.4 \pm 1.3$ and $12.9 \pm 1.1$\,\% per
100\,sfu for the two channels centred on 1.64 and 2.06\,$\mu$m. These global
averages only deviate by a few tenths per cent from the values for a latitude
of 25$^{\circ}$\,S. Although the latitude of Cerro Paranal is close to a minimum
for the southern hemisphere, this similarity in the results is possible due to
systematically lower effects in the northern hemisphere. The solar cycle
sensitivity derived by \citet{gao16} is only about 1\% per 100\,sfu lower than
the outcome of our regression analysis. This deviation is not significant. The
measurement of peak emission rates, the differences in the local time coverage,
zonal averages, and the use of annual mean values by \citet{gao16} could
contribute to the small difference. The regression analysis was performed with
and without linear long-term trend. However, the results for the solar cycle
effect were almost identical, which is consistent with our findings. 

Ground-based studies of OH intensities are less frequent than those of OH 
$T_\mathrm{rot}$. In most cases, both quantities were discussed together
(c.f.~Section~\ref{sec:comptrot}). For Cachoeira Paulista and \mbox{OH(6-2)},
\citet{clemesha05} found an amplitude for the 11-year oscillation of
$14 \pm 4$\,\%, which corresponds to about $+21 \pm 6$\,\% per 100\,sfu if we
assume the same conversion as \citet{beig08} used for $T_\mathrm{rot}$. The
effect tends to be higher than for Cerro Paranal but the difference is not
significant. The corresponding long-term trend is $+2.2 \pm 0.5$\,\% per
decade. The amount is within the range of our OH measurements. However, with
the relatively small uncertainties given, there would be a weak positive trend
for the years from 1987 to 2000. \citet{scheer05} studied the long-term
\mbox{OH(6-2)} intensity variations at El Leoncito. Since their data set only
covers the years from 1998 to 2002, we focus on their result for a regression
analysis without long-term trend, which is $-1.8 \pm 1.5$\,\% per 100\,sfu.
This lack of a significant solar cycle effect is similar to the situation for
$T_\mathrm{rot}$ described in Section~\ref{sec:comptrot}. It also appears to be
an apparent effect due to the limitations in the data set without periods of
low solar activity. In order to prove this assumption, it would be necessary to
perform a detailed analysis of the mostly unpublished \mbox{OH(6-2)} intensity
data taken since 2006 \citep[c.f.][]{scheer13}, which cover periods of high
and low solar activity. \citet{reid14} studied \mbox{OH(8-3)} intensity data
taken between 1995 and 2010 at Buckland Park (35$^{\circ}$\,S) near Adelaide in
Australia. Their multiple harmonic analysis resulted in an amplitude of
$6 \pm 5$\,\% for the solar $11.2$-year period, which suggests a linear solar
cycle effect of less than 10\% per 100\,sfu. However, the very large
uncertainty does not allow us to detect a significant difference in comparison
to our and the other measurements.

For the northern mid-latitude station Zvenigorod, \citet{pertsev08} report a
solar cycle sensitivity of $+30 \pm 4$\,\% per 100\,sfu within the period from
2000 to 2006. This large effect was obtained by means of \mbox{OH(6-2)} but it
would be similar for other OH bands which were measured in their spectroscopic
data. The lack of a $v'$ dependence is consistent with our UVES-based results.
The large discrepancy between the solar cycle effects for Zvenigorod and Cerro
Paranal could be explained by the large latitude difference. The SABER-based
analysis by \citet{gao16} shows a maximum-to-minimum ratio of 2 for the
latitude range from 50$^{\circ}$\,S to 50$^{\circ}$\,N. Zvenigorod at
56$^{\circ}$\,N is outside this range. However, the steep increase of the solar
cycle effect towards higher latitudes could be sufficient to significantly
reduce the discrepancy between the \citet{pertsev08} and our results.

Our discussion of the few studies on the OH intensity long-term variations in
the last two solar cycles shows that our knowledge is still poor. Therefore,
our combined study of UVES and SABER data provides important constraints
especially on the OH intensity solar cycle effect at low latitudes.

\subsection{Non-LTE contributions to OH rotational temperatures}
\label{sec:resnlte}

\begin{figure*}[t]
\begin{center}
\mbox{    
\includegraphics[width=90mm]{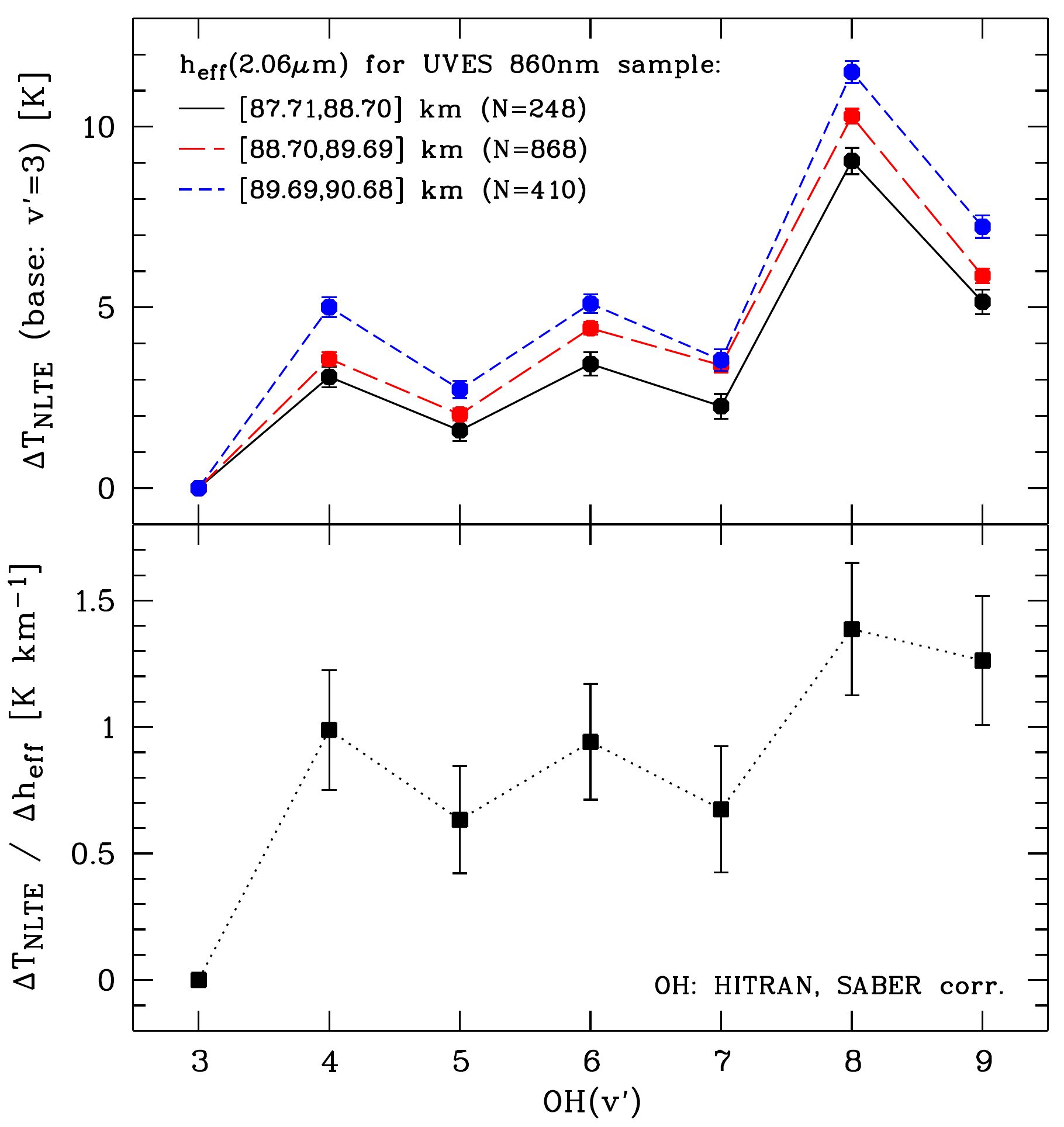}
\includegraphics[width=90mm]{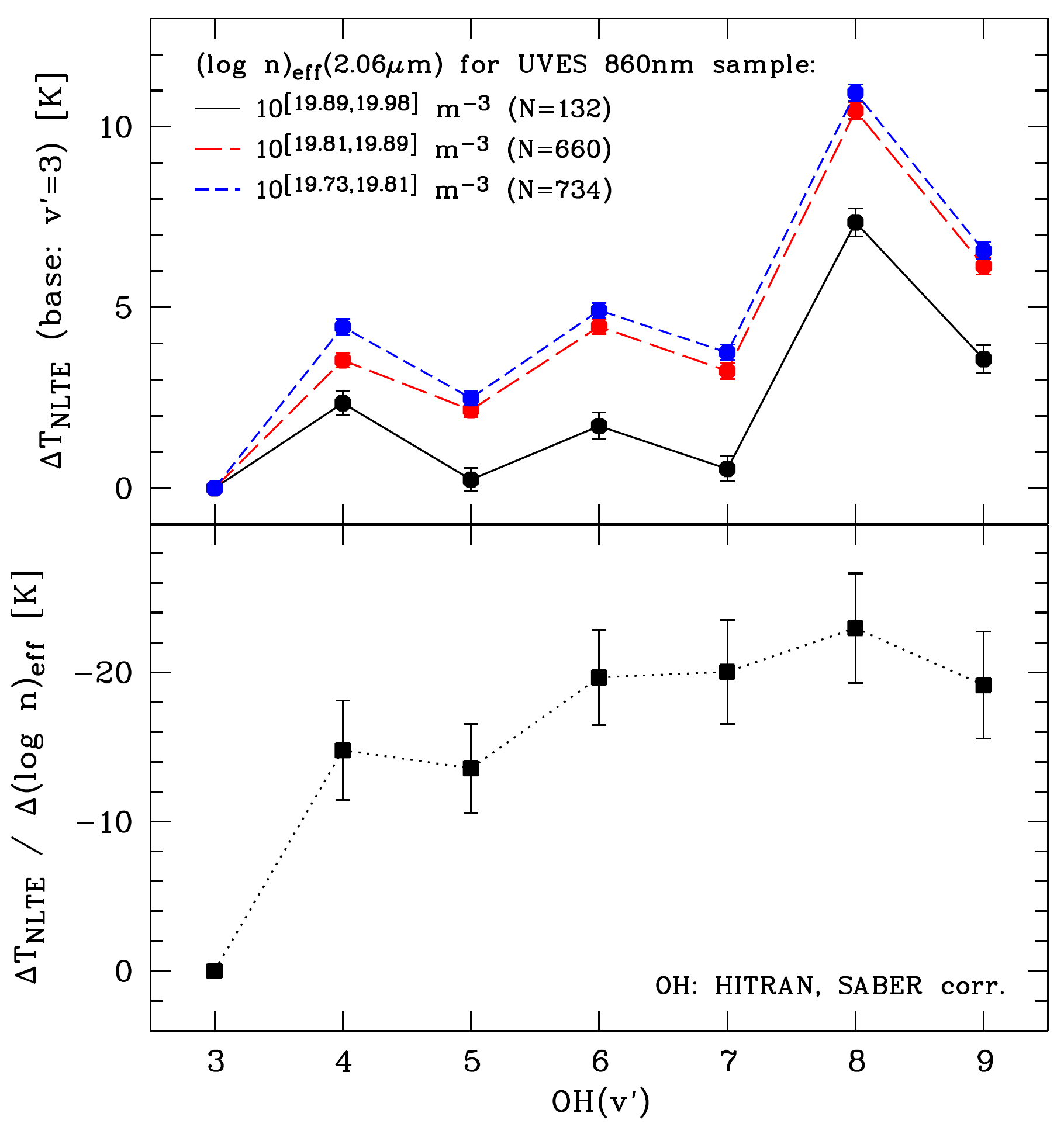}
}
\end{center}
\caption{Change of non-LTE contributions to OH $T_\mathrm{rot}$
  ($\Delta T_\mathrm{NLTE}$) by variations of effective height (left) and
  logarithmic air density (right) of the OH emission layer for the 860\,nm UVES
  sample (see Fig.~\ref{fig:sample_uves}) and different $v'$. The upper panels
  show $\Delta T_\mathrm{NLTE}$ in K with a zero line defined by the
  $T_\mathrm{rot}$ for OH($v' = 3$). The values for the other $v'$ were
  corrected for differences in the SABER-related emission profiles and
  temperature offsets depending on OH band and line set (see
  Sect.~\ref{sec:estimnlte}). The $v'$-related $\Delta T_\mathrm{NLTE}$ are
  shown for three subsamples that cover the full range of effective height and
  logarithmic density values by intervals of equal size (see legend). The given
  error bars indicate the mean errors and should be considered as lower limits
  since the $T_\mathrm{rot}$ distribution cannot be described by a single
  Gaussian (see Sect.~\ref{sec:uves}). The lower panels show the results of a
  linear regression analysis for the relation between $\Delta T_\mathrm{NLTE}$
  and the effective height (left) and logarithmic density (right),
  respectively. The error bars are related to the uncertainties of the
  regression analysis. Note that the ordinate for the right plot is reversed in
  order to consider that height and density are anti-correlated. For the same
  reason, the order of the intervals used in the upper panels is different.
\label{fig:delTrot_vp_hn}}
\end{figure*}

The difference of $0.4$\,K per 100\,sfu between the mean solar cycle effects
for the UVES OH $T_\mathrm{rot}$ and SABER OH $T_\mathrm{eff}$ (see
Section~\ref{sec:resvar}) could significantly be affected by long-term
variations in the non-LTE contributions to OH $T_\mathrm{rot}$,
$\Delta T_\mathrm{NLTE}$. On nocturnal and seasonal time scales, the variability
of these effects can be very crucial for interpreting the total $T_\mathrm{rot}$
variations, as shown by \citet{noll16}. It seems that these variations are
correlated with changes in the height and shape of the OH emission layer since
this affects the ratio of OH production, thermalisation by N$_2$ and O$_2$
collisions, and OH deactivation and destruction by atomic oxygen. Our sample of
UVES data allowed us to study this relation in detail.

\subsubsection{Relation to effective emission height and density}
\label{sec:releffpar}

\citet{noll16} showed that the quantification of the varying non-LTE effects
works best if the $T_\mathrm{rot}$ of an OH band with low $v'$ is used as
reference since the corresponding variations appear to be distinctly weaker
than those for bands with higher $v'$. \citet{noll16} could use \mbox{OH(2-0)}
for this purpose. The band with the lowest $v'$ in the UVES wavelength range is
\mbox{OH(3-0)}. The \mbox{X-shooter}-based study of the nocturnal
$\Delta T_\mathrm{NLTE}$ variations indicated that the changes related to
$v' = 3$ are only about one quarter of those related to $v' = 8$. This amount
seems to be small enough in order to use \mbox{OH(3-0)} as a reference in this
study. Since this band and also \mbox{OH(4-0)} are only present in data taken
with the UVES 860\,nm set-up, the study of non-LTE effects is focused on the
corresponding sample with 1,526 spectra (see Section~\ref{sec:corrtemp}).
Despite the lower size and time coverage compared to the main sample used for
the study of long-term variations in Section~\ref{sec:resvar}, it is well
suited to study the solar cycle effect since it covers the maximum of cycle 23
and the subsequent deep activity minimum (see Fig.~\ref{fig:sample_uves}). For
the analysis, the $T_\mathrm{rot}$ of the seven selected bands with $v'$ from 3
to 9 were corrected for constant offsets related to systematic measurement
errors (see Section~\ref{sec:corrtemp}). Moreover, we corrected the
$T_\mathrm{rot}$ for $T_\mathrm{kin}$ differences caused by the $v'$ dependence of
the OH emission height using $v' = 3$ as the reference. This was performed by
means of SABER $T_\mathrm{eff}$ data, which were linked with the UVES
observations as described in Section~\ref{sec:linking}. The latter was also
done for the effective emission height $h_\mathrm{eff}$ and the VER-weighted
decadal logarithm of the number density of air molecules in m$^{-3}$,
$(\log n)_\mathrm{eff}$.

At first, we checked whether we can confirm that the contribution of non-LTE
effects to OH $T_\mathrm{rot}$ depends on $h_\mathrm{eff}$. For this purpose, we
primarily used the $h_\mathrm{eff}$ values for the SABER OH channel centred on
$2.06$\,$\mu$m since the effective $v'$ of about 8.3 better represents those
OH bands where the largest effect is expected. Nevertheless, we tested the
other OH channel as well. The results agree within the errors. The
$h_\mathrm{eff}$ for the $2.06$\,$\mu$m channel vary from $87.7$ to $90.7$\,km.
Note that this variation is smaller than for individual SABER profiles since
the connection with UVES data was only possible by an averaging procedure (see
Section~\ref{sec:linking}). We divided this range in three intervals of equal
size with 248, 868, and 410 UVES spectra. The bin with the lowest
$h_\mathrm{eff}$ therefore exhibits the highest statistical uncertainties. The
$\Delta T_\mathrm{NLTE}$ results for the three intervals are given as a function
of $v'$ in Fig.~\ref{fig:delTrot_vp_hn}. The values for $v' = 3$ are always at
zero since it is the reference for the calculation of the non-LTE effects. The
curves for the three $h_\mathrm{eff}$ ranges convincingly confirm the expected
trend. Higher $\Delta T_\mathrm{NLTE}$ are related to higher $h_\mathrm{eff}$.

We also performed a linear regression analysis for these two parameters and the
individual UVES measurements. The results can also be found in
Fig.~\ref{fig:delTrot_vp_hn}. The $h_\mathrm{eff}$ dependence of
$\Delta T_\mathrm{NLTE}$ varies from $0.63$\,K per km for $v' = 5$ to $1.39$\,K
per km for $v' = 8$. With an average mean error of $0.24$\,K per km, all
regression slopes appear to be significant except those for $v' = 5$ and 7.
There is no clear dependence on $v'$. However, there seems to be a weak
positive correlation between the slope and $\Delta T_\mathrm{NLTE}$, i.e.~those
bands which show strong non-LTE effects in $T_\mathrm{rot}$ on average are also
those which tend to be more sensitive to changes in the OH emission height.
Note that the significance of our results is also affected by the sample
selection. As described for our main sample in Section~\ref{sec:uvessample},
we also performed the whole analysis based on a more restrictive data set with
1,386 spectra. This selection reduced the regression slopes in
Fig.~\ref{fig:delTrot_vp_hn} by $0.27$\,K per km on average, which is about the
mean uncertainty of the regression analysis. This result suggests that the true
total uncertainties are somewhat higher. They cannot be neglected as it was
possible for the analysis of the main sample of 3,113 spectra.

We also checked whether the combination of individual measurements
($T_\mathrm{rot}$) and weighted averages ($T_\mathrm{eff}$, $h_\mathrm{eff}$) could
affect our regression results. For this purpose, we first derived averages for
these quantities from the individual UVES and SABER measurements for the fixed
grid of observing times and days of year discussed in
Section~\ref{sec:detrend}. Then, the $\Delta T_\mathrm{NLTE}$ were calculated
and the regression analysis was performed for the nighttime grid points.
This procedure resulted in regression slopes which were slightly but not
significantly lower, which justifies our approach used for
Fig.~\ref{fig:delTrot_vp_hn}. Nevertheless, the derived regression slopes
should be taken with caution since geographical and temporal mismatches of the
UVES and SABER data and the limited vertical resolution of the SABER profiles
could still cause additional systematics.

The thermalisation of the OH rotational level populations essentially depends
on collisions with N$_2$ and O$_2$ \citep[e.g.][]{dodd94,kliner99}. Therefore,
the effective air density in the OH emission layer appears to be an even more
promising tracer of $\Delta T_\mathrm{NLTE}$ than $h_\mathrm{eff}$. Therefore, we
performed the same analysis for $(\log n)_\mathrm{eff}$. The logarithm of the
density was chosen since it results in weights of the different parts of the
emission layer which are comparable to those for the altitude. It also avoids
that the most critical heights for the non-LTE contributions have negligible
weight. Fig.~\ref{fig:delTrot_vp_hn} shows the results for three
$(\log n)_\mathrm{eff}$ bins with 132, 660, and 734 spectra. Observations at
high density are relatively rare. Since $\Delta T_\mathrm{NLTE}$ tends to
increase during the night at Cerro Paranal \citep{noll16}, the high-density
spectra are usually related to the early evening. The mean local time is 19:57.
The non-LTE contributions are remarkably low for this subsample.
$\Delta T_\mathrm{NLTE}$ of several Kelvins are only present for $v' = 8$ and 9,
i.e.~the levels which are directly populated by the hydrogen--ozone reaction.
For the other two bins, the $\Delta T_\mathrm{NLTE}$ are distinctly higher. The
mean difference for the lowest and highest density intervals is $2.9$\,K if the
results for $v'$ from 4 to 9 are averaged. The corresponding mean difference
for $h_\mathrm{eff}$ is only $1.8$\,K. These values confirm our assumption that
$(\log n)_\mathrm{eff}$ is the better tracer for $\Delta T_\mathrm{NLTE}$.

This can also be seen for the results of the corresponding regression analysis,
which are also displayed in Fig.~\ref{fig:delTrot_vp_hn}. The slopes are highly
significant with an average confidence of $5.4\,\sigma$. The differences
between the $v'$-dependent data points are relatively small and not
significant. Compared to the mean value of $-18.4$\,K, the scatter is only
$3.5$\,K. The characteristic $\Delta T_\mathrm{NLTE}$($v'$) structure is only
mar\-ginally present. The large step from the reference $v'$ of 3 to 4
challenges the assumption of comparably small $\Delta T_\mathrm{NLTE}$
variations for the lowest $v'$. It might be that a significant fraction of the
total effect could not be detected. The differences in the band-specific line
sets used for this study \citep[especially \mbox{OH(3-0)} with five and
\mbox{OH(4-0)} with four lines, and only three common emissions; see][]{noll15}
might also play a role if each OH line shows a different sensitivity to non-LTE
effects. If the SABER OH channel with $v' \approx 4.6$ is used for the
analysis, the result is very similar to the one for $v' \approx 8.3$ except for
a less negative mean slope of $-15.2$\,K. The difference can be almost fully
explained by a higher variability in $(\log n)_\mathrm{eff}$. For the SABER
high-$v'$ channel, the study of the alternative UVES sample with 1,386 spectra
also gives a mean slope of $-15.2$\,K. Its significance is 4.6\,$\sigma$.
Therefore, the sample selection uncertainty does not critically decrease the
reliability of our result on the existence of strong variations of the
$T_\mathrm{rot}$ non-LTE effects as a function of the effective air density.
Finally, as already found for $h_\mathrm{eff}$, the regression slopes do not
appear to significantly depend on our approach for linking the SABER with the
UVES data.

\subsubsection{Non-LTE long-term variations}\label{sec:nltelongvar}

\begin{figure}[t]
\begin{center}
\includegraphics[width=88mm]{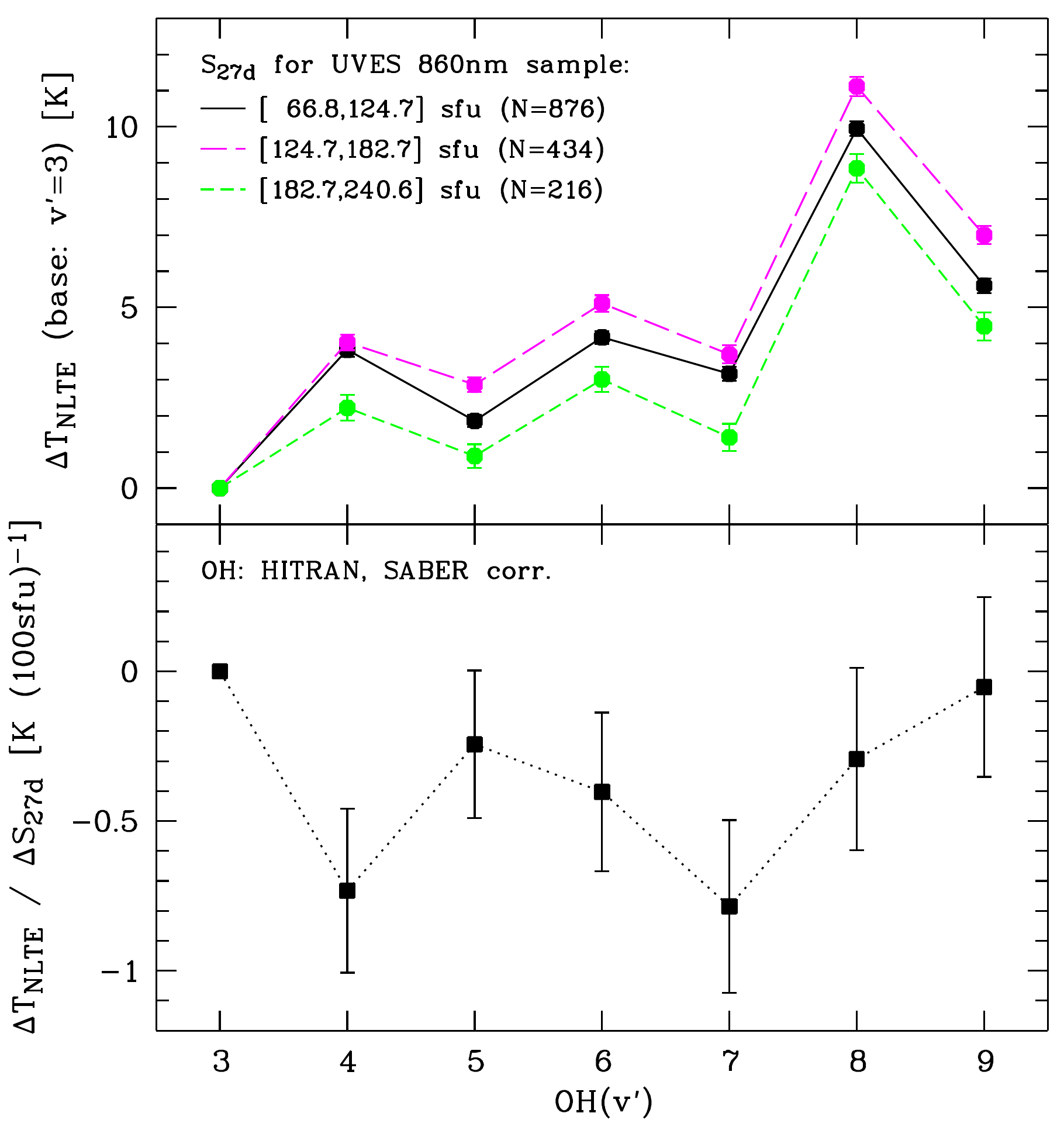}
\end{center}
\caption{Change of non-LTE contributions to OH $T_\mathrm{rot}$
  ($\Delta T_\mathrm{NLTE}$) by variations of the solar radio flux
  $S_\mathrm{27d}$ for the 860\,nm UVES sample and different $v'$. For more
  details, see the legend and Fig.~\ref{fig:delTrot_vp_hn}.
\label{fig:delTrot_vp_srf}}
\end{figure}

\begin{figure*}[t]
\begin{center}
\mbox{     
\includegraphics[width=90mm]{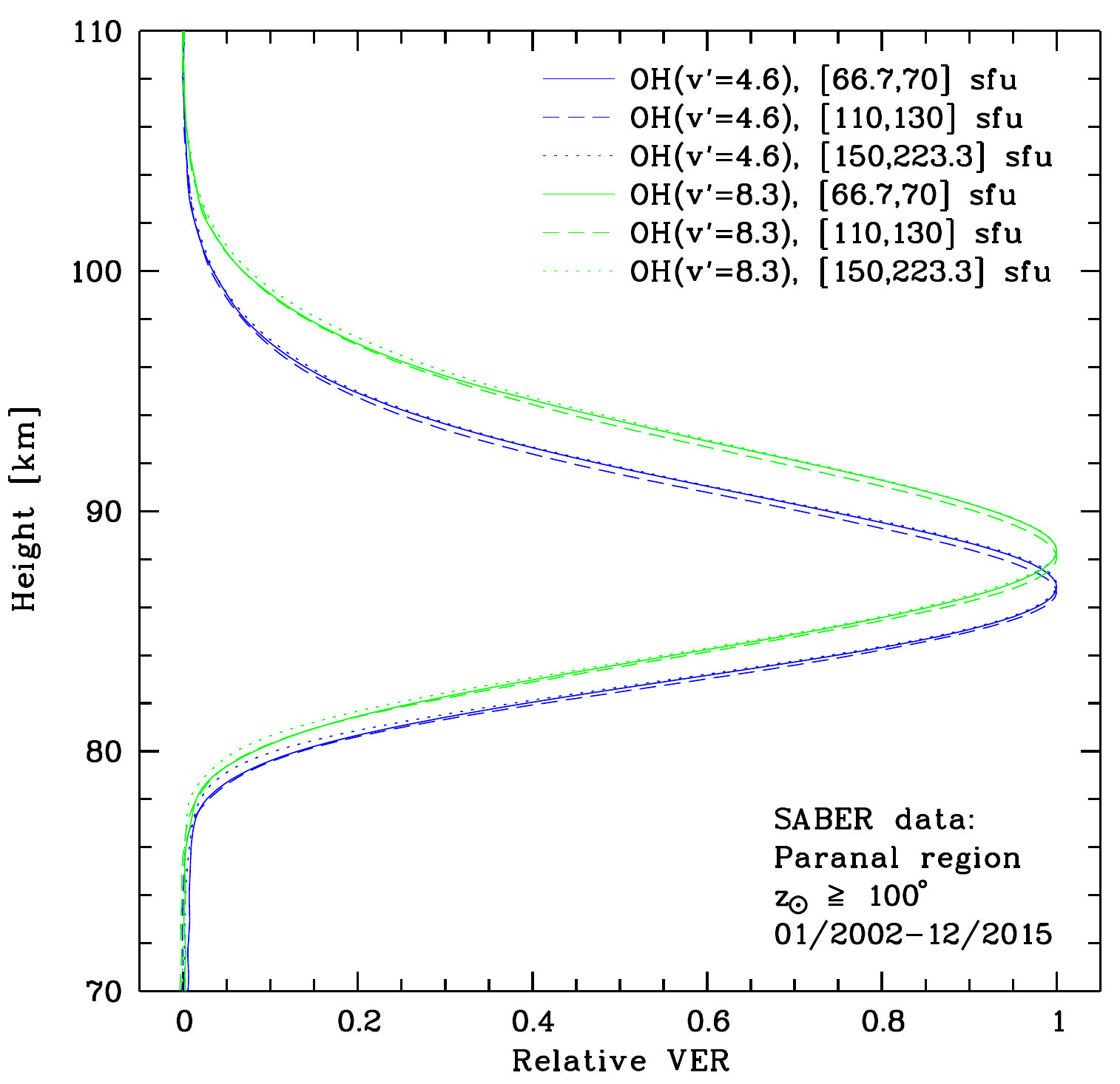}
\includegraphics[width=90mm]{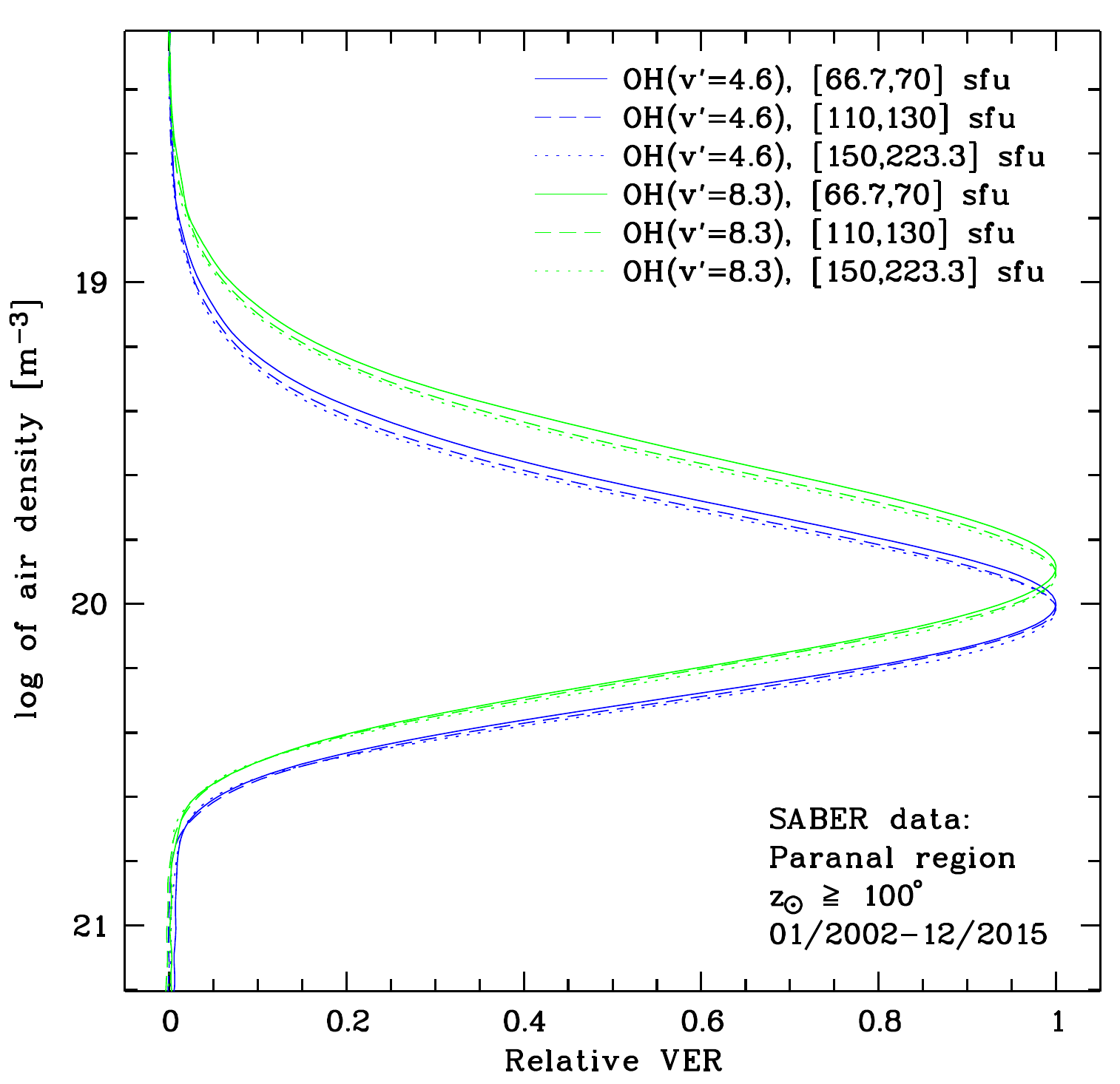}}
\end{center}
\caption{OH relative volume emission rate profiles of the two SABER OH
  channels centred on 1.64 ($v' \approx 4.6$; blue) and 2.06\,$\mu$m
  ($v' \approx 8.3$; green) for Cerro Paranal-related SABER subsamples with low
  ($< 70$\,sfu; $N = 495$; solid), intermediate (between 110 and 130\,sfu;
  $N = 918$; dashed), and high ($> 150$\,sfu; $N = 455$; dotted) solar radio
  flux $S_\mathrm{27d}$. The profiles are shown for height (left) and
  logarithmic air density (right).
\label{fig:rverOH_hn_srf}}
\end{figure*}

The discussion in the previous section has shown that changes in the OH
emission layer can cause variations in OH $T_\mathrm{rot}$ of the order of
several Kelvins. This seems to be especially critical for time scales of hours
to months \citep{noll16}. Now, the crucial question is: how large is the impact
of non-LTE effects on the observed $T_\mathrm{rot}$ long-term variations? For
this, we studied the dependence of $\Delta T_\mathrm{NLTE}$ on the solar radio
flux measured by $S_\mathrm{27d}$. We performed the same analysis as described
in Section~\ref{sec:releffpar} for $h_\mathrm{eff}$ and $(\log n)_\mathrm{eff}$.

The results for our UVES 860\,nm sample are given in
Fig.~\ref{fig:delTrot_vp_srf}. The $T_\mathrm{rot}$ data set was divided into
three subsamples with 876, 434, 216 members, which cover $S_\mathrm{27d}$
intervals of equal size in the range from $66.8$ to $240.6$\,sfu. The smallest
population is found for $S_\mathrm{27d}$ greater than $182.7$\,sfu, which only
occurred during the maximum of solar cycle 23. The shape of the three
$\Delta T_\mathrm{NLTE}$($v'$) functions is relatively similar. However, the
dependence of the non-LTE effects on $S_\mathrm{27d}$ is remarkable. The highest
solar fluxes cause the lowest $\Delta T_\mathrm{NLTE}$, whereas data taken at
intermediate $S_\mathrm{27d}$ indicates the strongest non-LTE contributions.
This result is inconsistent with a linear relation. Therefore, a linear
regression does not seem to be well suited to describe the data.
Nevertheless, we also show the results of such an analysis in
Fig.~\ref{fig:delTrot_vp_srf}. The slopes range from $-0.79$ to $-0.05$\,K per
100\,sfu with a mean value and error of $-0.42 \pm 0.28$\,K per 100\,sfu. Using
the alternative sample with 1,386 instead of 1,526 spectra, we obtain
$-0.19 \pm 0.27$\,K per 100\,sfu, which is even closer to zero. These results
suggest that non-LTE effects seem to be almost negligible for the
interpretation of the total effect of $+4.7 \pm 0.4$\,K per 100\,sfu discussed
in Section~\ref{sec:resvar}. However, the real situation is more complex.   
If we only use the data in the two intervals with the lowest $S_\mathrm{27d}$,
the regression analysis returns a mean slope of $+0.8 \pm 0.4$\,K per 100\,sfu.
For the two intervals with the highest $S_\mathrm{27d}$, the corresponding
result is $-3.5 \pm 0.7$\,K per 100\,sfu. Although these results should be
taken with care since the involved samples are relatively small and the
regression analysis is less robust, the large discrepancies in the slopes
imply that the contribution of non-LTE effects could significantly depend on
the sample properties. They could be an explanation for the sample-specific
differences in the solar cycle effects discussed in Section~\ref{sec:resvar}.

In view of the uncertainties in the interpretation of the relation between
$\Delta T_\mathrm{NLTE}$ and the solar radio flux, we also investigated the
$S_\mathrm{27d}$ dependence of the OH emission profile. For this purpose, we
used our SABER sample with 4,496 profiles (Section~\ref{sec:saber}) and derived
mean profiles for the two OH channels in three very different $S_\mathrm{27d}$
ranges with mean values of 68, 120, and 169\,sfu. All profiles for solar fluxes
below 70\,sfu ($N = 495$), beween 110 to 130\,sfu ($N = 918$), and above
150\,sfu ($N = 455$) were considered. The resulting relative VER profiles are
shown in Fig.~\ref{fig:rverOH_hn_srf} as a function of the altitude and
$\log n$. The profiles for $v' \approx 4.6$ and $8.3$ with sample-averaged
$h_\mathrm{eff}$ of $87.8$ and $89.2$\,km show only a weak variation of a few
$10^{-1}$\,km and $10^{-2}$ logarithmic units. The profile for the highest solar
activity shows the highest effective emission height and density. This apparent
contradiction can be explained by an expansion of the atmosphere which is
stronger than the rise of the OH emission layer. The expansion is expected due
to the warming effect (see Sect.~\ref{sec:resvar}) by the increased solar
energy input at high solar activity. This effect becomes especially strong at
higher altitudes in the thermosphere \citep[e.g.][]{emmert10}.
Fig.~\ref{fig:rverOH_hn_srf} also suggests that the response of the OH emission
layer on the solar activity is not a linear function since the profile for
intermediate $S_\mathrm{27d}$ is the lowest in terms of altitude. Another aspect
of the weak profile variations is the corresponding negligible change in the
$T_\mathrm{kin}$-related $T_\mathrm{eff}$. For the mean OH VER and $T_\mathrm{kin}$
profile of our SABER sample, the sensitivity is only about $-0.2$\,K per km.

In order to better understand the effect of the solar cycle on the full SABER
sample, we performed a linear regression analysis for $h_\mathrm{eff}$ and
$(\log n)_\mathrm{eff}$ similar to the ones for the SABER-related OH intensities
and temperatures (Section~\ref{sec:regress}). The $h_\mathrm{eff}$-related slopes
for the solar cycle effect are $0.00 \pm 0.06$ and $+0.07 \pm 0.05$\,km per
100\,sfu for the channels related to $v' \approx 4.6$ and $8.3$. This is not
significant. The linear long-term trend also turned out to be consistent with
zero. For $(\log n)_\mathrm{eff}$, there could be a weak negative trend as
indicated by a mean value of $-0.007 \pm 0.003$ per decade for both OH
channels. The solar cycle sensitivity is more significant but still weak. We
derived $+0.019 \pm 0.004$ and $+0.016 \pm 0.004$ per 100\,sfu for the channels
related to $v' \approx 4.6$ and $8.3$. Hence, there is a positive correlation
of $(\log n)_\mathrm{eff}$ and $S_\mathrm{27d}$, as already suggested by
Fig.~\ref{fig:rverOH_hn_srf}.

As described in Section~\ref{sec:compint}, \citet{gao16} also used SABER data
to study long-term variations in the OH emission. This analysis included a
latitude-dependent regression analysis for the peak height of the emission in
both channels. They did not find any significant effect, which we can confirm
based on an analysis also using the emission peak. As illustrated by
Fig.~\ref{fig:rverOH_hn_srf}, changes due to high solar activity appear to be
limited to the faint wings of the emission profile, which cannot be measured by
the peak height and where also $h_\mathrm{eff}$ does not seem to be sensitive
enough. Based on \mbox{OH(3-1)} VER profiles from SCIAMACHY on Envisat,
\citet{savigny15} performed a regression analysis for the long-term trends in
$h_\mathrm{eff}$ for the years 2003 to 2011, 22:00 local solar time, and several
10$^{\circ}$ latitude bins. Linear trends are negligible. Only for the range
from 20 to 30$^{\circ}$\,N, there could be a weak trend of $-0.26 \pm 0.07$\,km
per decade. There is no significant solar cycle effect in the investigated
latitude range from 10$^{\circ}$\,S to 30$^{\circ}$\,N. For the range from 20 to
30$^{\circ}$\,N, the result is $+0.19 \pm 0.13$\,km per 100\,sfu, which agrees
well with our findings. Finally, \citet{liu06} used \mbox{OH(8-3)} VER profiles
related to WINDII onboard UARS to investigate the peak emission altitude for
the years from 1991 to 1997 for several latitude intervals. For 20 to
30$^{\circ}$\,S, they found a significant slope of about $+1.4$\,km per
100\,sfu. It is not clear how this large deviation from the other results can
be explained \citep{savigny15}. 

As a final alternative estimation of the contributions of varying non-LTE
effects to the OH $T_\mathrm{rot}$ long-term variations, we can combine the
slopes for $\Delta T_\mathrm{NLTE}$ versus $h_\mathrm{eff}$ or
$(\log n)_\mathrm{eff}$ (see Section~\ref{sec:releffpar}) with those for
$h_\mathrm{eff}$ or $(\log n)_\mathrm{eff}$ versus $S_\mathrm{27d}$. The resulting
amounts can be considered as lower limits for the solar cycle effect since a
single parameter like $h_\mathrm{eff}$ cannot trace all variations related to
the impact of the solar activity on the OH $T_\mathrm{rot}$. As input for the
slopes that were derived from our UVES 860\,nm sample, we used the
$v'$-averaged mean values, i.e.~$+0.98 \pm 0.24$\,K per km and
$-18.4 \pm 3.4$\,K for the SABER $2.06$\,$\mu$m OH channel data and
$+0.94 \pm 0.21$\,K per km and $-15.2 \pm 2.8$\,K for the $1.64$\,$\mu$m data.
These slopes were then multiplied by the corresponding ones for the
$h_\mathrm{eff}$ and $(\log n)_\mathrm{eff}$ solar cycle effects listed above. In
the end, we obtained four estimates which range from $-0.35$ to $+0.07$\,K per
100\,sfu. The most negative value is related to $(\log n)_\mathrm{eff}$ and
$v' \approx 8.3$. The average of the four slopes is $-0.13 \pm 0.12$\,K per
100\,sfu. For completeness, we also performed this calculation for the
long-term trend, which resulted in $+0.06 \pm 0.06$\,K per decade and is
extremely small. These results suggest that the long-term variations of OH
$T_\mathrm{rot}$ at Cerro Paranal for 2000 to 2015 are only slightly affected by
non-LTE effects. The same conclusion can be drawn from the comparison of the
UVES $T_\mathrm{rot}$ and SABER $T_\mathrm{eff}$ trends in
Section~\ref{sec:resvar} and the direct regression analysis for
$\Delta T_\mathrm{NLTE}$ and $S_\mathrm{27d}$. The main reason for this behaviour
appears to be the long-term stability of the OH emission layer.

\section{Conclusions}\label{sec:conclusions}

We used 3,113 spectra from the UVES echelle spectrograph at the VLT in Chile
to study the long-term variations in OH intensity and OH-related rotational
temperature $T_\mathrm{rot}$ for the period from April 2000 to March 2015.
Focusing on the strongest P-branch lines of the five OH bands \mbox{(5-1)},
\mbox{(6-2)}, \mbox{(7-3)}, \mbox{(8-3)}, and \mbox{(9-4)}, our linear
regression analysis resulted in a strong solar cycle effect in both parameters,
which does not significantly depend on the vibrational level of the upper state
$v'$. The mean effect for all bands is $+16.1 \pm 1.9$\,\% per 100\,sfu for the
relative intensity and $+4.7 \pm 0.4$\,K per 100\,sfu for $T_\mathrm{rot}$. We
do not see significant long-term trends.

For a comparison, we also analysed 4,496 nighttime limb-sounding observations
of the multi-channel radiometer SABER on TIMED for the Cerro Paranal region and
the period from January 2002 to December 2015. The vertically-integrated volume
emission rates (VERs) for the two OH-related channels with effective
$v' \approx 4.6$ and $8.3$ showed a mean solar cycle effect of
$+13.3 \pm 1.9$\,\% per 100\,sfu. In the case of the VER-weighted kinetic
temperature profiles, we obtained an effect of $+4.3 \pm 0.4$\,K per 100\,sfu.
There is no significant $v'$ dependence and the corresponding long-term trends
are consistent with zero. Our UVES- and SABER-related trend parameters agree
well with each other. Other results from the literature for SABER data and
ground-based OH measurements at low and mid-latitudes are generally consistent
with our findings. For the few long OH $T_\mathrm{rot}$ time series at low
latitudes, we could even find that the agreement appears to be much better than
previously expected from the published values, which mainly deviate by
differences in the analysis and limitations in the data sets.

Measurements of kinetic temperatures in the mesopause region by means of OH
$T_\mathrm{rot}$ are affected by varying contributions of non-LTE effects
$\Delta T_\mathrm{NLTE}$. They are caused by variations in the frequency of the
crucial thermalising OH collisions with air molecules compared to radiative
processes and deactivating or destructing collisions with atomic oxygen. They
can be estimated by comparisons of $T_\mathrm{rot}$ with large $v'$ differences
corrected for deviations in the OH emission profiles. We studied their impact
on the long-term variations by different approaches.

The small difference in the temperatures from UVES and SABER data is already a
good indication that the non-LTE contributions must be low. A regression
analysis for $\Delta T_\mathrm{NLTE}$ and the solar radio flux for a sample of
1,526 UVES spectra where the OH bands \mbox{(3-0)} and \mbox{(4-0)} could also
be measured revealed a mean solar cycle effect for all $v'$ from 4 to 9 of
$-0.4 \pm 0.3$\,K per 100\,sfu. A more indirect approach included an
intermediate step involving the $\Delta T_\mathrm{NLTE}$ dependence on the
SABER-related effective emission height $h_\mathrm{eff}$ and the effective
decadal logarithm of the air density for the emission layer
$(\log n)_\mathrm{eff}$, which are good indicators for the change of the non-LTE
effects. This is demonstrated by $v'$-averaged $\Delta T_\mathrm{NLTE}$
sensitivities for $h_\mathrm{eff}$ and $(\log n)_\mathrm{eff}$ from the two SABER
OH channels of $+1.0 \pm 0.2$\,K per km and $-17 \pm 3$\,K relative to a
reference $v'$ of 3. Using the relation between $h_\mathrm{eff}$ or
$(\log n)_\mathrm{eff}$ and the solar radio flux for both SABER OH channels and
the full sample, the resulting mean solar cycle effect for
$\Delta T_\mathrm{NLTE}$ is $-0.1 \pm 0.1$\,K, which agrees with our other
estimates. This approach also provides a result for the long-term term trend,
which is consistent with zero. The small contribution of non-LTE effects to the
solar cycle variability is caused by only weak long-term variations of the OH
emission profile, which is in accordance with other studies. The negative sign
in our estimates for the non-LTE effects seems to be related to an increase in
the air density in the OH emission layer for higher solar activity.

Our results show that $\Delta T_\mathrm{NLTE}$ contributions to OH
$T_\mathrm{rot}$, which can be critical on shorter time scales, do not appear to
significantly affect the strong solar cycle effects found in our data. This is
good news for studies of $T_\mathrm{rot}$ long-term variations. The global
application of our findings depends on the stability of the OH emission layer.
So far, the corresponding results are promising. Our study also shows that it
is necessary to cover a sufficiently long period with a wide range of solar
activity levels. Otherwise significant systematic deviations could be found for
the total effect as well as the non-LTE contributions. In part, this might be
related to non-linearities in the response of the OH emission layer to solar
forcing.

\section*{Acknowledgments}

This project made use of the ESO Science Archive Facility. UVES Phase\,3
spectra from different observing programmes of the period from April 2000 to
March 2015 were analysed. We thank the SABER team for providing the data
products used in this paper. Moreover, we thank the two anonymous referees for
their detailed and very helpful comments.
S.~Noll receives and B.~Proxauf and S.~Unterguggenberger received funding from
the project P26130 of the Austrian Science Fund (FWF). W.~Kausch was funded by
the project IS538003 (Hochschulraumstrukturmittel) provided by the Austrian
Ministry for Research (BMWFW).

\section*{References}

\bibliography{Nolletal2017a}

\begin{thebibliography}{62}
\expandafter\ifx\csname natexlab\endcsname\relax\def\natexlab#1{#1}\fi
\providecommand{\url}[1]{\texttt{#1}}
\providecommand{\href}[2]{#2}
\providecommand{\path}[1]{#1}
\providecommand{\DOIprefix}{doi:}
\providecommand{\ArXivprefix}{arXiv:}
\providecommand{\URLprefix}{URL: }
\providecommand{\Pubmedprefix}{pmid:}
\providecommand{\doi}[1]{\href{http://dx.doi.org/#1}{\path{#1}}}
\providecommand{\Pubmed}[1]{\href{pmid:#1}{\path{#1}}}
\providecommand{\bibinfo}[2]{#2}
\ifx\xfnm\relax \def\xfnm[#1]{\unskip,\space#1}\fi
\bibitem[{{Adler-Golden}(1997)}]{adler97}
\bibinfo{author}{{Adler-Golden}, S.}, \bibinfo{year}{1997}.
\newblock \bibinfo{title}{{Kinetic parameters for OH nightglow modeling
  consistent with recent laboratory measurements}}.
\newblock \bibinfo{journal}{J. Geophys. Res.} \bibinfo{volume}{102},
  \bibinfo{pages}{19969--19976}.
\newblock \DOIprefix\doi{10.1029/97JA01622}.
\bibitem[{{Baker} and {Stair}(1988)}]{baker88}
\bibinfo{author}{{Baker}, D.J.}, \bibinfo{author}{{Stair}, Jr., A.T.},
  \bibinfo{year}{1988}.
\newblock \bibinfo{title}{{Rocket measurements of the altitude distributions of
  the hydroxyl airglow}}.
\newblock \bibinfo{journal}{Phys. Scripta} \bibinfo{volume}{37},
  \bibinfo{pages}{611--622}.
\newblock \DOIprefix\doi{10.1088/0031-8949/37/4/021}.
\bibitem[{{Bates} and {Nicolet}(1950)}]{bates50}
\bibinfo{author}{{Bates}, D.R.}, \bibinfo{author}{{Nicolet}, M.},
  \bibinfo{year}{1950}.
\newblock \bibinfo{title}{{The Photochemistry of Atmospheric Water Vapor}}.
\newblock \bibinfo{journal}{J. Geophys. Res.} \bibinfo{volume}{55},
  \bibinfo{pages}{301--327}.
\newblock \DOIprefix\doi{10.1029/JZ055i003p00301}.
\bibitem[{{Beig}(2011a)}]{beig11a}
\bibinfo{author}{{Beig}, G.}, \bibinfo{year}{2011}a.
\newblock \bibinfo{title}{{Long-term trends in the temperature of the
  mesosphere/lower thermosphere region: 1. Anthropogenic influences}}.
\newblock \bibinfo{journal}{J. Geophys. Res.} \bibinfo{volume}{116},
  \bibinfo{pages}{A00H11}.
\newblock \DOIprefix\doi{10.1029/2011JA016646}.
\bibitem[{{Beig}(2011b)}]{beig11b}
\bibinfo{author}{{Beig}, G.}, \bibinfo{year}{2011}b.
\newblock \bibinfo{title}{{Long-term trends in the temperature of the
  mesosphere/lower thermosphere region: 2. Solar response}}.
\newblock \bibinfo{journal}{J. Geophys. Res.} \bibinfo{volume}{116},
  \bibinfo{pages}{A00H12}.
\newblock \DOIprefix\doi{10.1029/2011JA016766}.
\bibitem[{{Beig} et~al.(2003){Beig}, {Keckhut}, {Lowe}, {Roble}, {Mlynczak},
  {Scheer}, {Fomichev}, {Offermann}, {French}, {Shepherd}, {Semenov},
  {Remsberg}, {She}, {L{\"u}bken}, {Bremer}, {Clemesha}, {Stegman}, {Sigernes}
  and {Fadnavis}}]{beig03}
\bibinfo{author}{{Beig}, G.}, \bibinfo{author}{{Keckhut}, P.},
  \bibinfo{author}{{Lowe}, R.P.}, \bibinfo{author}{{Roble}, R.G.},
  \bibinfo{author}{{Mlynczak}, M.G.}, \bibinfo{author}{{Scheer}, J.},
  \bibinfo{author}{{Fomichev}, V.I.}, \bibinfo{author}{{Offermann}, D.},
  \bibinfo{author}{{French}, W.J.R.}, \bibinfo{author}{{Shepherd}, M.G.},
  \bibinfo{author}{{Semenov}, A.I.}, \bibinfo{author}{{Remsberg}, E.E.},
  \bibinfo{author}{{She}, C.Y.}, \bibinfo{author}{{L{\"u}bken}, F.J.},
  \bibinfo{author}{{Bremer}, J.}, \bibinfo{author}{{Clemesha}, B.R.},
  \bibinfo{author}{{Stegman}, J.}, \bibinfo{author}{{Sigernes}, F.},
  \bibinfo{author}{{Fadnavis}, S.}, \bibinfo{year}{2003}.
\newblock \bibinfo{title}{{Review of mesospheric temperature trends}}.
\newblock \bibinfo{journal}{Rev. Geophys.} \bibinfo{volume}{41},
  \bibinfo{pages}{RG1015}.
\newblock \DOIprefix\doi{10.1029/2002RG000121}.
\bibitem[{{Beig} et~al.(2008){Beig}, {Scheer}, {Mlynczak} and
  {Keckhut}}]{beig08}
\bibinfo{author}{{Beig}, G.}, \bibinfo{author}{{Scheer}, J.},
  \bibinfo{author}{{Mlynczak}, M.G.}, \bibinfo{author}{{Keckhut}, P.},
  \bibinfo{year}{2008}.
\newblock \bibinfo{title}{{Overview of the temperature response in the
  mesosphere and lower thermosphere to solar activity}}.
\newblock \bibinfo{journal}{Rev. Geophys.} \bibinfo{volume}{46},
  \bibinfo{pages}{RG3002}.
\newblock \DOIprefix\doi{10.1029/2007RG000236}.
\bibitem[{{Brooke} et~al.(2016){Brooke}, {Bernath}, {Western}, {Sneden}, {Af{\c
  s}ar}, {Li} and {Gordon}}]{brooke16}
\bibinfo{author}{{Brooke}, J.S.A.}, \bibinfo{author}{{Bernath}, P.F.},
  \bibinfo{author}{{Western}, C.M.}, \bibinfo{author}{{Sneden}, C.},
  \bibinfo{author}{{Af{\c s}ar}, M.}, \bibinfo{author}{{Li}, G.},
  \bibinfo{author}{{Gordon}, I.E.}, \bibinfo{year}{2016}.
\newblock \bibinfo{title}{{Line strengths of rovibrational and rotational
  transitions in the X$^{2}${$\Pi$} ground state of OH}}.
\newblock \bibinfo{journal}{J. Quant. Spectrosc. Radiat. Transf.}
  \bibinfo{volume}{168}, \bibinfo{pages}{142--157}.
\newblock \DOIprefix\doi{10.1016/j.jqsrt.2015.07.021}.
\bibitem[{{Clemesha} et~al.(2005){Clemesha}, {Takahashi}, {Simonich}, {Gobbi}
  and {Batista}}]{clemesha05}
\bibinfo{author}{{Clemesha}, B.}, \bibinfo{author}{{Takahashi}, H.},
  \bibinfo{author}{{Simonich}, D.}, \bibinfo{author}{{Gobbi}, D.},
  \bibinfo{author}{{Batista}, P.}, \bibinfo{year}{2005}.
\newblock \bibinfo{title}{{Experimental evidence for solar cycle and long-term
  change in the low-latitude MLT region}}.
\newblock \bibinfo{journal}{J. Atmos. Sol.-Terr. Physics} \bibinfo{volume}{67},
  \bibinfo{pages}{191--196}.
\newblock \DOIprefix\doi{10.1016/j.jastp.2004.07.027}.
\bibitem[{{Clough} et~al.(2005){Clough}, {Shephard}, {Mlawer}, {Delamere},
  {Iacono}, {Cady-Pereira}, {Boukabara} and {Brown}}]{clough05}
\bibinfo{author}{{Clough}, S.A.}, \bibinfo{author}{{Shephard}, M.W.},
  \bibinfo{author}{{Mlawer}, E.J.}, \bibinfo{author}{{Delamere}, J.S.},
  \bibinfo{author}{{Iacono}, M.J.}, \bibinfo{author}{{Cady-Pereira}, K.},
  \bibinfo{author}{{Boukabara}, S.}, \bibinfo{author}{{Brown}, P.D.},
  \bibinfo{year}{2005}.
\newblock \bibinfo{title}{{Atmospheric radiative transfer modeling: a summary
  of the AER codes}}.
\newblock \bibinfo{journal}{J. Quant. Spectrosc. Radiat. Transf.}
  \bibinfo{volume}{91}, \bibinfo{pages}{233--244}.
\newblock \DOIprefix\doi{10.1016/j.jqsrt.2004.05.058}.
\bibitem[{{Cosby} et~al.(2006){Cosby}, {Sharpee}, {Slanger}, {Huestis} and
  {Hanuschik}}]{cosby06}
\bibinfo{author}{{Cosby}, P.C.}, \bibinfo{author}{{Sharpee}, B.D.},
  \bibinfo{author}{{Slanger}, T.G.}, \bibinfo{author}{{Huestis}, D.L.},
  \bibinfo{author}{{Hanuschik}, R.W.}, \bibinfo{year}{2006}.
\newblock \bibinfo{title}{{High-resolution terrestrial nightglow emission line
  atlas from UVES/VLT: Positions, intensities, and identifications for 2808
  lines at 314-1043 nm}}.
\newblock \bibinfo{journal}{J. Geophys. Res.} \bibinfo{volume}{111},
  \bibinfo{pages}{A12307}.
\newblock \DOIprefix\doi{10.1029/2006JA012023}.
\bibitem[{{Cosby} and {Slanger}(2007)}]{cosby07}
\bibinfo{author}{{Cosby}, P.C.}, \bibinfo{author}{{Slanger}, T.G.},
  \bibinfo{year}{2007}.
\newblock \bibinfo{title}{{OH spectroscopy and chemistry investigated with
  astronomical sky spectra}}.
\newblock \bibinfo{journal}{Can. J. Phys.} \bibinfo{volume}{85},
  \bibinfo{pages}{77--99}.
\newblock \DOIprefix\doi{10.1139/P06-088}.
\bibitem[{{Dekker} et~al.(2000){Dekker}, {D'Odorico}, {Kaufer}, {Delabre} and
  {Kotzlowski}}]{dekker00}
\bibinfo{author}{{Dekker}, H.}, \bibinfo{author}{{D'Odorico}, S.},
  \bibinfo{author}{{Kaufer}, A.}, \bibinfo{author}{{Delabre}, B.},
  \bibinfo{author}{{Kotzlowski}, H.}, \bibinfo{year}{2000}.
\newblock \bibinfo{title}{{Design, construction, and performance of UVES, the
  echelle spectrograph for the UT2 Kueyen Telescope at the ESO Paranal
  Observatory}}, in: \bibinfo{editor}{{Iye}, M.}, \bibinfo{editor}{{Moorwood},
  A.F.} (Eds.), \bibinfo{booktitle}{Optical and IR Telescope Instrumentation
  and Detectors}, pp. \bibinfo{pages}{534--545}.
\newblock \DOIprefix\doi{10.1117/12.395512}.
\bibitem[{{Dodd} et~al.(1994){Dodd}, {Armstrong}, {Lipson}, {Lowell},
  {Blumberg}, {Nadile}, {Adler-Golden}, {Marinelli}, {Holtzclaw} and
  {Green}}]{dodd94}
\bibinfo{author}{{Dodd}, J.A.}, \bibinfo{author}{{Armstrong}, P.S.},
  \bibinfo{author}{{Lipson}, S.J.}, \bibinfo{author}{{Lowell}, J.R.},
  \bibinfo{author}{{Blumberg}, W.A.M.}, \bibinfo{author}{{Nadile}, R.M.},
  \bibinfo{author}{{Adler-Golden}, S.M.}, \bibinfo{author}{{Marinelli}, W.J.},
  \bibinfo{author}{{Holtzclaw}, K.W.}, \bibinfo{author}{{Green}, B.D.},
  \bibinfo{year}{1994}.
\newblock \bibinfo{title}{{Analysis of hydroxyl earthlimb airglow emissions:
  Kinetic model for state-to-state dynamics of OH(v,N)}}.
\newblock \bibinfo{journal}{J. Geophys. Res.} \bibinfo{volume}{99},
  \bibinfo{pages}{3559--3586}.
\newblock \DOIprefix\doi{10.1029/93JD03338}.
\bibitem[{{Dudok de Wit} et~al.(2008){Dudok de Wit}, {Kretzschmar},
  {Aboudarham}, {Amblard}, {Auch{\`e}re} and {Lilensten}}]{dudok08}
\bibinfo{author}{{Dudok de Wit}, T.}, \bibinfo{author}{{Kretzschmar}, M.},
  \bibinfo{author}{{Aboudarham}, J.}, \bibinfo{author}{{Amblard}, P.O.},
  \bibinfo{author}{{Auch{\`e}re}, F.}, \bibinfo{author}{{Lilensten}, J.},
  \bibinfo{year}{2008}.
\newblock \bibinfo{title}{{Which solar EUV indices are best for reconstructing
  the solar EUV irradiance?}}
\newblock \bibinfo{journal}{Adv. Space Res.} \bibinfo{volume}{42},
  \bibinfo{pages}{903--911}.
\newblock \DOIprefix\doi{10.1016/j.asr.2007.04.019}.
\bibitem[{{Dudok de Wit} et~al.(2009){Dudok de Wit}, {Kretzschmar}, {Lilensten}
  and {Woods}}]{dudok09}
\bibinfo{author}{{Dudok de Wit}, T.}, \bibinfo{author}{{Kretzschmar}, M.},
  \bibinfo{author}{{Lilensten}, J.}, \bibinfo{author}{{Woods}, T.},
  \bibinfo{year}{2009}.
\newblock \bibinfo{title}{{Finding the best proxies for the solar UV
  irradiance}}.
\newblock \bibinfo{journal}{Geophys. Res. Lett.} \bibinfo{volume}{36},
  \bibinfo{pages}{L10107}.
\newblock \DOIprefix\doi{10.1029/2009GL037825}.
\bibitem[{{Emmert} et~al.(2010){Emmert}, {Lean} and {Picone}}]{emmert10}
\bibinfo{author}{{Emmert}, J.T.}, \bibinfo{author}{{Lean}, J.L.},
  \bibinfo{author}{{Picone}, J.M.}, \bibinfo{year}{2010}.
\newblock \bibinfo{title}{{Record-low thermospheric density during the 2008
  solar minimum}}.
\newblock \bibinfo{journal}{Geophys. Res. Lett.} \bibinfo{volume}{37},
  \bibinfo{pages}{L12102}.
\newblock \DOIprefix\doi{10.1029/2010GL043671}.
\bibitem[{{Gao} et~al.(2016){Gao}, {Xu} and {Chen}}]{gao16}
\bibinfo{author}{{Gao}, H.}, \bibinfo{author}{{Xu}, J.},
  \bibinfo{author}{{Chen}, G.M.}, \bibinfo{year}{2016}.
\newblock \bibinfo{title}{{The responses of the nightglow emissions observed by
  the TIMED/SABER satellite to solar radiation}}.
\newblock \bibinfo{journal}{J. Geophys. Res.} \bibinfo{volume}{121},
  \bibinfo{pages}{1627--1642}.
\newblock \DOIprefix\doi{10.1002/2015JA021624}.
\bibitem[{{Gelinas} et~al.(2008){Gelinas}, {Hecht}, {Walterscheid}, {Roble} and
  {Woithe}}]{gelinas08}
\bibinfo{author}{{Gelinas}, L.J.}, \bibinfo{author}{{Hecht}, J.H.},
  \bibinfo{author}{{Walterscheid}, R.L.}, \bibinfo{author}{{Roble}, R.G.},
  \bibinfo{author}{{Woithe}, J.M.}, \bibinfo{year}{2008}.
\newblock \bibinfo{title}{{A seasonal study of mesospheric temperatures and
  emission intensities at Adelaide and Alice Springs}}.
\newblock \bibinfo{journal}{J. Geophys. Res.} \bibinfo{volume}{113},
  \bibinfo{pages}{A01304}.
\newblock \DOIprefix\doi{10.1029/2007JA012587}.
\bibitem[{{Goldman} et~al.(1998){Goldman}, {Schoenfeld}, {Goorvitch},
  {Chackerian}, {Dothe}, {M{\'e}len}, {Abrams} and {Selby}}]{goldman98}
\bibinfo{author}{{Goldman}, A.}, \bibinfo{author}{{Schoenfeld}, W.G.},
  \bibinfo{author}{{Goorvitch}, D.}, \bibinfo{author}{{Chackerian}, Jr., C.},
  \bibinfo{author}{{Dothe}, H.}, \bibinfo{author}{{M{\'e}len}, F.},
  \bibinfo{author}{{Abrams}, M.C.}, \bibinfo{author}{{Selby}, J.E.A.},
  \bibinfo{year}{1998}.
\newblock \bibinfo{title}{{Updated line parameters for OH X$^{2}$II-X$^{2}$II
  ($v'$,$v''$) transitions.}}
\newblock \bibinfo{journal}{J. Quant. Spectrosc. Radiat. Transf.}
  \bibinfo{volume}{59}, \bibinfo{pages}{453--469}.
\newblock \DOIprefix\doi{10.1016/S0022-4073(97)00112-X}.
\bibitem[{{Hamuy} et~al.(1994){Hamuy}, {Suntzeff}, {Heathcote}, {Walker},
  {Gigoux} and {Phillips}}]{hamuy94}
\bibinfo{author}{{Hamuy}, M.}, \bibinfo{author}{{Suntzeff}, N.B.},
  \bibinfo{author}{{Heathcote}, S.R.}, \bibinfo{author}{{Walker}, A.R.},
  \bibinfo{author}{{Gigoux}, P.}, \bibinfo{author}{{Phillips}, M.M.},
  \bibinfo{year}{1994}.
\newblock \bibinfo{title}{{Southern spectrophotometric standards, 2}}.
\newblock \bibinfo{journal}{Publ. Astron. Soc. Pac.} \bibinfo{volume}{106},
  \bibinfo{pages}{566--589}.
\newblock \DOIprefix\doi{10.1086/133417}.
\bibitem[{{Hanuschik}(2003)}]{hanuschik03}
\bibinfo{author}{{Hanuschik}, R.W.}, \bibinfo{year}{2003}.
\newblock \bibinfo{title}{{A flux-calibrated, high-resolution atlas of optical
  sky emission from UVES}}.
\newblock \bibinfo{journal}{Astron. Astrophys.} \bibinfo{volume}{407},
  \bibinfo{pages}{1157--1164}.
\newblock \DOIprefix\doi{10.1051/0004-6361:20030885}.
\bibitem[{{Huang} et~al.(2016){Huang}, {Mayr}, {Russell} and
  {Mlynczak}}]{huang16}
\bibinfo{author}{{Huang}, F.T.}, \bibinfo{author}{{Mayr}, H.G.},
  \bibinfo{author}{{Russell}, III, J.M.}, \bibinfo{author}{{Mlynczak}, M.G.},
  \bibinfo{year}{2016}.
\newblock \bibinfo{title}{{Ozone and temperature decadal responses to solar
  variability in the stratosphere and lower mesosphere, based on measurements
  from SABER on TIMED}}.
\newblock \bibinfo{journal}{Ann. Geophys.} \bibinfo{volume}{34},
  \bibinfo{pages}{801--813}.
\newblock \DOIprefix\doi{10.5194/angeo-34-801-2016}.
\bibitem[{{Kalicinsky} et~al.(2016){Kalicinsky}, {Knieling}, {Koppmann},
  {Offermann}, {Steinbrecht} and {Wintel}}]{kalicinsky16}
\bibinfo{author}{{Kalicinsky}, C.}, \bibinfo{author}{{Knieling}, P.},
  \bibinfo{author}{{Koppmann}, R.}, \bibinfo{author}{{Offermann}, D.},
  \bibinfo{author}{{Steinbrecht}, W.}, \bibinfo{author}{{Wintel}, J.},
  \bibinfo{year}{2016}.
\newblock \bibinfo{title}{{Long-term dynamics of OH$^{*}$ temperatures over
  central Europe: trends and solar correlations}}.
\newblock \bibinfo{journal}{Atmos. Chem. Phys.} \bibinfo{volume}{16},
  \bibinfo{pages}{15033--15047}.
\newblock \DOIprefix\doi{10.5194/acp-16-15033-2016}.
\bibitem[{{Kalogerakis} et~al.(2016){Kalogerakis}, {Matsiev}, {Sharma} and
  {Wintersteiner}}]{kalogerakis16}
\bibinfo{author}{{Kalogerakis}, K.S.}, \bibinfo{author}{{Matsiev}, D.},
  \bibinfo{author}{{Sharma}, R.D.}, \bibinfo{author}{{Wintersteiner}, P.P.},
  \bibinfo{year}{2016}.
\newblock \bibinfo{title}{{Resolving the mesospheric nighttime 4.3 $\mu$m
  emission puzzle: Laboratory demonstration of new mechanism for OH($v$)
  relaxation}}.
\newblock \bibinfo{journal}{Geophys. Res Lett.} \bibinfo{volume}{43},
  \bibinfo{pages}{8835--8843}.
\newblock \DOIprefix\doi{10.1002/2016GL069645}.
\bibitem[{{Kausch} et~al.(2015){Kausch}, {Noll}, {Smette}, {Kimeswenger},
  {Barden}, {Szyszka}, {Jones}, {Sana}, {Horst} and {Kerber}}]{kausch15}
\bibinfo{author}{{Kausch}, W.}, \bibinfo{author}{{Noll}, S.},
  \bibinfo{author}{{Smette}, A.}, \bibinfo{author}{{Kimeswenger}, S.},
  \bibinfo{author}{{Barden}, M.}, \bibinfo{author}{{Szyszka}, C.},
  \bibinfo{author}{{Jones}, A.M.}, \bibinfo{author}{{Sana}, H.},
  \bibinfo{author}{{Horst}, H.}, \bibinfo{author}{{Kerber}, F.},
  \bibinfo{year}{2015}.
\newblock \bibinfo{title}{{Molecfit: A general tool for telluric absorption
  correction. II. Quantitative evaluation on ESO-VLT/X-Shooter spectra}}.
\newblock \bibinfo{journal}{Astron. Astrophys.} \bibinfo{volume}{576},
  \bibinfo{pages}{A78}.
\newblock \DOIprefix\doi{10.1051/0004-6361/201423909}.
\bibitem[{{Khomich} et~al.(2008){Khomich}, {Semenov} and {Shefov}}]{khomich08}
\bibinfo{author}{{Khomich}, V.Y.}, \bibinfo{author}{{Semenov}, A.I.},
  \bibinfo{author}{{Shefov}, N.N.}, \bibinfo{year}{2008}.
\newblock \bibinfo{title}{{Airglow as an Indicator of Upper Atmospheric
  Structure and Dynamics}}.
\newblock \bibinfo{publisher}{Springer, Berlin}.
\bibitem[{{Kliner} and {Farrow}(1999)}]{kliner99}
\bibinfo{author}{{Kliner}, D.A.V.}, \bibinfo{author}{{Farrow}, R.L.},
  \bibinfo{year}{1999}.
\newblock \bibinfo{title}{{Measurements of ground-state OH rotational
  energy-transfer rates}}.
\newblock \bibinfo{journal}{J. Chem. Phys.} \bibinfo{volume}{110},
  \bibinfo{pages}{412--422}.
\newblock \DOIprefix\doi{10.1063/1.478073}.
\bibitem[{{Liu} and {Shepherd}(2006)}]{liu06}
\bibinfo{author}{{Liu}, G.}, \bibinfo{author}{{Shepherd}, G.G.},
  \bibinfo{year}{2006}.
\newblock \bibinfo{title}{{An empirical model for the altitude of the OH
  nightglow emission}}.
\newblock \bibinfo{journal}{Geophys. Res. Lett.} \bibinfo{volume}{33},
  \bibinfo{pages}{L09805}.
\newblock \DOIprefix\doi{10.1029/2005GL025297}.
\bibitem[{{Liu} et~al.(2015){Liu}, {Xu}, {Smith} and {Yuan}}]{liu15}
\bibinfo{author}{{Liu}, W.}, \bibinfo{author}{{Xu}, J.},
  \bibinfo{author}{{Smith}, A.K.}, \bibinfo{author}{{Yuan}, W.},
  \bibinfo{year}{2015}.
\newblock \bibinfo{title}{{Comparison of rotational temperature derived from
  ground-based OH airglow observations with TIMED/SABER to evaluate the
  Einstein coefficients}}.
\newblock \bibinfo{journal}{J. Geophys. Res.} \bibinfo{volume}{120},
  \bibinfo{pages}{10,069--10,082}.
\newblock \DOIprefix\doi{10.1002/2015JA021886}.
\bibitem[{{Meinel}(1950)}]{meinel50}
\bibinfo{author}{{Meinel}, I.A.B.}, \bibinfo{year}{1950}.
\newblock \bibinfo{title}{{OH Emission Bands in the Spectrum of the Night
  Sky.}}
\newblock \bibinfo{journal}{Astrophys. J.} \bibinfo{volume}{111},
  \bibinfo{pages}{555}.
\newblock \DOIprefix\doi{10.1086/145296}.
\bibitem[{{Mies}(1974)}]{mies74}
\bibinfo{author}{{Mies}, F.H.}, \bibinfo{year}{1974}.
\newblock \bibinfo{title}{{Calculated vibrational transition probabilities of
  OH( X$^{2}${$\Pi$})}}.
\newblock \bibinfo{journal}{J. Molec. Spectrosc.} \bibinfo{volume}{53},
  \bibinfo{pages}{150--188}.
\newblock \DOIprefix\doi{10.1016/0022-2852(74)90125-8}.
\bibitem[{{Moehler} et~al.(2014){Moehler}, {Modigliani}, {Freudling},
  {Giammichele}, {Gianninas}, {Gonneau}, {Kausch}, {Lan{\c c}on}, {Noll},
  {Rauch} and {Vinther}}]{moehler14}
\bibinfo{author}{{Moehler}, S.}, \bibinfo{author}{{Modigliani}, A.},
  \bibinfo{author}{{Freudling}, W.}, \bibinfo{author}{{Giammichele}, N.},
  \bibinfo{author}{{Gianninas}, A.}, \bibinfo{author}{{Gonneau}, A.},
  \bibinfo{author}{{Kausch}, W.}, \bibinfo{author}{{Lan{\c c}on}, A.},
  \bibinfo{author}{{Noll}, S.}, \bibinfo{author}{{Rauch}, T.},
  \bibinfo{author}{{Vinther}, J.}, \bibinfo{year}{2014}.
\newblock \bibinfo{title}{{Flux calibration of medium-resolution spectra from
  300 nm to 2500 nm: Model reference spectra and telluric correction}}.
\newblock \bibinfo{journal}{Astron. Astrophys.} \bibinfo{volume}{568},
  \bibinfo{pages}{A9}.
\newblock \DOIprefix\doi{10.1051/0004-6361/201423790}.
\bibitem[{{Nath} and {Sridharan}(2014)}]{nath14}
\bibinfo{author}{{Nath}, O.}, \bibinfo{author}{{Sridharan}, S.},
  \bibinfo{year}{2014}.
\newblock \bibinfo{title}{{Long-term variabilities and tendencies in zonal mean
  TIMED-SABER ozone and temperature in the middle atmosphere at
  10-15$^{\circ}$N}}.
\newblock \bibinfo{journal}{J. Atmos. Sol.-Terr. Phys.} \bibinfo{volume}{120},
  \bibinfo{pages}{1--8}.
\newblock \DOIprefix\doi{10.1016/j.jastp.2014.08.010}.
\bibitem[{{Noll} et~al.(2012){Noll}, {Kausch}, {Barden}, {Jones}, {Szyszka},
  {Kimeswenger} and {Vinther}}]{noll12}
\bibinfo{author}{{Noll}, S.}, \bibinfo{author}{{Kausch}, W.},
  \bibinfo{author}{{Barden}, M.}, \bibinfo{author}{{Jones}, A.M.},
  \bibinfo{author}{{Szyszka}, C.}, \bibinfo{author}{{Kimeswenger}, S.},
  \bibinfo{author}{{Vinther}, J.}, \bibinfo{year}{2012}.
\newblock \bibinfo{title}{{An atmospheric radiation model for Cerro Paranal. I.
  The optical spectral range}}.
\newblock \bibinfo{journal}{Astron. Astrophys.} \bibinfo{volume}{543},
  \bibinfo{pages}{A92}.
\newblock \DOIprefix\doi{10.1051/0004-6361/201219040}.
\bibitem[{{Noll} et~al.(2015){Noll}, {Kausch}, {Kimeswenger},
  {Unterguggenberger} and {Jones}}]{noll15}
\bibinfo{author}{{Noll}, S.}, \bibinfo{author}{{Kausch}, W.},
  \bibinfo{author}{{Kimeswenger}, S.}, \bibinfo{author}{{Unterguggenberger},
  S.}, \bibinfo{author}{{Jones}, A.M.}, \bibinfo{year}{2015}.
\newblock \bibinfo{title}{{OH populations and temperatures from simultaneous
  spectroscopic observations of 25 bands}}.
\newblock \bibinfo{journal}{Atmos. Chem. Phys.} \bibinfo{volume}{15},
  \bibinfo{pages}{3647--3669}.
\newblock \DOIprefix\doi{10.5194/acp-15-3647-2015}.
\bibitem[{{Noll} et~al.(2016){Noll}, {Kausch}, {Kimeswenger},
  {Unterguggenberger} and {Jones}}]{noll16}
\bibinfo{author}{{Noll}, S.}, \bibinfo{author}{{Kausch}, W.},
  \bibinfo{author}{{Kimeswenger}, S.}, \bibinfo{author}{{Unterguggenberger},
  S.}, \bibinfo{author}{{Jones}, A.M.}, \bibinfo{year}{2016}.
\newblock \bibinfo{title}{{Comparison of VLT/X-shooter OH and O$_{2}$
  rotational temperatures with consideration of TIMED/SABER emission and
  temperature profiles}}.
\newblock \bibinfo{journal}{Atmos. Chem. Phys.} \bibinfo{volume}{16},
  \bibinfo{pages}{5021--5042}.
\newblock \DOIprefix\doi{10.5194/acp-16-5021-2016}.
\bibitem[{{Parihar} et~al.(2017){Parihar}, {Singh} and {Gurubaran}}]{parihar17}
\bibinfo{author}{{Parihar}, N.}, \bibinfo{author}{{Singh}, D.},
  \bibinfo{author}{{Gurubaran}, S.}, \bibinfo{year}{2017}.
\newblock \bibinfo{title}{{A comparison of ground-based hydroxyl airglow
  temperatures with SABER/TIMED measurements over 23$^{\circ}$ N, India}}.
\newblock \bibinfo{journal}{Ann. Geophys.} \bibinfo{volume}{35},
  \bibinfo{pages}{353--363}.
\newblock \DOIprefix\doi{10.5194/angeo-35-353-2017}.
\bibitem[{{Pendleton} and {Taylor}(2002)}]{pendleton02}
\bibinfo{author}{{Pendleton}, Jr., W.R.}, \bibinfo{author}{{Taylor}, M.J.},
  \bibinfo{year}{2002}.
\newblock \bibinfo{title}{{The impact of L-uncoupling on Einstein coefficients
  for the OH Meinel (6,2) band: implications for Q-branch rotational
  temperatures}}.
\newblock \bibinfo{journal}{J. Atmos. Sol.-Terr. Phys.} \bibinfo{volume}{64},
  \bibinfo{pages}{971--983}.
\newblock \DOIprefix\doi{10.1016/S1364-6826(02)00051-2}.
\bibitem[{{Pertsev} and {Perminov}(2008)}]{pertsev08}
\bibinfo{author}{{Pertsev}, N.}, \bibinfo{author}{{Perminov}, V.},
  \bibinfo{year}{2008}.
\newblock \bibinfo{title}{{Response of the mesopause airglow to solar activity
  inferred from measurements at Zvenigorod, Russia}}.
\newblock \bibinfo{journal}{Ann. Geophys.} \bibinfo{volume}{26},
  \bibinfo{pages}{1049--1056}.
\newblock \DOIprefix\doi{10.5194/angeo-26-1049-2008}.
\bibitem[{{Reid} et~al.(2014){Reid}, {Spargo} and {Woithe}}]{reid14}
\bibinfo{author}{{Reid}, I.M.}, \bibinfo{author}{{Spargo}, A.J.},
  \bibinfo{author}{{Woithe}, J.M.}, \bibinfo{year}{2014}.
\newblock \bibinfo{title}{{Seasonal variations of the nighttime O($^{1}$S) and
  OH(8-3) airglow intensity at Adelaide, Australia}}.
\newblock \bibinfo{journal}{J. Geophys. Res.} \bibinfo{volume}{119},
  \bibinfo{pages}{6991--7013}.
\newblock \DOIprefix\doi{10.1002/2013JD020906}.
\bibitem[{{Reisin} and {Scheer}(2002)}]{reisin02}
\bibinfo{author}{{Reisin}, E.R.}, \bibinfo{author}{{Scheer}, J.},
  \bibinfo{year}{2002}.
\newblock \bibinfo{title}{{Evidence of change after 2001 in the seasonal
  behaviour of the mesopause region from airglow data at El Leoncito}}.
\newblock \bibinfo{journal}{Adv. Space Res.} \bibinfo{volume}{44},
  \bibinfo{pages}{401--412}.
\newblock \DOIprefix\doi{10.1016/j.asr.2009.04.007}.
\bibitem[{{Reisin} et~al.(2014){Reisin}, {Scheer}, {Dyrland}, {Sigernes},
  {Deehr}, {Schmidt}, {H{\"o}ppner}, {Bittner}, {Ammosov}, {Gavrilyeva},
  {Stegman}, {Perminov}, {Semenov}, {Knieling}, {Koppmann}, {Shiokawa}, {Lowe},
  {L{\'o}pez-Gonz{\'a}lez}, {Rodr{\'{\i}}guez}, {Zhao}, {Taylor}, {Buriti},
  {Espy}, {French}, {Eichmann}, {Burrows} and {von Savigny}}]{reisin14}
\bibinfo{author}{{Reisin}, E.R.}, \bibinfo{author}{{Scheer}, J.},
  \bibinfo{author}{{Dyrland}, M.E.}, \bibinfo{author}{{Sigernes}, F.},
  \bibinfo{author}{{Deehr}, C.S.}, \bibinfo{author}{{Schmidt}, C.},
  \bibinfo{author}{{H{\"o}ppner}, K.}, \bibinfo{author}{{Bittner}, M.},
  \bibinfo{author}{{Ammosov}, P.P.}, \bibinfo{author}{{Gavrilyeva}, G.A.},
  \bibinfo{author}{{Stegman}, J.}, \bibinfo{author}{{Perminov}, V.I.},
  \bibinfo{author}{{Semenov}, A.I.}, \bibinfo{author}{{Knieling}, P.},
  \bibinfo{author}{{Koppmann}, R.}, \bibinfo{author}{{Shiokawa}, K.},
  \bibinfo{author}{{Lowe}, R.P.}, \bibinfo{author}{{L{\'o}pez-Gonz{\'a}lez},
  M.J.}, \bibinfo{author}{{Rodr{\'{\i}}guez}, E.}, \bibinfo{author}{{Zhao},
  Y.}, \bibinfo{author}{{Taylor}, M.J.}, \bibinfo{author}{{Buriti}, R.A.},
  \bibinfo{author}{{Espy}, P.J.}, \bibinfo{author}{{French}, W.J.R.},
  \bibinfo{author}{{Eichmann}, K.U.}, \bibinfo{author}{{Burrows}, J.P.},
  \bibinfo{author}{{von Savigny}, C.}, \bibinfo{year}{2014}.
\newblock \bibinfo{title}{{Traveling planetary wave activity from mesopause
  region airglow temperatures determined by the Network for the Detection of
  Mesospheric Change (NDMC)}}.
\newblock \bibinfo{journal}{J. Atmos. Sol.-Terr. Phys.} \bibinfo{volume}{119},
  \bibinfo{pages}{71--82}.
\newblock \DOIprefix\doi{10.1016/j.jastp.2014.07.002}.
\bibitem[{{Rezac} et~al.(2015){Rezac}, {Kutepov}, {Russell}, {Feofilov}, {Yue}
  and {Goldberg}}]{rezac15}
\bibinfo{author}{{Rezac}, L.}, \bibinfo{author}{{Kutepov}, A.},
  \bibinfo{author}{{Russell}, III, J.M.}, \bibinfo{author}{{Feofilov}, A.G.},
  \bibinfo{author}{{Yue}, J.}, \bibinfo{author}{{Goldberg}, R.A.},
  \bibinfo{year}{2015}.
\newblock \bibinfo{title}{{Simultaneous retrieval of T(p) and CO$_{2}$ VMR from
  two-channel non-LTE limb radiances and application to daytime SABER/TIMED
  measurements}}.
\newblock \bibinfo{journal}{J. Atmos. Sol.-Terr. Phys.} \bibinfo{volume}{130},
  \bibinfo{pages}{23--42}.
\newblock \DOIprefix\doi{10.1016/j.jastp.2015.05.004}.
\bibitem[{{Rothman} et~al.(2013){Rothman}, {Gordon}, {Babikov}, {Barbe}, {Chris
  Benner}, {Bernath}, {Birk}, {Bizzocchi}, {Boudon}, {Brown}, {Campargue},
  {Chance}, {Cohen}, {Coudert}, {Devi}, {Drouin}, {Fayt}, {Flaud}, {Gamache},
  {Harrison}, {Hartmann}, {Hill}, {Hodges}, {Jacquemart}, {Jolly}, {Lamouroux},
  {Le Roy}, {Li}, {Long}, {Lyulin}, {Mackie}, {Massie}, {Mikhailenko},
  {M{\"u}ller}, {Naumenko}, {Nikitin}, {Orphal}, {Perevalov}, {Perrin},
  {Polovtseva}, {Richard}, {Smith}, {Starikova}, {Sung}, {Tashkun}, {Tennyson},
  {Toon}, {Tyuterev} and {Wagner}}]{rothman13}
\bibinfo{author}{{Rothman}, L.S.}, \bibinfo{author}{{Gordon}, I.E.},
  \bibinfo{author}{{Babikov}, Y.}, \bibinfo{author}{{Barbe}, A.},
  \bibinfo{author}{{Chris Benner}, D.}, \bibinfo{author}{{Bernath}, P.F.},
  \bibinfo{author}{{Birk}, M.}, \bibinfo{author}{{Bizzocchi}, L.},
  \bibinfo{author}{{Boudon}, V.}, \bibinfo{author}{{Brown}, L.R.},
  \bibinfo{author}{{Campargue}, A.}, \bibinfo{author}{{Chance}, K.},
  \bibinfo{author}{{Cohen}, E.A.}, \bibinfo{author}{{Coudert}, L.H.},
  \bibinfo{author}{{Devi}, V.M.}, \bibinfo{author}{{Drouin}, B.J.},
  \bibinfo{author}{{Fayt}, A.}, \bibinfo{author}{{Flaud}, J.M.},
  \bibinfo{author}{{Gamache}, R.R.}, \bibinfo{author}{{Harrison}, J.J.},
  \bibinfo{author}{{Hartmann}, J.M.}, \bibinfo{author}{{Hill}, C.},
  \bibinfo{author}{{Hodges}, J.T.}, \bibinfo{author}{{Jacquemart}, D.},
  \bibinfo{author}{{Jolly}, A.}, \bibinfo{author}{{Lamouroux}, J.},
  \bibinfo{author}{{Le Roy}, R.J.}, \bibinfo{author}{{Li}, G.},
  \bibinfo{author}{{Long}, D.A.}, \bibinfo{author}{{Lyulin}, O.M.},
  \bibinfo{author}{{Mackie}, C.J.}, \bibinfo{author}{{Massie}, S.T.},
  \bibinfo{author}{{Mikhailenko}, S.}, \bibinfo{author}{{M{\"u}ller}, H.S.P.},
  \bibinfo{author}{{Naumenko}, O.V.}, \bibinfo{author}{{Nikitin}, A.V.},
  \bibinfo{author}{{Orphal}, J.}, \bibinfo{author}{{Perevalov}, V.},
  \bibinfo{author}{{Perrin}, A.}, \bibinfo{author}{{Polovtseva}, E.R.},
  \bibinfo{author}{{Richard}, C.}, \bibinfo{author}{{Smith}, M.A.H.},
  \bibinfo{author}{{Starikova}, E.}, \bibinfo{author}{{Sung}, K.},
  \bibinfo{author}{{Tashkun}, S.}, \bibinfo{author}{{Tennyson}, J.},
  \bibinfo{author}{{Toon}, G.C.}, \bibinfo{author}{{Tyuterev}, V.G.},
  \bibinfo{author}{{Wagner}, G.}, \bibinfo{year}{2013}.
\newblock \bibinfo{title}{{The HITRAN2012 molecular spectroscopic database}}.
\newblock \bibinfo{journal}{J. Quant. Spectrosc. Radiat. Transf.}
  \bibinfo{volume}{130}, \bibinfo{pages}{4--50}.
\newblock \DOIprefix\doi{10.1016/j.jqsrt.2013.07.002}.
\bibitem[{{Rousselot} et~al.(2000){Rousselot}, {Lidman}, {Cuby}, {Moreels} and
  {Monnet}}]{rousselot00}
\bibinfo{author}{{Rousselot}, P.}, \bibinfo{author}{{Lidman}, C.},
  \bibinfo{author}{{Cuby}, J.G.}, \bibinfo{author}{{Moreels}, G.},
  \bibinfo{author}{{Monnet}, G.}, \bibinfo{year}{2000}.
\newblock \bibinfo{title}{{Night-sky spectral atlas of OH emission lines in the
  near-infrared}}.
\newblock \bibinfo{journal}{Astron. Astrophys.} \bibinfo{volume}{354},
  \bibinfo{pages}{1134--1150}.
\bibitem[{{Rozenberg}(1966)}]{rozenberg66}
\bibinfo{author}{{Rozenberg}, G.V.}, \bibinfo{year}{1966}.
\newblock \bibinfo{title}{{Twilight: a Study in Atmospheric Optics}}.
\newblock \bibinfo{publisher}{Plenum Press, New York}.
\bibitem[{{Russell} et~al.(1999){Russell}, {Mlynczak}, {Gordley}, {Tansock} and
  {Esplin}}]{russell99}
\bibinfo{author}{{Russell}, III, J.M.}, \bibinfo{author}{{Mlynczak}, M.G.},
  \bibinfo{author}{{Gordley}, L.L.}, \bibinfo{author}{{Tansock}, J.},
  \bibinfo{author}{{Esplin}, R.}, \bibinfo{year}{1999}.
\newblock \bibinfo{title}{{Overview of the SABER experiment and preliminary
  calibration results}}, in: \bibinfo{editor}{{Larar}, A.M.} (Ed.),
  \bibinfo{booktitle}{Optical Spectroscopic Techniques and Instrumentation for
  Atmospheric and Space Research III}, pp. \bibinfo{pages}{277--288}.
\newblock \DOIprefix\doi{10.1117/12.366382}.
\bibitem[{{Scheer} and {Reisin}(2013)}]{scheer13}
\bibinfo{author}{{Scheer}, J.}, \bibinfo{author}{{Reisin}, E.R.},
  \bibinfo{year}{2013}.
\newblock \bibinfo{title}{{Simpson's paradox in trend analysis: An example from
  El Leoncito airglow data}}.
\newblock \bibinfo{journal}{J. Geophys. Res.} \bibinfo{volume}{118},
  \bibinfo{pages}{5223--5229}.
\newblock \DOIprefix\doi{10.1002/jgra.50461}.
\bibitem[{{Scheer} et~al.(2005){Scheer}, {Reisin} and {Mandrini}}]{scheer05}
\bibinfo{author}{{Scheer}, J.}, \bibinfo{author}{{Reisin}, E.R.},
  \bibinfo{author}{{Mandrini}, C.H.}, \bibinfo{year}{2005}.
\newblock \bibinfo{title}{{Solar activity signatures in mesopause region
  temperatures and atomic oxygen related airglow brightness at El Leoncito,
  Argentina}}.
\newblock \bibinfo{journal}{J. Atmos. Sol.-Terr. Phys.} \bibinfo{volume}{67},
  \bibinfo{pages}{145--154}.
\newblock \DOIprefix\doi{10.1016/j.jastp.2004.07.023}.
\bibitem[{{Schmidt} et~al.(2013){Schmidt}, {H{\"o}ppner} and
  {Bittner}}]{schmidt13}
\bibinfo{author}{{Schmidt}, C.}, \bibinfo{author}{{H{\"o}ppner}, K.},
  \bibinfo{author}{{Bittner}, M.}, \bibinfo{year}{2013}.
\newblock \bibinfo{title}{{A ground-based spectrometer equipped with an InGaAs
  array for routine observations of OH(3-1) rotational temperatures in the
  mesopause region}}.
\newblock \bibinfo{journal}{J. Atmos. Sol.-Terr. Phys.} \bibinfo{volume}{102},
  \bibinfo{pages}{125--139}.
\newblock \DOIprefix\doi{10.1016/j.jastp.2013.05.001}.
\bibitem[{{Smette} et~al.(2015){Smette}, {Sana}, {Noll}, {Horst}, {Kausch},
  {Kimeswenger}, {Barden}, {Szyszka}, {Jones}, {Gallenne}, {Vinther},
  {Ballester} and {Taylor}}]{smette15}
\bibinfo{author}{{Smette}, A.}, \bibinfo{author}{{Sana}, H.},
  \bibinfo{author}{{Noll}, S.}, \bibinfo{author}{{Horst}, H.},
  \bibinfo{author}{{Kausch}, W.}, \bibinfo{author}{{Kimeswenger}, S.},
  \bibinfo{author}{{Barden}, M.}, \bibinfo{author}{{Szyszka}, C.},
  \bibinfo{author}{{Jones}, A.M.}, \bibinfo{author}{{Gallenne}, A.},
  \bibinfo{author}{{Vinther}, J.}, \bibinfo{author}{{Ballester}, P.},
  \bibinfo{author}{{Taylor}, J.}, \bibinfo{year}{2015}.
\newblock \bibinfo{title}{{Molecfit: A general tool for telluric absorption
  correction. I. Method and application to ESO instruments}}.
\newblock \bibinfo{journal}{Astron. Astrophys.} \bibinfo{volume}{576},
  \bibinfo{pages}{A77}.
\newblock \DOIprefix\doi{10.1051/0004-6361/201423932}.
\bibitem[{{Takahashi} et~al.(1995){Takahashi}, {Clemesha} and
  {Batista}}]{takahashi95}
\bibinfo{author}{{Takahashi}, H.}, \bibinfo{author}{{Clemesha}, B.R.},
  \bibinfo{author}{{Batista}, P.P.}, \bibinfo{year}{1995}.
\newblock \bibinfo{title}{{Predominant semi-annual oscillation of the upper
  mesospheric airglow intensities and temperatures in the equatorial region}}.
\newblock \bibinfo{journal}{J. Atmos. Terr. Phys.} \bibinfo{volume}{57},
  \bibinfo{pages}{407--414}.
\newblock \DOIprefix\doi{10.1016/0021-9169(94)E0006-9}.
\bibitem[{{Takahashi} et~al.(1998){Takahashi}, {Gobbi}, {Batista}, {Melo},
  {Teixeira} and {Buriti}}]{takahashi98}
\bibinfo{author}{{Takahashi}, H.}, \bibinfo{author}{{Gobbi}, D.},
  \bibinfo{author}{{Batista}, P.P.}, \bibinfo{author}{{Melo}, S.M.L.},
  \bibinfo{author}{{Teixeira}, N.R.}, \bibinfo{author}{{Buriti}, R.A.},
  \bibinfo{year}{1998}.
\newblock \bibinfo{title}{{Dynamical influence on the equatorial airglow
  observed from the south american sector}}.
\newblock \bibinfo{journal}{Adv. Space Res.} \bibinfo{volume}{21},
  \bibinfo{pages}{817--825}.
\newblock \DOIprefix\doi{10.1016/S0273-1177(97)00680-7}.
\bibitem[{{Tapping}(2013)}]{tapping13}
\bibinfo{author}{{Tapping}, K.F.}, \bibinfo{year}{2013}.
\newblock \bibinfo{title}{{The 10.7 cm solar radio flux (F$_{10.7}$)}}.
\newblock \bibinfo{journal}{Space Weather} \bibinfo{volume}{11},
  \bibinfo{pages}{394--406}.
\newblock \DOIprefix\doi{10.1002/swe.20064}.
\bibitem[{{T{\"u}g}(1977)}]{tueg77}
\bibinfo{author}{{T{\"u}g}, H.}, \bibinfo{year}{1977}.
\newblock \bibinfo{title}{{Vertical Extinction on La Silla}}.
\newblock \bibinfo{journal}{The Messenger} \bibinfo{volume}{11},
  \bibinfo{pages}{7--8}.
\newblock \URLprefix
  \url{https://www.eso.org/sci/publications/messenger/archive/no.11-dec77/messenger-no11.pdf}.
\bibitem[{{van Rhijn}(1921)}]{vanrhijn21}
\bibinfo{author}{{van Rhijn}, P.J.}, \bibinfo{year}{1921}.
\newblock \bibinfo{title}{{On the brightness of the sky at night and the total
  amount of starlight}}.
\newblock \bibinfo{journal}{Publ. Kapteyn Astron. Lab. Groningen}
  \bibinfo{volume}{31}, \bibinfo{pages}{1--83}.
\bibitem[{{Vernet} et~al.(2011){Vernet}, {Dekker}, {D'Odorico}, {Kaper},
  {Kjaergaard}, {Hammer}, {Randich}, {Zerbi}, {Groot}, {Hjorth}, {Guinouard},
  {Navarro}, {Adolfse}, {Albers}, {Amans}, {Andersen}, {Andersen}, {Binetruy},
  {Bristow}, {Castillo}, {Chemla}, {Christensen}, {Conconi}, {Conzelmann},
  {Dam}, {de Caprio}, {de Ugarte Postigo}, {Delabre}, {di Marcantonio},
  {Downing}, {Elswijk}, {Finger}, {Fischer}, {Flores}, {Fran{\c c}ois},
  {Goldoni}, {Guglielmi}, {Haigron}, {Hanenburg}, {Hendriks}, {Horrobin},
  {Horville}, {Jessen}, {Kerber}, {Kern}, {Kiekebusch}, {Kleszcz}, {Klougart},
  {Kragt}, {Larsen}, {Lizon}, {Lucuix}, {Mainieri}, {Manuputy}, {Martayan},
  {Mason}, {Mazzoleni}, {Michaelsen}, {Modigliani}, {Moehler}, {M{\o}ller},
  {Norup S{\o}rensen}, {N{\o}rregaard}, {P{\'e}roux}, {Patat}, {Pena}, {Pragt},
  {Reinero}, {Rigal}, {Riva}, {Roelfsema}, {Royer}, {Sacco}, {Santin},
  {Schoenmaker}, {Spano}, {Sweers}, {Ter Horst}, {Tintori}, {Tromp}, {van
  Dael}, {van der Vliet}, {Venema}, {Vidali}, {Vinther}, {Vola}, {Winters},
  {Wistisen}, {Wulterkens} and {Zacchei}}]{vernet11}
\bibinfo{author}{{Vernet}, J.}, \bibinfo{author}{{Dekker}, H.},
  \bibinfo{author}{{D'Odorico}, S.}, \bibinfo{author}{{Kaper}, L.},
  \bibinfo{author}{{Kjaergaard}, P.}, \bibinfo{author}{{Hammer}, F.},
  \bibinfo{author}{{Randich}, S.}, \bibinfo{author}{{Zerbi}, F.},
  \bibinfo{author}{{Groot}, P.J.}, \bibinfo{author}{{Hjorth}, J.},
  \bibinfo{author}{{Guinouard}, I.}, \bibinfo{author}{{Navarro}, R.},
  \bibinfo{author}{{Adolfse}, T.}, \bibinfo{author}{{Albers}, P.W.},
  \bibinfo{author}{{Amans}, J.P.}, \bibinfo{author}{{Andersen}, J.J.},
  \bibinfo{author}{{Andersen}, M.I.}, \bibinfo{author}{{Binetruy}, P.},
  \bibinfo{author}{{Bristow}, P.}, \bibinfo{author}{{Castillo}, R.},
  \bibinfo{author}{{Chemla}, F.}, \bibinfo{author}{{Christensen}, L.},
  \bibinfo{author}{{Conconi}, P.}, \bibinfo{author}{{Conzelmann}, R.},
  \bibinfo{author}{{Dam}, J.}, \bibinfo{author}{{de Caprio}, V.},
  \bibinfo{author}{{de Ugarte Postigo}, A.}, \bibinfo{author}{{Delabre}, B.},
  \bibinfo{author}{{di Marcantonio}, P.}, \bibinfo{author}{{Downing}, M.},
  \bibinfo{author}{{Elswijk}, E.}, \bibinfo{author}{{Finger}, G.},
  \bibinfo{author}{{Fischer}, G.}, \bibinfo{author}{{Flores}, H.},
  \bibinfo{author}{{Fran{\c c}ois}, P.}, \bibinfo{author}{{Goldoni}, P.},
  \bibinfo{author}{{Guglielmi}, L.}, \bibinfo{author}{{Haigron}, R.},
  \bibinfo{author}{{Hanenburg}, H.}, \bibinfo{author}{{Hendriks}, I.},
  \bibinfo{author}{{Horrobin}, M.}, \bibinfo{author}{{Horville}, D.},
  \bibinfo{author}{{Jessen}, N.C.}, \bibinfo{author}{{Kerber}, F.},
  \bibinfo{author}{{Kern}, L.}, \bibinfo{author}{{Kiekebusch}, M.},
  \bibinfo{author}{{Kleszcz}, P.}, \bibinfo{author}{{Klougart}, J.},
  \bibinfo{author}{{Kragt}, J.}, \bibinfo{author}{{Larsen}, H.H.},
  \bibinfo{author}{{Lizon}, J.L.}, \bibinfo{author}{{Lucuix}, C.},
  \bibinfo{author}{{Mainieri}, V.}, \bibinfo{author}{{Manuputy}, R.},
  \bibinfo{author}{{Martayan}, C.}, \bibinfo{author}{{Mason}, E.},
  \bibinfo{author}{{Mazzoleni}, R.}, \bibinfo{author}{{Michaelsen}, N.},
  \bibinfo{author}{{Modigliani}, A.}, \bibinfo{author}{{Moehler}, S.},
  \bibinfo{author}{{M{\o}ller}, P.}, \bibinfo{author}{{Norup S{\o}rensen}, A.},
  \bibinfo{author}{{N{\o}rregaard}, P.}, \bibinfo{author}{{P{\'e}roux}, C.},
  \bibinfo{author}{{Patat}, F.}, \bibinfo{author}{{Pena}, E.},
  \bibinfo{author}{{Pragt}, J.}, \bibinfo{author}{{Reinero}, C.},
  \bibinfo{author}{{Rigal}, F.}, \bibinfo{author}{{Riva}, M.},
  \bibinfo{author}{{Roelfsema}, R.}, \bibinfo{author}{{Royer}, F.},
  \bibinfo{author}{{Sacco}, G.}, \bibinfo{author}{{Santin}, P.},
  \bibinfo{author}{{Schoenmaker}, T.}, \bibinfo{author}{{Spano}, P.},
  \bibinfo{author}{{Sweers}, E.}, \bibinfo{author}{{Ter Horst}, R.},
  \bibinfo{author}{{Tintori}, M.}, \bibinfo{author}{{Tromp}, N.},
  \bibinfo{author}{{van Dael}, P.}, \bibinfo{author}{{van der Vliet}, H.},
  \bibinfo{author}{{Venema}, L.}, \bibinfo{author}{{Vidali}, M.},
  \bibinfo{author}{{Vinther}, J.}, \bibinfo{author}{{Vola}, P.},
  \bibinfo{author}{{Winters}, R.}, \bibinfo{author}{{Wistisen}, D.},
  \bibinfo{author}{{Wulterkens}, G.}, \bibinfo{author}{{Zacchei}, A.},
  \bibinfo{year}{2011}.
\newblock \bibinfo{title}{{X-shooter, the new wide band intermediate resolution
  spectrograph at the ESO Very Large Telescope}}.
\newblock \bibinfo{journal}{Astron. Astrophys.} \bibinfo{volume}{536},
  \bibinfo{pages}{A105}.
\newblock \DOIprefix\doi{10.1051/0004-6361/201117752}.
\bibitem[{{von Savigny}(2015)}]{savigny15}
\bibinfo{author}{{von Savigny}, C.}, \bibinfo{year}{2015}.
\newblock \bibinfo{title}{{Variability of OH(3-1) emission altitude from 2003
  to 2011: Long-term stability and universality of the emission rate-altitude
  relationship}}.
\newblock \bibinfo{journal}{J. Atmos. Sol.-Terr. Phys.} \bibinfo{volume}{127},
  \bibinfo{pages}{120--128}.
\newblock \DOIprefix\doi{10.1016/j.jastp.2015.02.001}.
\bibitem[{{von Savigny} et~al.(2012){von Savigny}, {McDade}, {Eichmann} and
  {Burrows}}]{savigny12}
\bibinfo{author}{{von Savigny}, C.}, \bibinfo{author}{{McDade}, I.C.},
  \bibinfo{author}{{Eichmann}, K.U.}, \bibinfo{author}{{Burrows}, J.P.},
  \bibinfo{year}{2012}.
\newblock \bibinfo{title}{{On the dependence of the OH$^{*}$ Meinel emission
  altitude on vibrational level: SCIAMACHY observations and model
  simulations}}.
\newblock \bibinfo{journal}{Atmos. Chem. Phys.} \bibinfo{volume}{12},
  \bibinfo{pages}{8813--8828}.
\newblock \DOIprefix\doi{10.5194/acp-12-8813-2012}.
\bibitem[{{Wintoft}(2011)}]{wintoft11}
\bibinfo{author}{{Wintoft}, P.}, \bibinfo{year}{2011}.
\newblock \bibinfo{title}{{The variability of solar EUV: A multiscale
  comparison between sunspot number, 10.7 cm flux, LASP MgII index, and
  SOHO/SEM EUV flux}}.
\newblock \bibinfo{journal}{J. Atmos. Sol.-Terr. Phys.} \bibinfo{volume}{73},
  \bibinfo{pages}{1708--1714}.
\newblock \DOIprefix\doi{10.1016/j.jastp.2011.03.009}.
\bibitem[{{Xu} et~al.(2012){Xu}, {Gao}, {Smith} and {Zhu}}]{xu12}
\bibinfo{author}{{Xu}, J.}, \bibinfo{author}{{Gao}, H.},
  \bibinfo{author}{{Smith}, A.K.}, \bibinfo{author}{{Zhu}, Y.},
  \bibinfo{year}{2012}.
\newblock \bibinfo{title}{{Using TIMED/SABER nightglow observations to
  investigate hydroxyl emission mechanisms in the mesopause region}}.
\newblock \bibinfo{journal}{J. Geophys. Res.} \bibinfo{volume}{117},
  \bibinfo{pages}{D02301}.
\newblock \DOIprefix\doi{10.1029/2011JD016342}.

\end{thebibliography}

\end{document}